\documentclass[10pt]{article}
\usepackage{graphicx}
\usepackage{amsmath}
\usepackage{amssymb}
\usepackage{caption2}
\begin{document}
\begin{center}
{\large {\bf \sc{Analysis of the strong decays of the $Y(4660)$ in tetraquark scenario via the QCD sum rules}}} \\[2mm]
Xiao-Song Yang$^{*\dag}$, Zhi-Gang Wang$^*$\footnote{E-mail: zgwang@aliyun.com.}     \\
Department of Physics, North China Electric Power University, Baoding 071003, P. R. China$^*$
School of Nuclear Science and Engineering, North China Electric Power University, Beijing 102206, P. R. China$^\dag$
\end{center}

\begin{abstract}
Motivated by the enigmatic vector charmonium-like states, we investigate  the strong decay behaviors of four kinds of vector tetraquark states, which are possible candidates for the $Y(4660)$, within the framework of three-point QCD sum rules based on rigorous quark-hadron duality. We take into account the  vacuum condensates  up to dimension 5 on the QCD side, and  obtain the hadronic coupling constants therefore the partial decay widths of those states. The predicted total width $61.5\pm7.3\,\rm{MeV}$ is in excellent  agreement with the experimental data for the  $Y(4660)$, which supports its interpretation as a $[sc][\bar{s}\bar{c}]$ tetraquark state with the $J^{PC}=1^{--}$.
\end{abstract}



\section{Introduction}
In past years, a number of charmonium-like states have been observed \cite{PDG}, which cannot be accommodated in the traditional quark model comfortably. They play an important role in understanding the QCD long distance dynamics and have inspired extensive researches, especially the complicated $Y$ states. In the present study, we will focus on the $Y(4660)$ in the tetraquark scenario.

The $Y(4660)$ was  observed firstly by the Belle collaboration in the process $e^+e^-\rightarrow\pi^+\pi^-\psi(\rm 2S)$ in 2007 \cite{Belle-Y4660-2007} and later confirmed by the Belle, BaBar and BESIII collaborations \cite{BaBar-Y4660-2012,Belle-Y4660-2014,BESIII-Y4660-2021,BESIII-Y4660-2023}. The masses and widths of the $Y(4660)$ and related states from different experiments are presented in Table \ref{experiment}.

In 2008, the Belle collaboration reported a measurement of the exclusive process $e^+e^-\rightarrow\Lambda^+_c\Lambda^-_c$ and observed a significant structure, which was denoted as $Y(4630)$ \cite{Belle-Y4630-2008}. Due to the analogous masses and widths, the $Y(4630)$ and $Y(4660)$ are taken as the same state by the Particle Data Group \cite{PDG} and several work groups \cite{FKGuo-4630,Polosa-4660,DVBugg-4630}.
However, the BESIII collaboration studied the process $e^+e^-\rightarrow\Lambda^+_c\bar{\Lambda}^-_c$ with more statistics in 2023 \cite{BESIII-Y4630-2023},  and the measured cross section indicated no enhancement around the $Y(4630)$ structure, which is different significantly from the Belle collaboration \cite{Belle-Y4630-2008}.

In 2019, the Belle collaboration studied the $e^+e^-\rightarrow D^+_s D_{s1}(2536)^-$ cross section and observed a charmonium-like state $Y(4626)$ having  the measured mass and width close to those of the $Y(4660)$ \cite{Belle-Y4660-2019}, which is the first discovery of the $Y$ state around $4.6\, \rm{GeV}$ in an open charm channel. Later, the $Y(4626)$ was also confirmed in the $e^+e^-\rightarrow D^+_s D^*_{s2}(2573)^-$ channel \cite{Belle-Y4660-2020}.
If the $Y(4626)$ and $Y(4660)$ are the same state \cite{PDG}, it could be assigned as a $[sc][\bar{s}\bar{c}]$ state according to the decay to a $D^+_s D_{s1}^-$ pair.

On the theoretical side, after its discovery, the $Y(4660)$ was  interpreted as
$\psi^\prime f_0(980)$ molecular state \cite{FKGuo-4630,FKGuo-4660,Nielsen-4660-mole,Wang-CTP-4660},
tetraquark state \cite{Polosa-4660,Nielsen-4660,Ebert-4660,ChenZhu,ZhangHuang-PRD,ESantopinto-PRD,Azizi-4660, Maiani-4660,WangY4360Y4660-1803,Wang-tetra-formula,WangEPJC-1601,WZG-NPB-cucd-Vector, WZG-NPB-cscs-Vector,NLi-CPC},
hadro-charmonium state \cite{Hadro-Charm},
$\psi(\rm 5S)$ \cite{MLYan-4660,BSZou-PRD,XHZhong-PRD,LuZhongZhu-4660},
$\psi(\rm 6S)$ \cite{KTChao-4660,XLiu-4660} and so on.

In Ref.\cite{CFQiao-4660}, the authors studied  the mass spectrum  of the prospective hidden-bottom and hidden-charm hexaquark states via the QCD sum rules, and found that the $Y(4660)$ is close in magnitude to the $\Lambda_c \bar{\Lambda}_c$-type baryonium state. The $\Lambda_c \bar{\Lambda}_c$-type baryonium states have been studied by several works \cite{CFQiao-4260,CFQiao-baryonium,JHe-baryonium,XLiu-baryonium}.
In Ref.\cite{WZG-baryonium}, we also studied $\Lambda_c \bar{\Lambda}_c$-type baryonium states via the QCD sum rules, and obtained  conclusion consistent with that of Ref.\cite{CFQiao-4660}. The interpretation of the $Y(4660)$ in the baryonium scenario serves as a meaningful insight in understanding  the $Y$ states.

In Table \ref{QCDSR}, we list out the predictions of the masses of the $Y(4660)$ in the framework of the QCD sum rules, which have made several successful achievements on the exotic states \cite{WZG-review,WangZG-landau-PRD,Nielsen-JPG-Review}. It is distinct that the experimental mass of the $Y(4660)$ can be reproduced in different structures, and thus we should explore the decay widths to shed light on its nature.  Or the $Y(4660)$ might have several important Fock states, such as molecules, tetraquarks, etc, and embodies the collective effects.

Considering  the $Y(4260)$ and $Y(4660)$ decaying into the $J/\psi\pi^+\pi^-$ and $\psi({\rm 2S})\pi^+\pi^-$ respectively \cite{Belle-Y4660-2007,BaBar-Y4260-2005}, we might  assign the $Y(4660)$ as the radial excitation of the $Y(4260)$ \cite{Maiani-4660}, unfortunately, the process $Y(4230)\to \psi({\rm 2S}) \pi^+\pi^-$ was observed by the BESIII collaboration \cite{BESIII-Y4260pis23-2017,BESIII-Y4260pis23-2021}.
In Ref.\cite{WZG-CPC-cucd-Pwave}, we investigated the ground states and first radial excitations of the vector hidden-charm tetraquark states with an explicit P-wave via the QCD sum rules. The results support assigning the $Y(4260)$ as the 1P tetraquark state, and there is no room to accommodate the $Y(4660)$ in the first radial excitations.

In Refs.\cite{WangY4360Y4660-1803,Wang-tetra-formula,WangEPJC-1601,WZG-CPC-cucd-Pwave,WZG-4260-Pwave,WZG-vector-Pwave}, we studied the mass spectrum of the vector hidden-charm tetraquark states with or without an explicit P-wave via the QCD sum rules, which support assigning the $Y(4660)$ to be the $[sc]_P[\bar{s}\bar{c}]_A-[sc]_A[\bar{s}\bar{c}]_P$ type tetraquark state. In Ref.\cite{WZG-Y4660-Decay}, we studied the decay behaviors of the $[sc]_P[\bar{s}\bar{c}]_A-[sc]_A[\bar{s}\bar{c}]_P$ type tetraquark state.
Later, in Refs.\cite{WZG-NPB-cucd-Vector,WZG-NPB-cscs-Vector}, we took  the scalar $(S)$, pseudoscalar $(P)$, vector $(V)$, axial-vector $(A)$ and tensor $(\widetilde{A},\widetilde{V})$ (anti)diquark operators as the elementary building blocks to construct local four-quark currents with or without hidden-strangeness in a comprehensive and consistent way, and updated the old calculations. The $\widetilde{A}$ and $\widetilde{V}$ denote the $J^P=1^+$ and $J^P=1^-$ components of the tensor diquarks $\varepsilon^{ijk}q^T_j C\sigma_{\mu\nu}c_k$ or $\varepsilon^{ijk}q^T_j C\sigma_{\mu\nu}\gamma_5 c_k$.
We studied the mass spectrum of those tetraquark states with the $J^{PC}=1^{--}$ and $1^{-+}$ in details and revisited the assignments of the $Y$ states. The results prefer assigning the $Y(4660)$ to be the $[sc]_{\widetilde{A}}[\bar{s}\bar{c}]_V+[sc]_V[\bar{s}\bar{c}]_{\widetilde{A}}$, $[sc]_S[\bar{s}\bar{c}]_{\widetilde{V}}-[sc]_{\widetilde{V}}[\bar{s}\bar{c}]_S$, $[uc]_P[\bar{u}\bar{c}]_A+[dc]_P[\bar{d}\bar{c}]_A-[uc]_A[\bar{u}\bar{c}]_P-[dc]_A[\bar{d}\bar{c}]_P$ or $[uc]_A[\bar{u}\bar{c}]_A+[dc]_A[\bar{d}\bar{c}]_A$ type tetraquark states.

In the present work, we take account of those  four tetraquark configurations to explore the strong decay behaviors of the $Y(4660)$. We study the hadronic coupling constants and also partial decay widths of the two-body strong decays, which can take place through the Okubo-Zweig-Iizuka super-allowed fall-apart mechanism without annihilating and creating a quark-antiquark pair.
In detail, we take account of the channels $Y\to \bar{D}_{(s)}D_{(s)}$, $\bar{D}^*_{(s)}D_{(s)}$, $\bar{D}^*_{(s)}D^*_{(s)}$, $\bar{D}_{(s)0}D^*_{(s)}$, $\bar{D}_{(s)1}D_{(s)}$, $\eta_c\omega(\phi(1020))$, $J/\psi\omega(\phi(1020))$, $\chi_{c0}\omega(\phi(1020))$, $\chi_{c1}\omega(\phi(1020))$ and $J/\psi f_0(500)(f_0(980))$.

The article is organized as follows: in Section 2, we obtain the hadronic coupling constants in the two-body strong decays of the four vector tetraquark states via QCD sum rules; in Section 3, we present the numerical results and discussions; finally, the conclusion  is presented in Section 4.

\begin{table}
\begin{center}
\begin{tabular}{|c|c|c|c|c|c|c|c|}\hline\hline

Year  &           &Mass (MeV)                    &Width (MeV)                     &channel      &Experiment     \\ \hline\hline

2007  &$Y(4660)$  &$4664\pm 11\pm 5$             &$48\pm 15\pm 3$                 &$e^+e^-\rightarrow\pi^+\pi^-\psi(2{\rm S})$      &Belle\cite{Belle-Y4660-2007}    \\  \hline

2008  &$Y(4630)$  &$4634^{+8}_{-7}{}^{+5}_{-8}$  &$92^{+40}_{-24}{}^{+10}_{-21}$  &$e^+e^-\rightarrow\Lambda^+_c\Lambda^-_c$      &Belle\cite{Belle-Y4630-2008}    \\
2023  &           &not seen                      &not seen                        &           &BESIII\cite{BESIII-Y4630-2023} \\ \hline
2012  &$Y(4660)$  &$4669\pm 21\pm 3$              &$104\pm 48\pm 10 $                &$e^+e^-\rightarrow\psi(2{\rm S})\pi^+\pi^-$      &BaBar\cite{BaBar-Y4660-2012}     \\ \hline
2014  &$Y(4660)$  &$4652\pm10\pm 8$              &$68\pm 11\pm 1$                 &$e^+e^-\rightarrow\pi^+\pi^-\psi(2{\rm S})$      &Belle\cite{Belle-Y4660-2014}    \\  \hline
2019  &$Y(4626)$  &$4625.9^{+6.2}_{-6.0}\pm0.4$  &$49.8^{+13.9}_{-11.5}\pm4.0$    &$e^+e^-\rightarrow D^+_s D_{s1}(2536)^-$      &Belle \cite{Belle-Y4660-2019}    \\  \hline
2020  &$Y(4626)$  &$4619.8^{+8.9}_{-8.0}\pm2.3$  &$47.0^{+31.3}_{-14.8}\pm4.6$    &$e^+e^-\rightarrow D^+_s D^*_{s2}(2573)^-$      &Belle \cite{Belle-Y4660-2020}    \\  \hline
2021  &$Y(4660)$  &$4651.0\pm37.8\pm2.1$  &$155.4\pm24.8\pm0.8$    &$e^+e^-\rightarrow \pi^+\pi^-\psi(3686)$      &BESIII\cite{BESIII-Y4660-2021}    \\  \hline
2023  &$Y(4660)$  &$4675.3\pm29.5$  &$218.3\pm72.9$    &$e^+e^-\rightarrow D^{*0} D^{*-}\pi^+$      &BESIII\cite{BESIII-Y4660-2023}    \\  \hline\hline
\end{tabular}
\end{center}
\caption{ The masses, widths and channels of different experiments. }\label{experiment}
\end{table}

\begin{table}
\begin{center}
\begin{tabular}{|c|c|c|c|c|c|c|c|}\hline\hline
           &Structures                                                      &OPE\,(No)    &mass(GeV)      &References   \\ \hline

$Y(4660)$  &$\psi^\prime f_0(980)$                                          &$10$         &$4.71$        &\cite{Wang-CTP-4660}  \\
     &                                          &$6$          &$4.67$    &\cite{Nielsen-4660-mole}  \\ \hline

$Y(4660)$  &$[sc]_S[\bar{s}\bar{c}]_V+[sc]_V[\bar{s}\bar{c}]_S$             &$8\,(7)$     &$4.65$         &\cite{Nielsen-4660}  \\
  &             &$10$         &$4.68$         &\cite{Azizi-4660}  \\  \hline

$Y(4660)$  &$[sc]_{\widetilde{V}}[\bar{s}\bar{c}]_A-[sc]_A[\bar{s}\bar{c}]_{\widetilde{V}}$             &$8\,(7)$     &$4.64$         &\cite{ChenZhu}  \\ \hline

$Y(4660)$  &$[qc]_S[\bar{q}\bar{c}]_V+[qc]_V[\bar{q}\bar{c}]_S$             &$8\,(7)$     &$4.64$         &\cite{ChenZhu}  \\
$Y(4360)$  &$[qc]_S[\bar{q}\bar{c}]_V+[qc]_V[\bar{q}\bar{c}]_S$             &$10$         &$4.34$  &\cite{WangY4360Y4660-1803}  \\ \hline

$Y(4660)$  &$[sc]_P[\bar{s}\bar{c}]_A-[sc]_A[\bar{s}\bar{c}]_P$             &$10$         &$4.70$   &\cite{Wang-tetra-formula}  \\
  &             &         &$4.66$  &\cite{WangY4360Y4660-1803}  \\ \hline

  &             &         &$4.66$   &\cite{Wang-tetra-formula}   \\
$Y(4660)$  &$[qc]_P[\bar{q}\bar{c}]_A-[qc]_A[\bar{q}\bar{c}]_P$             &$10$         &$4.59$  &\cite{WangY4360Y4660-1803}  \\
  &             &          &$4.66$ &\cite{WZG-NPB-cucd-Vector} \\  \hline

$Y(4660)$  &$[qc]_A[\bar{q}\bar{c}]_A$                                      &$10$         &$4.66$        &\cite{WangEPJC-1601}  \\
  &                                     &          &$4.69$ &\cite{WZG-NPB-cucd-Vector}  \\ \hline

$Y(4660)$  &$[sc]_S[\bar{s}\bar{c}]_S$                                      &$6$          &$4.69$       &\cite{ZhangHuang-PRD}  \\ \hline

$Y(4660)$  &$[sc]_{\widetilde{A}}[\bar{s}\bar{c}]_V+[sc]_V[\bar{s}\bar{c}]_{\widetilde{A}}$             &$10$          &$4.65$ &\cite{WZG-NPB-cscs-Vector} \\ \hline
$Y(4660)$  &$[sc]_S[\bar{s}\bar{c}]_{\widetilde{V}}-[sc]_{\widetilde{V}}[\bar{s}\bar{c}]_S$                                     &$10$          &$4.68$ &\cite{WZG-NPB-cscs-Vector}  \\ \hline
$Y(4660)$  &$\Lambda_c \bar{\Lambda}_c$      &$12$    &$4.78$ &\cite{CFQiao-4660} \\ \hline
$Y(4660)$  &$\Lambda_c \bar{\Lambda}_c$      &$16$    &$4.68$ &\cite{WZG-baryonium} \\ \hline
\hline
\end{tabular}
\end{center}
\caption{ The  masses from the QCD sum rules with different quark structures, where the OPE denotes  truncations of  the operator product expansion up to the vacuum condensates of dimension $n$, the No denotes that the vacuum condensates of dimension $n^\prime$ are not included.   }\label{QCDSR}
\end{table}

\section{QCD sum rules for the hadronic coupling constants}
We choose the following four tetraquark currents with the quantum numbers $J^{PC}=1^{--}$ to study the $Y(4660)$,
\begin{eqnarray}\label{PA-Current}
J_{\mu}^{PA}(x)&=&\frac{\varepsilon^{ijk}\varepsilon^{imn}}{2}
\Big[u^{T}_j(x)C c_k(x)\bar{u}_m(x)\gamma_\mu C \bar{c}^{T}_n(x)
+d^{T}_j(x)C c_k(x)\bar{d}_m(x)\gamma_\mu C \bar{c}^{T}_n(x)\nonumber\\
&&
-u^{T}_j(x)C\gamma_\mu c_k(x)\bar{u}_m(x) C \bar{c}^{T}_n(x)
-d^{T}_j(x)C\gamma_\mu c_k(x)\bar{d}_m(x) C \bar{c}^{T}_n(x) \Big] \, ,
\end{eqnarray}
\begin{eqnarray}\label{AA-Current}
J_{\mu\nu}^{AA}(x)&=&\frac{\varepsilon^{ijk}\varepsilon^{imn}}{2}
\Big[u^{T}_j(x)C\gamma_\mu c_k(x)\bar{u}_m(x)\gamma_\nu C \bar{c}^{T}_n(x)
+d^{T}_j(x)C\gamma_\mu c_k(x)\bar{d}_m(x)\gamma_\nu C \bar{c}^{T}_n(x)\nonumber\\
&&
-u^{T}_j(x)C\gamma_\nu c_k(x)\bar{u}_m(x)\gamma_\mu C \bar{c}^{T}_n(x)
-d^{T}_j(x)C\gamma_\nu c_k(x)\bar{d}_m(x)\gamma_\mu C \bar{c}^{T}_n(x) \Big] \, ,
\end{eqnarray}
\begin{eqnarray}\label{AV-Current}
J_{\mu}^{\widetilde{A}V}(x)&=&\frac{\varepsilon^{ijk}\varepsilon^{imn}}{\sqrt{2}}
\Big[s^{T}_j(x)C\sigma_{\mu\nu}\gamma_5 c_k(x)\bar{s}_m(x)\gamma_5\gamma^\nu C \bar{c}^{T}_n(x) +s^{T}_j(x)C\gamma^\nu\gamma_5 c_k(x)\bar{s}_m(x)\gamma_5\sigma_{\mu\nu} C \bar{c}^{T}_n(x) \Big] \, ,\nonumber \\
&&
\end{eqnarray}
\begin{eqnarray}\label{SV-Current}
J^{S\widetilde{V}}_{\mu\nu}(x)&=&\frac{\varepsilon^{ijk}\varepsilon^{imn}}{\sqrt{2}}
\Big[s^{T}_j(x)C\gamma_5 c_k(x)  \bar{s}_m(x)\sigma_{\mu\nu} C \bar{c}^{T}_n(x) - s^{T}_j(x)C\sigma_{\mu\nu} c_k(x)  \bar{s}_m(x)\gamma_5 C \bar{c}^{T}_n(x) \Big] \, ,\nonumber \\
&&
\end{eqnarray}
where the $i$, $j$, $k$, $m$ and $n$ are color indices, the superscripts  $S$, $P$, $A$($\widetilde{A}$) and $V$($\widetilde{V}$) denote the scalar, pseudoscalar, axialvector and vector diquarks or antidiquarks, respectively.
For the conventional mesons, we adopt the currents,
\begin{eqnarray}
J^{\bar{D}}(x)&=&\bar{c}(x)i\gamma_{5} u(x)  \, ,\nonumber \\
J^{D}(y)&=&\bar{u}(y)i\gamma_{5} c(y) \, ,\nonumber \\
J_{\alpha}^{\bar{D}^*}(x)&=&\bar{c}(x)\gamma_{\alpha} u(x)  \, ,\nonumber \\
J_{\beta}^{D^*}(y)&=&\bar{u}(y)\gamma_{\beta} c(y) \, ,\nonumber \\
J^{\bar{D}_0}(x)&=&\bar{c}(x) u(x) \, ,\nonumber \\
J_{\alpha}^{\bar{D}_1}(x)&=&\bar{c}(x)\gamma_\alpha \gamma_5 u(x)\, ,\nonumber \\
J^{\bar{D}_s}(x)&=&\bar{c}(x)i\gamma_{5} s(x)  \, ,\nonumber \\
J^{D_s}(y)&=&\bar{s}(y)i\gamma_{5} c(y) \, ,\nonumber \\
J_{\alpha}^{\bar{D}^*_s}(x)&=&\bar{c}(x)\gamma_{\alpha} s(x)  \, ,\nonumber \\
J_{\beta}^{D^*_s}(y)&=&\bar{s}(y)\gamma_{\beta} c(y) \, ,\nonumber \\
J^{\bar{D}_{s0}}(x)&=&\bar{c}(x) s(x) \, ,\nonumber \\
J_{\alpha}^{\bar{D}_{s1}}(x)&=&\bar{c}(x)\gamma_\alpha \gamma_5 s(x) \, ,\nonumber \\
J^{\eta_c}(x)&=&\bar{c}(x)i \gamma_5 c(x) \, ,\nonumber \\
J_{\alpha}^{J/\psi}(x)&=&\bar{c}(x)\gamma_{\alpha} c(x)  \, ,\nonumber \\
J^{\chi_{c0}}(x)&=&\bar{c}(x) c(x) \, ,\nonumber \\
J_{\alpha}^{\chi_{c1}}(x)&=&\bar{c}(x)\gamma_{\alpha}\gamma_5 c(x)  \, ,\nonumber \\
J_{\alpha}^{\omega}(y)&=&\frac{\bar{u}(y) \gamma_\alpha u(y)+\bar{d}(y) \gamma_\alpha d(y)}{\sqrt{2}} \, ,\nonumber \\
J_{\alpha}^{\phi(1020)}(y)&=&\bar{s}(y) \gamma_\alpha s(y) \, ,\nonumber \\
J^{f_0(500)}(y)&=&\frac{\bar{u}(y)  u(y)+\bar{d}(y)  d(y)}{\sqrt{2}} \, ,\nonumber \\
J^{f_0(980)}(y)&=& \bar{s}(y)  s(y) \, .
\end{eqnarray}

Now, we introduce the following three-point correlation functions to study the hadronic coupling constants in the QCD sum rules,
\begin{eqnarray}
\Pi^{\bar{D}D PA}_{\mu}(p,q)&=&i^2\int d^4xd^4y \, e^{ip\cdot x}e^{iq\cdot y}\, \langle 0|T\left\{J^{\bar{D}}(x)J^{D}(y)J_{\mu}^{PA}{}^\dagger(0)\right\}|0\rangle\, ,
\end{eqnarray}

\begin{eqnarray}
\Pi^{\bar{D}^*D PA}_{\alpha\mu}(p,q)&=&i^2\int d^4xd^4y \, e^{ip\cdot x}e^{iq\cdot y}\, \langle 0|T\left\{J_{\alpha}^{\bar{D}^*}(x)J^{D}(y)J_{\mu}^{PA}{}^\dagger(0)\right\}|0\rangle\, ,
\end{eqnarray}

\begin{eqnarray}
\Pi^{\bar{D}^*D^* PA}_{\alpha\beta\mu}(p,q)&=&i^2\int d^4xd^4y \, e^{ip\cdot x}e^{iq\cdot y}\, \langle 0|T\left\{J_{\alpha}^{\bar{D}^*}(x)J_\beta^{D^*}(y)J_{\mu}^{PA}{}^\dagger(0)\right\}|0\rangle\, ,
\end{eqnarray}

\begin{eqnarray}
\Pi^{\bar{D}_0D^* PA}_{\alpha\mu}(p,q)&=&i^2\int d^4xd^4y \, e^{ip\cdot x}e^{iq\cdot y}\, \langle 0|T\left\{J^{\bar{D}_0}(x)J_\alpha^{D^*}(y)J_{\mu}^{PA}{}^\dagger(0)\right\}|0\rangle\, ,
\end{eqnarray}

\begin{eqnarray}
\Pi^{\bar{D}_1D PA}_{\alpha\mu}(p,q)&=&i^2\int d^4xd^4y \, e^{ip\cdot x}e^{iq\cdot y}\, \langle 0|T\left\{J_{\alpha}^{\bar{D}_1}(x)J^{D}(y)J_{\mu}^{PA}{}^\dagger(0)\right\}|0\rangle\, ,
\end{eqnarray}

\begin{eqnarray}
\Pi^{\eta_c\omega PA}_{\alpha\mu}(p,q)&=&i^2\int d^4xd^4y \, e^{ip\cdot x}e^{iq\cdot y}\, \langle 0|T\left\{J^{\eta_c}(x)J_\alpha^{\omega}(y)J_{\mu}^{PA}{}^\dagger(0)\right\}|0\rangle\, ,
\end{eqnarray}

\begin{eqnarray}
\Pi^{J/\psi\omega PA}_{\alpha\beta\mu}(p,q)&=&i^2\int d^4xd^4y \, e^{ip\cdot x}e^{iq\cdot y}\, \langle 0|T\left\{J_{\alpha}^{J/\psi}(x)J_\beta^{\omega}(y)J_{\mu}^{PA}{}^\dagger(0)\right\}|0\rangle\, ,
\end{eqnarray}

\begin{eqnarray}
\Pi^{\chi_{c0}\omega PA}_{\alpha\mu}(p,q)&=&i^2\int d^4xd^4y \, e^{ip\cdot x}e^{iq\cdot y}\, \langle 0|T\left\{J^{\chi_{c0}}(x)J_\alpha^{\omega}(y)J_{\mu}^{PA}{}^\dagger(0)\right\}|0\rangle\, ,
\end{eqnarray}

\begin{eqnarray}
\Pi^{\chi_{c1}\omega PA}_{\alpha\beta\mu}(p,q)&=&i^2\int d^4xd^4y \, e^{ip\cdot x}e^{iq\cdot y}\, \langle 0|T\left\{J_\alpha^{\chi_{c1}}(x)J_\beta^{\omega}(y)J_{\mu}^{PA}{}^\dagger(0)\right\}|0\rangle\, ,
\end{eqnarray}

\begin{eqnarray}
\Pi^{J/\psi f_0(500) PA}_{\alpha\mu}(p,q)&=&i^2\int d^4xd^4y \, e^{ip\cdot x}e^{iq\cdot y}\, \langle 0|T\left\{J_\alpha^{J/\psi}(x)J^{f_0(500)}(y)J_{\mu}^{PA}{}^\dagger(0)\right\}|0\rangle\, .
\end{eqnarray}

We can take the replacements $\mu \to \mu\nu$, $PA \to AA$  for the current $J_{\mu\nu}^{AA}$;
$(\bar{D}$, $D$, $\bar{D}^*$, $D^*$, $\bar{D}_0$, $\bar{D}_1$, $\omega$, $f_0(500)) \to (\bar{D}_s$, $D_s$, $\bar{D}^*_s$, $D^*_s$, $\bar{D}_{s0}$, $\bar{D}_{s1}$, $\phi(1020)$, $f_0(980))$, $PA \to \widetilde{A}V$ for the current $J_{\mu}^{\widetilde{A}V}$;
$\mu \to \mu\nu$, $(\bar{D}$, $D$, $\bar{D}^*$, $D^*$, $\bar{D}_0$, $\bar{D}_1$, $\omega$, $f_0(500)) \to (\bar{D}_s$, $D_s$, $\bar{D}^*_s$, $D^*_s$, $\bar{D}_{s0}$, $\bar{D}_{s1}$, $\phi(1020)$, $f_0(980))$, $PA \to S\widetilde{V}$ for the current $J_{\mu\nu}^{S\widetilde{V}}$ to obtain other  correlation functions.

On  the phenomenological side, we insert a complete set of intermediate hadronic states with same quantum numbers as the interpolating currents and take the following  definitions of the decay constants or pole residues,
\begin{eqnarray}
\langle 0| J^{S}(0)|S(p)\rangle&=&f_{S} m_{S}\, , \nonumber \\
\langle 0| J^{P}(0)|P(p)\rangle&=&\frac{f_P m_P^2}{m_c}\, , \nonumber \\
\langle 0| J_\alpha^{A}(0)|A(p)\rangle&=&f_{A} m_{A}\xi_\alpha\, , \nonumber \\
\langle 0| J_\alpha^{V}(0)|V(p)\rangle&=&f_{V} m_{V}\xi_\alpha\, ,
\end{eqnarray}
\begin{eqnarray}
\langle 0| J_{\mu}^{PA/\widetilde{A}V}(0)|Y_{PA/\widetilde{A}V}(p^\prime)\rangle&=& \lambda_{PA/\widetilde{A}V}\,\varepsilon_\mu\, ,\nonumber \\
\langle 0| J_{\mu\nu}^{S\widetilde{V}}(0)|Y_{S\widetilde{V}}(p^\prime)\rangle&=&
\lambda_{S\widetilde{V}}\,\varepsilon_{\mu\nu\alpha\beta}\,\varepsilon^\alpha p^{\prime\beta}\, , \nonumber\\
\langle 0| J_{\mu\nu}^{S\widetilde{V}}(0)|X_{S\widetilde{A}}(p^\prime)\rangle&=&
\bar{\lambda}_{S\widetilde{A}}\left(\varepsilon_{\mu}p^{\prime}_\nu -\varepsilon_{\nu}p^{\prime}_\mu\right)\, ,\nonumber\\
\langle 0| J_{\mu\nu}^{AA}(0)|Y_{AA}(p^\prime)\rangle&=&
\lambda_{AA}\left(\varepsilon_{\mu}p^{\prime}_\nu -\varepsilon_{\nu}p^{\prime}_\mu\right)\, , \nonumber\\
\langle 0| J_{\mu\nu}^{AA}(0)|X_{AA}(p^\prime)\rangle&=&
\bar{\lambda}_{AA}\,\varepsilon_{\mu\nu\alpha\beta}\,\varepsilon^\alpha p^{\prime\beta}\, ,
\end{eqnarray}
where the $\xi_\mu$ and $\varepsilon_{\mu}$ denote the polarization vectors of the corresponding mesons or tetraquark states; the $S$ denotes the scalar mesons, including $D_0$, $D_{s0}$, $\chi_{c0}$, $f_{0}(500)$ and $f_{0}(980)$; the $P$ denotes the pseudoscalar mesons, including $D$, $D_s$ and $\eta_c$ (note that we should take the replacement $m_c\to m_c+m_s$ for the $D_s$ meson and $m_c\to 2m_c$ for the $\eta_c$ meson); the $A$ denotes the axialvector mesons, including $D_1$, $D_{s1}$ and $\chi_{c1}$; the $V$ denotes the vector mesons, including $D^*$, $D_s^*$, $J/\psi$, $\omega$ and $\phi(1020)$. The $X_{AA}$ and $X_{S\widetilde{A}}$ denote the tetraquark states with the $J^{PC}=1^{+-}$, which can couple to the currents $J_{\mu\nu}^{AA}$ and $J_{\mu\nu}^{S\widetilde{V}}$, respectively.
Furthermore, the hadronic coupling constants are defined as follows,
\begin{eqnarray}\label{Coupling-1}
\langle \bar{D}(p)D(q)|Y_{PA}(p^\prime)\rangle&=&i(p-q)\cdot \varepsilon \,G_{\bar{D}D PA}\, , \nonumber\\
\langle \bar{D}(p)D(q)|Y_{AA}(p^\prime)\rangle&=&-i(p-q)\cdot \varepsilon \,G_{\bar{D}D AA}\, , \nonumber\\
\langle \bar{D}_s(p)D_s(q)|Y_{\widetilde{A}V}(p^\prime)\rangle&=&(p-q)\cdot \varepsilon \,G_{\bar{D}_s D_s\widetilde{A}V}\, , \nonumber\\
\langle \bar{D}_s(p)D_s(q)|Y_{S\widetilde{V}}(p^\prime)\rangle&=&i(p-q)\cdot \varepsilon \,G_{\bar{D}_s D_s S\widetilde{V}}\, ,
\end{eqnarray}
\begin{eqnarray}\label{Coupling-2}
\langle \bar{D}^*(p)D(q)|Y_{PA}(p^\prime)\rangle&=&-\varepsilon^{\lambda\tau\rho\sigma}
p_\lambda \xi^*_\tau p^\prime_\rho \varepsilon_\sigma \,G_{\bar{D}^*D PA}\, , \nonumber\\
\langle \bar{D}^*(p)D(q)|Y_{AA}(p^\prime)\rangle&=&-i\varepsilon^{\lambda\tau\rho\sigma}
p_\lambda \xi^*_\tau p^\prime_\rho \varepsilon_\sigma \,G_{\bar{D}^*D AA}\, , \nonumber\\
\langle \bar{D}_s^*(p)D_s(q)|Y_{\widetilde{A}V}(p^\prime)\rangle&=&-\varepsilon^{\lambda\tau\rho\sigma}
p_\lambda \xi^*_\tau p^\prime_\rho \varepsilon_\sigma \,G_{\bar{D}_s^* D_s\widetilde{A}V}\, , \nonumber\\
\langle \bar{D}_s^*(p)D_s(q)|Y_{S\widetilde{V}}(p^\prime)\rangle&=&-i
\varepsilon^{\lambda\tau\rho\sigma}
p_\lambda \xi^*_\tau p^\prime_\rho \varepsilon_\sigma \,G_{\bar{D}^*DS\widetilde{V}}\, .
\end{eqnarray}
For simplicity, the definitions for other hadronic coupling constants are given in Appendix \ref{app-couplingconstants}.

Then we isolate the ground state contributions to get the following correlation functions \cite{SVZ79,Reinders85},
\begin{eqnarray}
\Pi^{\bar{D}D PA}_{\mu}(p,q)&=&\Pi_{\bar{D}D PA}(p^{\prime2},p^2,q^2)
\,\left(q-p\right)_\mu+\cdots\, ,
\end{eqnarray}

\begin{eqnarray}
\Pi^{\bar{D}^*D PA}_{\alpha\mu}(p,q)&=&
\Pi_{\bar{D}^*D PA}(p^{\prime2},p^2,q^2)
\,\left(-i\varepsilon_{\alpha\mu\lambda\tau}p^\lambda q^\tau\right)+\cdots\, ,
\end{eqnarray}

\begin{eqnarray}
\Pi^{\bar{D}^*D^* PA}_{\alpha\beta\mu}(p,q)&=&
\Pi_{\bar{D}^*D^* PA}(p^{\prime2},p^2,q^2)
\,\left(-g_{\alpha\beta}p_\mu\right)+\cdots\, ,
\end{eqnarray}

\begin{eqnarray}
\Pi^{\bar{D}_0D^* PA}_{\alpha\mu}(p,q)&=&
\Pi_{\bar{D}_0D^* PA}(p^{\prime2},p^2,q^2)
\,\left(-ig_{\alpha\mu}p \cdot q\right)+\cdots\, ,
\end{eqnarray}

\begin{eqnarray}
\Pi^{\bar{D}_1D PA}_{\alpha\mu}(p,q)&=&
\Pi_{\bar{D}_1D PA}(p^{\prime2},p^2,q^2)
\,\left(ig_{\alpha\mu}\right)+\cdots\, ,
\end{eqnarray}

\begin{eqnarray}
\Pi^{\eta_c\omega PA}_{\alpha\mu}(p,q)&=&
\Pi_{\eta_c\omega PA}(p^{\prime2},p^2,q^2)
\,\left(\varepsilon_{\alpha\mu\lambda\tau}p^\lambda q^\tau\right)+\cdots\, ,
\end{eqnarray}

\begin{eqnarray}
\Pi^{J/\psi\omega PA}_{\alpha\beta\mu}(p,q)&=&
\Pi_{J/\psi\omega PA}(p^{\prime2},p^2,q^2)\,
\left(ig_{\alpha\beta}p_\mu\right)+\cdots\, ,
\end{eqnarray}

\begin{eqnarray}
\Pi^{\chi_{c0}\omega PA}_{\alpha\mu}(p,q)&=&
\Pi_{\chi_{c0}\omega PA}(p^{\prime2},p^2,q^2)
\,\left(g_{\alpha\mu}\right)+\cdots\, ,
\end{eqnarray}

\begin{eqnarray}
\Pi^{\chi_{c1}\omega PA}_{\alpha\beta\mu}(p,q)&=&
\Pi_{\chi_{c1}\omega PA}(p^{\prime2},p^2,q^2)
\,\left(-i\varepsilon_{\alpha\beta\mu\lambda}p^\lambda \,p \cdot q\right)+\cdots\, ,
\end{eqnarray}

\begin{eqnarray}
\Pi^{J/\psi f_0(500) PA}_{\alpha\mu}(p,q)&=&
\Pi_{J/\psi f_0(500) PA}(p^{\prime2},p^2,q^2)
\,\left(-g_{\alpha\mu}\right)+\cdots\, .
\end{eqnarray}
Other correlation functions for the currents $J_{\mu\nu}^{AA}$, $J_{\mu}^{\widetilde{A}V}$ and $J_{\mu\nu}^{S\widetilde{V}}$ are presented in Appendix \ref{app-correlationfunctions}. In the above equations, the scalar invariant components are expressed as follows,
\begin{eqnarray}
\Pi_{\bar{D}D PA}(p^{\prime2},p^2,q^2)&=&
\frac{\lambda_{\bar{D}D PA}}{(m_Y^2-p^{\prime2})(m_D^2-p^2)(m_D^2-q^2)}
+\frac{C_{\bar{D}D PA}}{(m_D^2-p^2)(m_D^2-q^2)}\nonumber\\
&&+\cdots\, ,
\end{eqnarray}
\begin{eqnarray}
\Pi_{\bar{D}^*D PA}(p^{\prime2},p^2,q^2)&=&
\frac{\lambda_{\bar{D}^*D PA}}{(m_Y^2-p^{\prime2})(m_{D^*}^2-p^2)(m_D^2-q^2)}
+\frac{C_{\bar{D}^*D PA}}{(m_{D^*}^2-p^2)(m_D^2-q^2)}\nonumber\\
&&+\cdots\, ,
\end{eqnarray}
\begin{eqnarray}
\Pi_{\bar{D}^*D^* PA}(p^{\prime2},p^2,q^2)&=&
\frac{\lambda_{\bar{D}^*D^* PA}}{(m_Y^2-p^{\prime2})(m_{D^*}^2-p^2)(m_{D^*}^2-q^2)}
+\frac{C_{\bar{D}^*D^* PA}}{(m_{D^*}^2-p^2)(m_{D^*}^2-q^2)}\nonumber\\
&&+\cdots\, ,
\end{eqnarray}
\begin{eqnarray}
\Pi_{\bar{D}_0D^* PA}(p^{\prime2},p^2,q^2)&=&
\frac{\lambda_{\bar{D}_0D^* PA}}{(m_Y^2-p^{\prime2})(m_{D_0}^2-p^2)(m_{D^*}^2-q^2)}
+\frac{C_{\bar{D}_0D^* PA}}{(m_{D_0}^2-p^2)(m_{D^*}^2-q^2)}\nonumber\\
&&+\cdots\, ,
\end{eqnarray}
\begin{eqnarray}
\Pi_{\bar{D}_1D PA}(p^{\prime2},p^2,q^2)&=&
\frac{\lambda_{\bar{D}_1D PA}}{(m_Y^2-p^{\prime2})(m_{D_1}^2-p^2)(m_{D}^2-q^2)}
+\frac{C_{\bar{D}_1D PA}}{(m_{D_1}^2-p^2)(m_{D}^2-q^2)}\nonumber\\
&&+\cdots\, ,
\end{eqnarray}

\begin{eqnarray}
\Pi_{\eta_c\omega  PA}(p^{\prime2},p^2,q^2)&=&
\frac{\lambda_{\eta_c\omega PA}}{(m_Y^2-p^{\prime2})
(m_{\eta_c}^2-p^2)(m_{\omega}^2-q^2)}
+\frac{C_{\eta_c\omega PA}}{(m_{\eta_c}^2-p^2)(m_{\omega}^2-q^2)}\nonumber\\
&&+\cdots\, ,
\end{eqnarray}

\begin{eqnarray}
\Pi_{J/\psi\omega  PA}(p^{\prime2},p^2,q^2)&=&
\frac{\lambda_{J/\psi\omega PA}}{(m_Y^2-p^{\prime2})
(m_{J/\psi}^2-p^2)(m_{\omega}^2-q^2)}
+\frac{C_{J/\psi\omega PA}}{(m_{J/\psi}^2-p^2)(m_{\omega}^2-q^2)}\nonumber\\
&&+\cdots\, ,
\end{eqnarray}
\begin{eqnarray}
\Pi_{\chi_{c0}\omega PA}(p^{\prime2},p^2,q^2)&=&
\frac{\lambda_{\chi_{c0}\omega PA}}{(m_Y^2-p^{\prime2})
(m_{\chi_{c0}}^2-p^2)(m_{\omega}^2-q^2)}
+\frac{C_{\chi_{c0}\omega PA}}{(m_{\chi_{c0}}^2-p^2)(m_{\omega}^2-q^2)}\nonumber\\
&&+\cdots\, ,
\end{eqnarray}
\begin{eqnarray}
\Pi_{\chi_{c1}\omega PA}(p^{\prime2},p^2,q^2)&=&
\frac{\lambda_{\chi_{c1}\omega PA}}{(m_Y^2-p^{\prime2})
(m_{\chi_{c1}}^2-p^2)(m_{\omega}^2-q^2)}
+\frac{C_{\chi_{c1}\omega PA}}{(m_{\chi_{c1}}^2-p^2)(m_{\omega}^2-q^2)}\nonumber\\
&&+\cdots\, ,
\end{eqnarray}
\begin{eqnarray}
\Pi_{J/\psi f_0(500) PA}(p^{\prime2},p^2,q^2)&=&
\frac{\lambda_{J/\psi f_0(500) PA}}{(m_Y^2-p^{\prime2})
(m_{J/\psi}^2-p^2)(m_{f_0(500)}^2-q^2)}
+\frac{C_{J/\psi f_0(500) PA}}{(m_{J/\psi}^2-p^2)(m_{f_0(500)}^2-q^2)}\nonumber\\
&&+\cdots\, .
\end{eqnarray}
With the replacement $PA \to AA$, we can get the corresponding scalar invariant components for the current $J_{\mu\nu}^{AA}$;
With the replacements $(\bar{D}$, $D$, $\bar{D}^*$, $D^*$, $\bar{D}_0$, $\bar{D}_1$, $\omega$, $f_0(500)) \to (\bar{D}_s$, $D_s$, $\bar{D}^*_s$, $D^*_s$, $\bar{D}_{s0}$, $\bar{D}_{s1}$, $\phi(1020)$, $f_0(980))$, $PA \to \widetilde{A}V (S\widetilde{V})$, we can get the corresponding scalar invariant components for the current $J_{\mu}^{\widetilde{A}V} (J_{\mu\nu}^{S\widetilde{V}})$.

As both the currents $J_{\mu\nu}^{AA}$ and $J_{\mu\nu}^{S\widetilde{V}}$ can couple potentially to the tetraquark states with the quantum numbers $J^{PC}=1^{--}$ and $J^{PC}=1^{+-}$, we cannot eliminate the contaminations from the axial-vector components clearly in the $Y_{AA}\to \bar{D}_0D^*$, $Y_{AA}\to \bar{D}_1D$, $Y_{AA}\to \chi_{c0}\omega$, $Y_{AA}\to J/\psi f_0(500)$ and $Y_{S\widetilde{V}}\to \eta_c\phi$ channels in calculations. Therefore we parameterize the contributions of the axial-vector components on  the hadron side as $\bar{\lambda}$ to obtain collective  QCD sum rules. The corresponding scalar invariant components are expressed as,
\begin{eqnarray}\label{D0DVAA}
\Pi_{\bar{D}_0D^* AA}(p^{\prime2},p^2,q^2)&=&
\frac{\lambda_{\bar{D}_0D^* AA}}{(m_Y^2-p^{\prime2})
(m_{D_0}^2-p^2)(m_{D^*}^2-q^2)}
+\frac{C_{\bar{D}_0D^* AA}}{(m_{D_0}^2-p^2)(m_{D^*}^2-q^2)}\nonumber\\
&&+\frac{\bar{\lambda}_{\bar{D}_0D^* AA}}{(m_X^2-p^{\prime2})
(m_{D_0}^2-p^2)(m_{D^*}^2-q^2)}+\cdots\, ,
\end{eqnarray}
\begin{eqnarray}\label{DADAA}
\Pi_{\bar{D}_1D AA}(p^{\prime2},p^2,q^2)&=&
\frac{\lambda_{\bar{D}_1D AA}}{(m_Y^2-p^{\prime2})
(m_{D_1}^2-p^2)(m_{D}^2-q^2)}
+\frac{C_{\bar{D}_1D AA}}{(m_{D_1}^2-p^2)(m_{D}^2-q^2)}\nonumber\\
&&+\frac{\bar{\lambda}_{\bar{D}_1D AA}}{(m_X^2-p^{\prime2})
(m_{D_1}^2-p^2)(m_{D}^2-q^2)}+\cdots\, ,
\end{eqnarray}
\begin{eqnarray}\label{chi0omegaAA}
\Pi_{\chi_{c0}\omega AA}(p^{\prime2},p^2,q^2)&=&
\frac{\lambda_{\chi_{c0}\omega AA}}{(m_Y^2-p^{\prime2})
(m_{\chi_{c0}}^2-p^2)(m_{\omega}^2-q^2)}
+\frac{C_{\chi_{c0}\omega AA}}{(m_{\chi_{c0}}^2-p^2)(m_{\omega}^2-q^2)}\nonumber\\
&&+\frac{\bar{\lambda}_{\chi_{c0}\omega AA}}{(m_X^2-p^{\prime2})
(m_{\chi_{c0}}^2-p^2)(m_{\omega}^2-q^2)}+\cdots\, ,
\end{eqnarray}
\begin{eqnarray}\label{JpsifAA}
\Pi_{J/\psi f_0(500) AA}(p^{\prime2},p^2,q^2)&=&
\frac{\lambda_{J/\psi f_0(500) AA}}{(m_Y^2-p^{\prime2})
(m_{J/\psi}^2-p^2)(m_{f_0(500)}^2-q^2)}
+\frac{C_{J/\psi f_0(500) AA}}{(m_{J/\psi}^2-p^2)(m_{f_0(500)}^2-q^2)}\nonumber\\
&&+\frac{\bar{\lambda}_{J/\psi f_0(500) AA}}{(m_X^2-p^{\prime2})
(m_{J/\psi}^2-p^2)(m_{f_0(500)}^2-q^2)}+\cdots\, ,
\end{eqnarray}
\begin{eqnarray}\label{etacphiSV}
\Pi_{\eta_c\phi S\widetilde{V}}(p^{\prime2},p^2,q^2)&=&
\frac{\lambda_{\eta_c\phi S\widetilde{V}}}{(m_Y^2-p^{\prime2})
(m_{\eta_c}^2-p^2)(m_{\phi}^2-q^2)}
+\frac{C_{\eta_c\phi S\widetilde{V}}}{(m_{\eta_c}^2-p^2)(m_{\phi}^2-q^2)}\nonumber\\
&&+\frac{\bar{\lambda}_{\eta_c\phi S\widetilde{A}}}{(m_X^2-p^{\prime2})
(m_{\eta_c}^2-p^2)(m_{\phi}^2-q^2)}+\cdots\, .
\end{eqnarray}
In above equations, we introduce the following notations to simplify the formula,
\begin{eqnarray}
\lambda_{\bar{D}D PA}&=&\frac{f_D^2m_D^4}{m_c^2}
\lambda_{PA}G_{\bar{D}D PA} \, , \nonumber \\
\lambda_{\bar{D}^*D PA}&=&\frac{f_{D^*}m_{D^*}f_D m_D^2}{m_c}
\lambda_{PA}G_{\bar{D}^*D PA} \, ,\nonumber \\
\lambda_{\bar{D}^*D^* PA}&=&f_{D^*}^2m_{D^*}^2
\lambda_{PA}G_{\bar{D}^*D^* PA}\, ,\nonumber \\
\lambda_{\bar{D}_0D^* PA}&=&f_{D_0}m_{D_0}f_{D^*}m_{D^*}
\lambda_{PA}G_{\bar{D}_0D^* PA} \, ,\nonumber \\
\lambda_{\bar{D}_1D PA}&=&\frac{f_{D_1}m_{D_1}f_{D}m_{D}^2}{m_c}
\lambda_{PA}G_{\bar{D}_1D PA} \, ,
\end{eqnarray}
\begin{eqnarray}
\lambda_{\eta_c\omega PA}&=&\frac{f_{\eta_c}m_{\eta_c}^2f_{\omega}
m_{\omega}}{2m_c}
\lambda_{PA}G_{\eta_c\omega PA} \, ,\nonumber \\
\lambda_{J/\psi\omega PA}&=&f_{J/\psi}m_{J/\psi}f_{\omega}
m_{\omega}\lambda_{PA}G_{J/\psi\omega PA}
\left(1+\frac{m_\omega^2}{m_Y^2}-\frac{m_{J/\psi}^2}{m_Y^2} \right)\, ,\nonumber \\
\lambda_{\chi_{c0}\omega PA}&=&f_{\chi_{c0}}m_{\chi_{c0}}f_{\omega}
m_{\omega}\lambda_{PA}G_{\chi_{c0}\omega PA}\, ,\nonumber \\
\lambda_{\chi_{c1}\omega PA}&=&f_{\chi_{c1}}m_{\chi_{c1}}f_{\omega}
m_{\omega}\lambda_{PA}G_{\chi_{c1}\omega PA}\, ,\nonumber \\
\lambda_{J/\psi f_0(500) PA}&=&f_{J/\psi}m_{J/\psi}f_{f_0(500)}
m_{f_0(500)}\lambda_{PA}G_{J/\psi f_0(500) PA}\, .
\end{eqnarray}
With similar replacements as those of the scalar invariant components, we can get the corresponding notations for the currents $J_{\mu\nu}^{AA}$, $J_{\mu}^{\widetilde{A}V}$ and $J_{\mu\nu}^{S\widetilde{V}}$, except for,
\begin{eqnarray}
\lambda_{J/\psi\omega AA}&=&f_{J/\psi}m_{J/\psi}f_{\omega}
m_{\omega}\lambda_{AA}G_{J/\psi\omega AA}\, ,\nonumber \\
\lambda_{\eta_c\phi S\widetilde{V}}&=&\frac{f_{\eta_c}m_{\eta_c}^2f_{\phi}
m_{\phi}^3}{2m_c}
\lambda_{S \widetilde{V}}G_{\eta_c\phi S \widetilde{V}} \, ,\nonumber \\
\lambda_{J/\psi\phi S\widetilde{V}}&=&f_{J/\psi}m_{J/\psi}f_{\phi}
m_{\phi}\lambda_{S\widetilde{V}}G_{J/\psi\phi S\widetilde{V}}\, , \nonumber \\
\lambda_{\chi_{c1}\phi S\widetilde{V}}&=&f_{\chi_{c1}}m_{\chi_{c1}}^3f_{\phi}
m_{\phi}\lambda_{S \widetilde{V}}G_{\chi_{c1}\phi S \widetilde{V}}\, ,
\end{eqnarray}
and the notations for the axial-vector components are,
\begin{eqnarray}
\bar{\lambda}_{\bar{D}_0D^* AA}&=&f_{D_0}m_{D_0}^3f_{D^*}m_{D^*} \bar{\lambda}_{AA}\bar{G}_{\bar{D}_0D^* AA}\, , \nonumber \\
\bar{\lambda}_{\bar{D}_1D AA}&=&-\frac{f_{D_1}m_{D_1}f_{D}m_{D}^4}{m_c} \bar{\lambda}_{AA}\bar{G}_{\bar{D}_1D AA}\, , \nonumber \\
\bar{\lambda}_{\chi_{c0}\omega AA}&=&f_{\chi_{c0}}m_{\chi_{c0}}^3f_{\omega}m_{\omega} \bar{\lambda}_{AA}\bar{G}_{\chi_{c0}\omega AA}\, , \nonumber \\
\bar{\lambda}_{J/\psi f_0(500) AA}&=&-f_{J/\psi}m_{J/\psi}f_{f_0(500)}m_{f_0(500)}^3 \bar{\lambda}_{AA}\bar{G}_{J/\psi f_0(500) AA}\, , \nonumber \\
\bar{\lambda}_{\eta_c\phi S\widetilde{A}}&=&\frac{f_{\eta_c}m_{\eta_c}^2f_{\phi}
m_{\phi}}{2m_c}\bar{\lambda}_{S\widetilde{A}}\bar{G}_{\eta_c\phi S\widetilde{A}}\, .
\end{eqnarray}

Performing triple dispersion relation, we can get the following equation,
\begin{eqnarray}\label{dispersion-3}
\Pi_{H}(p^{\prime2},p^2,q^2)&=&\int_{\Delta_s^{\prime2}}^\infty ds^{\prime} \int_{\Delta_s^2}^\infty ds \int_{\Delta_u^2}^\infty du \frac{\rho_{H}(s^\prime,s,u)}{(s^\prime-p^{\prime2})(s-p^2)(u-q^2)}\, ,
\end{eqnarray}
where the $\Delta_{s}^{\prime2}$, $\Delta_{s}^{2}$ and $\Delta_{u}^{2}$ are thresholds, we introduce the subscript $H$ to denote the hadron side.

On the QCD side, we carry out  the operator product expansion up to the vacuum condensates of dimension 5, and perform double dispersion relations to acquire,
\begin{eqnarray}\label{dispersion-2}
\Pi_{QCD}(p^{\prime2},p^2,q^2)&=& \int_{\Delta_s^2}^\infty ds \int_{\Delta_u^2}^\infty du \frac{\rho_{QCD}(p^{\prime2},s,u)}{(s-p^2)(u-q^2)}\, ,
\end{eqnarray}
as
\begin{eqnarray}
{\rm lim}_{\epsilon \to 0}\frac{{\rm Im}\,\Pi_{QCD}(s^\prime+i\epsilon,p^2,q^2)}{\pi}&=&0\, .
\end{eqnarray}

The triple dispersion relation in Eq.\eqref{dispersion-3} on the hadron side cannot match with the double dispersion relation in Eq.\eqref{dispersion-3} on  the QCD side, we should carry out the integral over $ds^\prime$ firstly and then match the hadron side with the QCD side below the continuum thresholds to acquire rigorous quark-hadron duality \cite{WZG-Y4660-Decay,WZG-ZJX-Zc-Decay},
\begin{eqnarray}\label{rigorous}
  \int_{\Delta_s^2}^{s_{0}}ds \int_{\Delta_u^2}^{u_0}du  \frac{\rho_{QCD}(p^{\prime2},s,u)}{(s-p^2)(u-q^2)}&=& \int_{\Delta_s^2}^{s_0}ds \int_{\Delta_u^2}^{u_0}du  \left[ \int_{\Delta_{s}^{\prime2}}^{\infty}ds^\prime  \frac{\rho_H(s^\prime,s,u)}{(s^\prime-p^{\prime2})(s-p^2)(u-q^2)} \right]\, ,
\end{eqnarray}
where the $s_0$ and $u_0$ are the continuum thresholds.
We introduce some free parameters $C$ to parameterize the contributions of transitions between the  higher resonances (continuum states) in the $s^\prime$ channel and the ground state conventional  meson pairs. We take the parameter $C_{\bar{D}DPA}$ in the $Y_{PA}\to \bar{D}D$ channel as an example,
\begin{eqnarray}
C_{\bar{D}DPA}&=&\int_{s_0^\prime}^{\infty} ds^\prime \frac{\rho_H(s^\prime,m_{\bar{D}}^2,m_D^2)}{\left(s^\prime-m_Y^2\right)
\left(p^2-m_{\bar{D}}^2 \right)\left(q^2-m_D^2 \right)}\, ,
\end{eqnarray}
where the $s_0^\prime$ is the continuum threshold parameter for the ground state,
the $\rho_H(s^\prime,m_{\bar{D}}^2,m_D^2)$ is the formal  hadronic spectral density for transitions between the  higher resonances (continuum states) in the $s^\prime$ channel and ground state meson pairs $D\bar{D}$. It is obvious that in the $s$ and $u$ channels, the hadron side and QCD side of the spectral densities have one to one correspondence below the continuum thresholds $s_0$ and $u_0$, respectively, while in the $s^\prime$ channel, there exists no correspondence on the QCD side.
Experimentally, the spectroscopies   of the hidden-charm tetraquark states have not been established yet, although some exotic states are excellent candidates for the tetraquark states, definite conclusion cannot be obtained. We introduce a free  parameter $C_{\bar{D}DPA}$ to parameterize the contributions involving the higher resonances (continuum states) in the $s^\prime$ channel, which results in model dependence. At the present time, we have no choice to avoid model dependence. For more detailed discussions, one is suggested to consult Sect.{\bf 7} in Ref.\cite{WZG-review}.

Then we set $p^{\prime2}=p^2$ in the correlation functions $\Pi(p^{\prime 2},p^2,q^2)$, and perform double Borel transformation in regard to the variables $P^2=-p^2$ and $Q^2=-q^2$ respectively, and set the Borel parameters $T_1^2=T_2^2=T^2$  to obtain the QCD sum rules,
\begin{eqnarray}\label{H-side-QCDSR}
&&\frac{\lambda_{\bar{D}DPA}}{m_Y^2-m_D^2}
\left[\exp\left(-\frac{m_D^2}{T^2} \right)-\exp\left(-\frac{m_Y^2}{T^2} \right) \right]\exp\left(-\frac{m_D^2}{T^2}\right)+C_{\bar{D}DPA}\exp\left( -\frac{m_D^2}{T^2}-\frac{m_D^2}{T^2}\right)\nonumber\\
&&=\Pi^{QCD}_{\bar{D}DPA}(T^2)\, , \nonumber\\
&&\cdots\, .
\end{eqnarray}

In the following,  we illustrate  the QCD side of the QCD sum rules  for the  $Y_{PA}\to \bar{D}D$ channel as an example,
\begin{eqnarray}	
\Pi^{QCD}_{\bar{D}D PA}(T^2)&=&\frac{3m_c}{128\pi^4}
\int_{m_c^2}^{s^0_{D}}ds	\int_{m_c^2}^{s^0_{D}}du \left(1-\frac{m_c^2}{s}\right)^2\left(1-\frac{m_c^2}{u}\right)^2 u \, \exp\left(-\frac{s+u}{T^2} \right) \nonumber\\
&&-\frac{\langle\bar{q}q\rangle}{16\pi^2}\int_{m_c^2}^{s^0_{D}}ds \left(1-\frac{m_c^2}{s}\right)^2 \left(s+m_c^2\right) \,\exp\left(-\frac{s+m_c^2}{T^2} \right)\nonumber\\
&&-\frac{\langle\bar{q}g_s\sigma G q\rangle}{384\pi^2} \int_{m_c^2}^{s^0_{D}}du\left(1-\frac{m_c^2}{u}\right) \left(3+\frac{5m_c^2}{u}\right)\, \exp\left(-\frac{u+m_c^2}{T^2} \right)\nonumber\\
&&+\frac{m_c^2\langle\bar{q}g_s\sigma G q\rangle}{64\pi^2 T^4} \int_{m_c^2}^{s^0_{D}}du \left(1-\frac{m_c^2}{u}\right)^2 u \, \exp\left(-\frac{u+m_c^2}{T^2} \right)\nonumber\\
&&-\frac{m_c^2\langle\bar{q}g_s\sigma G q\rangle}{64\pi^2 T^2} \left(2-\frac{m_c^2}{T^2}\right) \int_{m_c^2}^{s^0_{D}}ds \left(1-\frac{m_c^2}{s}\right)^2 \exp\left(-\frac{s+m_c^2}{T^2} \right)\, .
\end{eqnarray}
Other QCD spectral densities are shown explicitly in the Appendix \ref{app-spectraldensities}, while expressions of the hadron side of the QCD sum rules have the same form.

In calculations, we neglect the gluon condensates due to their tiny contributions \cite{WZG-Y4660-Decay,WZG-ZJX-Zc-Decay}. In addition, there appear endpoint divergences  in some channels at the thresholds  $s=m_c^2$, $s=4m_c^2$ and $u=m_c^2$ by the factors $s-m_c^2$, $s-4m_c^2$ and $u-m_c^2$ in the denominators.
The replacements $s-m_c^2\to s-m_c^2+\Delta^2$, $s-4m_c^2\to s-4m_c^2+\Delta^2$ and $u-m_c^2\to u-m_c^2+\Delta^2$ with $\Delta^2=m_c^2$ are adopted to eliminate the divergences as we have done in previous works \cite{WZG-YXS-cccc-AAPPS,WZG-YXS-cccc-CPC}.

\section{Numerical results and discussions}
The input parameters in this work are listed in Table \ref{Parameters}. We set $m_u=m_d=0$ and take the $\overline{MS}$ masses $m_{s}(2\,\rm{GeV})=(0.095\pm0.005)\,\rm{GeV}$ and $m_{c}(m_c)=(1.275\pm0.025)\,\rm{GeV}$ from the Particle Data Group \cite{PDG}, and take into account the energy-scale dependence of the vacuum condensates and the quark masses $m_s$ and $m_c$  from the  re-normalization group equation,
\begin{eqnarray}
\langle\bar{q}q \rangle(\mu)&=&\langle\bar{q}q \rangle({\rm 1GeV})\left[\frac{\alpha_{s}({\rm 1GeV})}{\alpha_{s}(\mu)}\right]^{\frac{12}{33-2n_f}}\, , \nonumber\\
\langle\bar{s}s \rangle(\mu)&=&\langle\bar{s}s \rangle({\rm 1GeV})\left[\frac{\alpha_{s}({\rm 1GeV})}{\alpha_{s}(\mu)}\right]^{\frac{12}{33-2n_f}}\, , \nonumber\\
\langle\bar{q}g_s \sigma Gq \rangle(\mu)&=&\langle\bar{q}g_s \sigma Gq \rangle({\rm 1GeV})\left[\frac{\alpha_{s}({\rm 1GeV})}{\alpha_{s}(\mu)}\right]^{\frac{2}{33-2n_f}}\, , \nonumber\\
\langle\bar{s}g_s \sigma Gs \rangle(\mu)&=&\langle\bar{s}g_s \sigma Gs \rangle({\rm 1GeV})\left[\frac{\alpha_{s}({\rm 1GeV})}{\alpha_{s}(\mu)}\right]^{\frac{2}{33-2n_f}}\, , \nonumber\\
m_c(\mu)&=&m_c(m_c)\left[\frac{\alpha_{s}(\mu)}{\alpha_{s}(m_c)}\right]^{\frac{12}{33-2n_f}} \, ,\nonumber\\
m_s(\mu)&=&m_s({\rm 2GeV} )\left[\frac{\alpha_{s}(\mu)}{\alpha_{s}({\rm 2GeV})}\right]^{\frac{12}{33-2n_f}}\, ,\nonumber\\
\alpha_s(\mu)&=&\frac{1}{b_0t}\left[1-\frac{b_1}{b_0^2}\frac{\log t}{t} +\frac{b_1^2(\log^2{t}-\log{t}-1)+b_0b_2}{b_0^4t^2}\right]\, ,
\end{eqnarray}
where $t=\log \frac{\mu^2}{\Lambda_{QCD}^2}$, $b_0=\frac{33-2n_f}{12\pi}$, $b_1=\frac{153-19n_f}{24\pi^2}$, $b_2=\frac{2857-\frac{5033}{9}n_f+\frac{325}{27}n_f^2}{128\pi^3}$, $\Lambda_{QCD}=210\,\rm{MeV}$, $292\,\rm{MeV}$ and  $332\,\rm{MeV}$ for the flavors $n_f=5$, $4$ and $3$, respectively \cite{PDG,Narison-mix}. In detail, we choose $n_f=4$ for the hidden-charm tetraquark states, and evolve all the values of input parameters to the typical energy scale $\mu=1\,\rm{GeV}$.

We intend to study the exotic states in our unique scheme step by step, firstly, we study the mass spectrum, then we study the two-body strong decays. In studying the mass spectrum of the hidden-charm (bottom) tetraquark (molecular) states, we compute the terms $g_s^2\langle\bar{q}q\rangle^2$ with $q=u$, $d$ or $s$ \cite{Wang-tetra-formula,WangZG-HuangT-PRD-2014,
WangZG-HuangT-EPJC-2014,WangZG-EPJC-2014,WangZG-HuangT-NPA-2014}.
 In the full light quark propagators, there are terms $\langle\bar{q}_j\gamma_{\mu}q_i\rangle$ \cite{Wang-tetra-formula,WangZG-HuangT-PRD-2014,
WangZG-HuangT-EPJC-2014,WangZG-EPJC-2014,WangZG-HuangT-NPA-2014}, which  absorb the gluons  emitted from the other quark lines to form $\langle\bar{q}_j\gamma_{\mu}q_ig_s D_\nu G_{\alpha\beta}\rangle$  to make contribution to the four-quark condensate $g_s^2\langle\bar{q}q\rangle^2$.
 The four-quark condensate $g_s^2\langle \bar{q}q\rangle^2$ comes from the terms
$\langle \bar{q}\gamma_\mu  q g_s D_\eta G_{\lambda\tau}\rangle$, $\langle\bar{q}_jD^{\dagger}_{\mu}D^{\dagger}_{\nu}D^{\dagger}_{\alpha}q_i\rangle$  and
$\langle\bar{q}_jD_{\mu}D_{\nu}D_{\alpha}q_i\rangle$  rather than comes from the radiative  $\mathcal{O}(\alpha_s)$ corrections for the four-quark condensate $\langle \bar{q}q\rangle^2$, where $D_\alpha=\partial_\alpha-ig_sG_\alpha $. The strong coupling constant $\alpha_s(\mu)=\frac{g_s^2(\mu)}{4\pi}$ appears  at the tree level, which is energy scale dependent.
 In fact, the contributions of such terms are tiny. In the present work (also other works on the two-body strong decays of the hidden-charm tetraquark states), we neglect such terms as we only take account of the vacuum condensates up to dimension 5 \cite{WZG-review}, but we consider the energy scale dependence of the input parameters for consistency according to our previous works \cite{Wang-tetra-formula,WangZG-HuangT-PRD-2014,
WangZG-HuangT-EPJC-2014,WangZG-EPJC-2014,WangZG-HuangT-NPA-2014}. In Ref.\cite{Nielsen-JPG-Review}, the vacuum condensates are taken at the energy scale $\mu=1\,\rm{GeV}$, while the $c$-quark mass is taken as the $\overline{MS}$ mass $m_c(m_c)$ (or approximate pole mass), it is another scheme to choose the input parameters. In two-point QCD sum rules for the hidden-charm tetraquark (molecular) states, the largest contributions do not come from the perturbative terms, but come from the vacuum condensates, we have to calculate the radiative $\mathcal{O}(\alpha_s)$ corrections to the perturbative terms and vacuum condensates (at least the $\langle\bar{q}q\rangle$) at the same time, if the next-to-leading contributions are required. Up to know, only the radiative $\mathcal{O}(\alpha_s)$ corrections to the perturbative terms have been studied partly \cite{Narison-SB-2024,Narison-SB-2023,KTChao-JHEP-2024}.

As we take the rigorous quark-hadron duality in the $s$ and $u$ channels (see Eq.\eqref{rigorous}), which correspond to the traditional mesons, it is reasonable to choose the typical energy scale $\mu=1\,\rm{GeV}$ as in the usual two-point QCD sum rules. If we choose a slightly larger energy scale, the integral ranges $m_c^2(\mu)-s_0/u_0$ and $4m_c^2(\mu)-s_0/u_0$ would be slightly  larger, therefore we expect a slightly larger hadronic coupling constant. As the light mesons $f_0(500)$, $\omega$, $f_0(980)$ and $\phi(1020)$ are concerned, we prefer to choose $\mu=1\,\rm{GeV}$ as the universal energy scale.

In order to obtain flat platforms, the free parameters are fitted as,
\begin{eqnarray}
C_{\bar{D}_sD_s\widetilde{A}V}&=&0.0023\,{\rm GeV^5}\times T^2\, , \nonumber\\
C_{\bar{D}_s^*D_s\widetilde{A}V}&=&-0.0000035\,{\rm GeV^4}\times T^2\, , \nonumber\\
C_{\bar{D}_s^*D_s^*\widetilde{A}V}&=&0.0004\,{\rm GeV^5}\times T^2\, , \nonumber\\
C_{\bar{D}_{s0}D_s^*\widetilde{A}V}&=&0.00009\,{\rm GeV^6}\times T^2\, , \nonumber\\
C_{\bar{D}_{s1}D_s\widetilde{A}V}&=&0.0126\,{\rm GeV^6}\times T^2\, , \nonumber\\
C_{\eta_c\phi\widetilde{A}V}&=&0.00017\,{\rm GeV^4}\times T^2\, , \nonumber\\
C_{J/\psi\phi\widetilde{A}V}&=&0.0\, , \nonumber\\
C_{\chi_{c0}\phi\widetilde{A}V}&=&0.002\,{\rm GeV^6}\times T^2\, , \nonumber\\
C_{\chi_{c1}\phi\widetilde{A}V}&=&0.000012\,{\rm GeV^5}\times T^2\, , \nonumber\\
C_{J/\psi f_0(980)\widetilde{A}V}&=&0.0027\,{\rm GeV^6}\times T^2\, ,
\end{eqnarray}

\begin{table}
\begin{center}
\begin{tabular}{|c|c|c|c|c|c|c|c|c|}\hline\hline
Parameters                &Values(\rm{GeV})                            &Parameters     &Values \cite{WZG-heavy-decay}  &Parameters               &Values    \\ \hline

$m_{J/\psi}$              &$3.0969$ \cite{PDG}                          &$m_{D_0}$      &$2.40\,\rm{GeV}$    &$f_{\omega}$,$f_{\rho}$    &$0.215\,\rm{GeV}$ \cite{PBall-decay-Kv}   \\

$m_{\eta_c}$              &$2.9839$ \cite{PDG}                          &$m_{D_{s0}}$   &$2.32\,\rm{GeV}$        &$f_{f_0(500)}$           &$0.350\,\rm{GeV}$ \cite{ChengHY-2022,WZG-EPJC-scalar}   \\

$m_{\phi}$                &$1.019460$ \cite{PDG}                         &$m_{D_1}$      &$2.42\,\rm{GeV}$        &$f_{f_0(980)}$           &$0.180\,\rm{GeV}$ \cite{Wang-f980-decay}    \\

$m_{\chi_{c0}}$           &$3.41471$ \cite{PDG}                         &$m_{D_{s1}}$   &$2.46\,\rm{GeV}$        &$\lambda_{PA}$           &$7.19\times 10^{-2}\, \rm{GeV}^5$ \cite{WZG-NPB-cucd-Vector}   \\

$m_{\chi_{c1}}$           &$3.51067$ \cite{PDG}                         &$f_{D}$        &$0.208\,\rm{GeV}$         &$\lambda_{AA}$           &$6.65\times 10^{-2}\, \rm{GeV}^5$ \cite{WZG-NPB-cucd-Vector} \\

$m_{D}$                   &$1.86484$ \cite{PDG}                         &$f_{D_s}$      &$0.240\,\rm{GeV}$     &$\lambda_{\widetilde{A}V}$ &$1.23\times 10^{-1}\, \rm{GeV}^5$ \cite{WZG-NPB-cscs-Vector} \\

$m_{D_s}$                 &$1.96835$\cite{PDG}                         &$f_{D^*}$      &$0.263\,\rm{GeV}$    &$\lambda_{S\widetilde{V}}$ &$6.22\times 10^{-2}\, \rm{GeV}^5$ \cite{WZG-NPB-cscs-Vector} \\

$m_{D^*}$                 &$2.00685$ \cite{PDG}                         &$f_{D_s^*}$    &$0.308\,\rm{GeV}$  &$s^0_\omega$,$s^0_{\rho}$  &$(1.2\,\rm{GeV})^2$ \cite{PBall-decay-Kv} \\

$m_{D_s^*}$               &$2.1066$ \cite{PDG}                          &$f_{D_0}$      &$0.373\,\rm{GeV}$       &$s^0_{f_0(500)}$         &$1.0\,\rm{GeV}^2$ \cite{ChengHY-2022,WZG-EPJC-scalar} \\

$m_\omega$                &$0.78266$ \cite{PDG}                         &$f_{D_{s0}}$   &$0.333\,\rm{GeV}$       &$s^0_{f_0(980)}$         &$(1.3\,\rm{GeV})^2$ \cite{Wang-f980-decay} \\

$m_{f_0(500)}$            &$0.550$ \cite{PDG}                           &$f_{D_1}$      &$0.332\,\rm{GeV}$         &$s^0_{J/\psi}$           &$(3.6\,\rm{GeV})^2$ \cite{PDG,Becirevic,Charmonium-PRT}  \\

$m_{f_0(980)}$            &$0.990$ \cite{PDG}                           &$f_{D_{s1}}$   &$0.345\,\rm{GeV}$          &$s^0_{\eta_c}$           &$(3.5\,\rm{GeV})^2$ \cite{PDG,Becirevic,Charmonium-PRT} \\

$M_{Y(PA)}$               &$4.66$ \cite{WZG-NPB-cucd-Vector}            &$s^0_{D}$      &$6.2\,\rm{GeV}^2$       &$s^0_{\chi_{c0}}$        &$(3.9\,\rm{GeV})^2$ \cite{PDG,Becirevic,Charmonium-PRT}    \\

$M_{Y(AA)}$              &$4.69$ \cite{WZG-NPB-cucd-Vector}             &$s^0_{D_s}$    &$7.3\,\rm{GeV}^2$      &$s^0_{\chi_{c1}}$        &$(4.0\,\rm{GeV})^2$ \cite{PDG,Becirevic,Charmonium-PRT}    \\

$M_{Y(\widetilde{A}V)}$   &$4.65$ \cite{WZG-NPB-cscs-Vector}            &$s^0_{D^*}$    &$6.4\,\rm{GeV}^2$             &$\langle\bar{q}q\rangle$ &$-(0.24\pm 0.01\, \rm{GeV})^3$ \cite{SVZ79,Reinders85,Colangelo-Review}  \\

$M_{Y(S\widetilde{V})}$  &$4.68$ \cite{WZG-NPB-cscs-Vector}             &$s^0_{D_s^*}$  &$7.5\,\rm{GeV}^2$       &$\langle\bar{s}s\rangle$ &$(0.8\pm0.1)\langle\bar{q}q\rangle$ \cite{SVZ79,Reinders85,Colangelo-Review}   \\

$f_{J/\psi}$              &$0.418$ \cite{Becirevic}                     &$s^0_{D_0}$    &$8.3\,\rm{GeV}^2$     &$\langle\bar{q}g_s\sigma Gq\rangle$  &$m_0^2\langle\bar{q}q\rangle$ \cite{SVZ79,Reinders85,Colangelo-Review} \\

$f_{\eta_c}$              &$0.387$ \cite{Becirevic}                     &$s^0_{D_{s0}}$ &$7.4\,\rm{GeV}^2$     &$\langle\bar{s}g_s\sigma Gs\rangle$  &$m_0^2\langle\bar{s}s\rangle$ \cite{SVZ79,Reinders85,Colangelo-Review} \\

$f_{\chi_{c0}}$           &$0.359$ \cite{Charmonium-PRT}                &$s^0_{D_1}$    &$8.6\,\rm{GeV}^2$              &$m_0^2$    &$(0.8\pm0.1)\,\rm{GeV}^2$ \cite{SVZ79,Reinders85,Colangelo-Review}  \\

$f_{\chi_{c1}}$           &$0.338$ \cite{Charmonium-PRT}                &$s^0_{D_{s1}}$ &$9.3\,\rm{GeV}^2$  &- &-  \\   \hline\hline
\end{tabular}
\end{center}
\caption{The input parameters used in numerical calculations, where the values of the vacuum condensates are taken at the energy scale $\mu=1\,\rm{GeV}$. }\label{Parameters}
\end{table}

\begin{eqnarray}
C_{\bar{D}D PA}&=&0.00016\,{\rm GeV^5}\times T^2\, , \nonumber\\
C_{\bar{D}^*D PA}&=&-0.0000018\,{\rm GeV^4}\times T^2\, , \nonumber\\
C_{\bar{D}^*D^* PA}&=&0.00013\,{\rm GeV^5}\times T^2\, , \nonumber\\
C_{\bar{D}_0D^* PA}&=&0.0001\,{\rm GeV^6}\times T^2\, , \nonumber\\
C_{\bar{D}_1D PA}&=&0.001\,{\rm GeV^6}\times T^2\, , \nonumber\\
C_{\eta_c\omega PA}&=&0.00009\,{\rm GeV^4}\times T^2\, , \nonumber\\
C_{J/\omega\omega PA}&=&0.0\, , \nonumber\\
C_{\chi_{c0}\omega PA}&=&0.00026\,{\rm GeV^6}\times T^2\, , \nonumber\\
C_{\chi_{c1}\omega PA}&=&0.000016\,{\rm GeV^5}\times T^2\, , \nonumber\\
C_{J/\psi f_0(500) PA}&=&0.00037\,{\rm GeV^6}\times T^2\, ,
\end{eqnarray}
\begin{eqnarray}
C_{\bar{D}D AA}&=&-0.0000013\,{\rm GeV^4}\times T^2\, , \nonumber\\
C_{\bar{D}^*D AA}&=&-0.0000012\,{\rm GeV^3}\times T^2\, , \nonumber\\
C_{\bar{D}^*D^* AA}&=&0.0\, , \nonumber\\
C_{\bar{D}_0D^* AA}&=&0.0003\,{\rm GeV^7}\, , \nonumber\\
\bar{\lambda}_{\bar{D}_0D^* AA}&=&0.0044\, {\rm GeV^7}\times T^2\, , \nonumber\\
C_{\bar{D}_1D AA}&=&0.00055\,{\rm GeV^7}\, , \nonumber\\
\bar{\lambda}_{\bar{D}_1D AA}&=&0.0037\, {\rm GeV^7}\times T^2\, , \nonumber\\
C_{\eta_c\omega AA}&=&0.0\, , \nonumber\\
C_{J/\omega\omega AA}&=&0.0\, , \nonumber\\
C_{\chi_{c0}\omega AA}&=&0.00013\,{\rm GeV^7}\, , \nonumber\\
\bar{\lambda}_{\chi_{c0}\omega AA}&=&0.0034\, {\rm GeV^7}\times T^2\, , \nonumber\\
C_{\chi_{c1}\omega AA}&=&0.043\,{\rm GeV^6}\, , \nonumber\\
C_{J/\psi f_0(500) AA}&=&0.000044\,{\rm GeV^7}\, ,\nonumber\\
\bar{\lambda}_{J/\psi f_0(500) AA}&=&0.0011\, {\rm GeV^7}\times T^2\, ,
\end{eqnarray}
\begin{eqnarray}
C_{\bar{D}_s D_s S\widetilde{V}}&=&0.000081\,{\rm GeV^4}\times T^2\, ,\nonumber\\
C_{\bar{D}_s^*D_s S\widetilde{V}}&=&-0.0000013\,{\rm GeV^5}\times T^2\, , \nonumber\\
C_{\bar{D}_s^*D_s^* S\widetilde{V}}&=&0.000025\,{\rm GeV^4}\times T^2\, , \nonumber\\
C_{\bar{D}_{s0}D_s^* S\widetilde{V}}&=&0.00013\,{\rm GeV^5}\times T^2\, , \nonumber\\
C_{\bar{D}_{s1}D_s S\widetilde{V}}&=&0.0007\,{\rm GeV^5}\times T^2\, , \nonumber\\
C_{\eta_c \phi S\widetilde{V}}&=&0.00015\, {\rm GeV^7}\, , \nonumber\\
\bar{\lambda}_{\eta_c\phi S\widetilde{A}}&=&0.0033\, {\rm GeV^7}\times T^2\, , \nonumber\\
C_{J/\psi\phi S\widetilde{V}}&=&0.00002\,{\rm GeV^4}\times T^2\, , \nonumber\\
C_{\chi_{c0}\phi S\widetilde{V}}&=&0.00032\,{\rm GeV^5}\times T^2\, , \nonumber\\
C_{\chi_{c1}\phi S\widetilde{V}}&=&0.0004\,{\rm GeV^6}\times T^2\, , \nonumber\\
C_{J/\psi f_0(980) S\widetilde{V}}&=&0.00022\,{\rm GeV^5}\times T^2\, .
\end{eqnarray}
Then we obtain uniform flat Borel windows $T^2_{max}-T^2_{min}=1\,\rm{GeV}^2$, which are presented explicitly in Table \ref{BorelP}.

\begin{figure}
\centering
\includegraphics[totalheight=5cm,width=7cm]{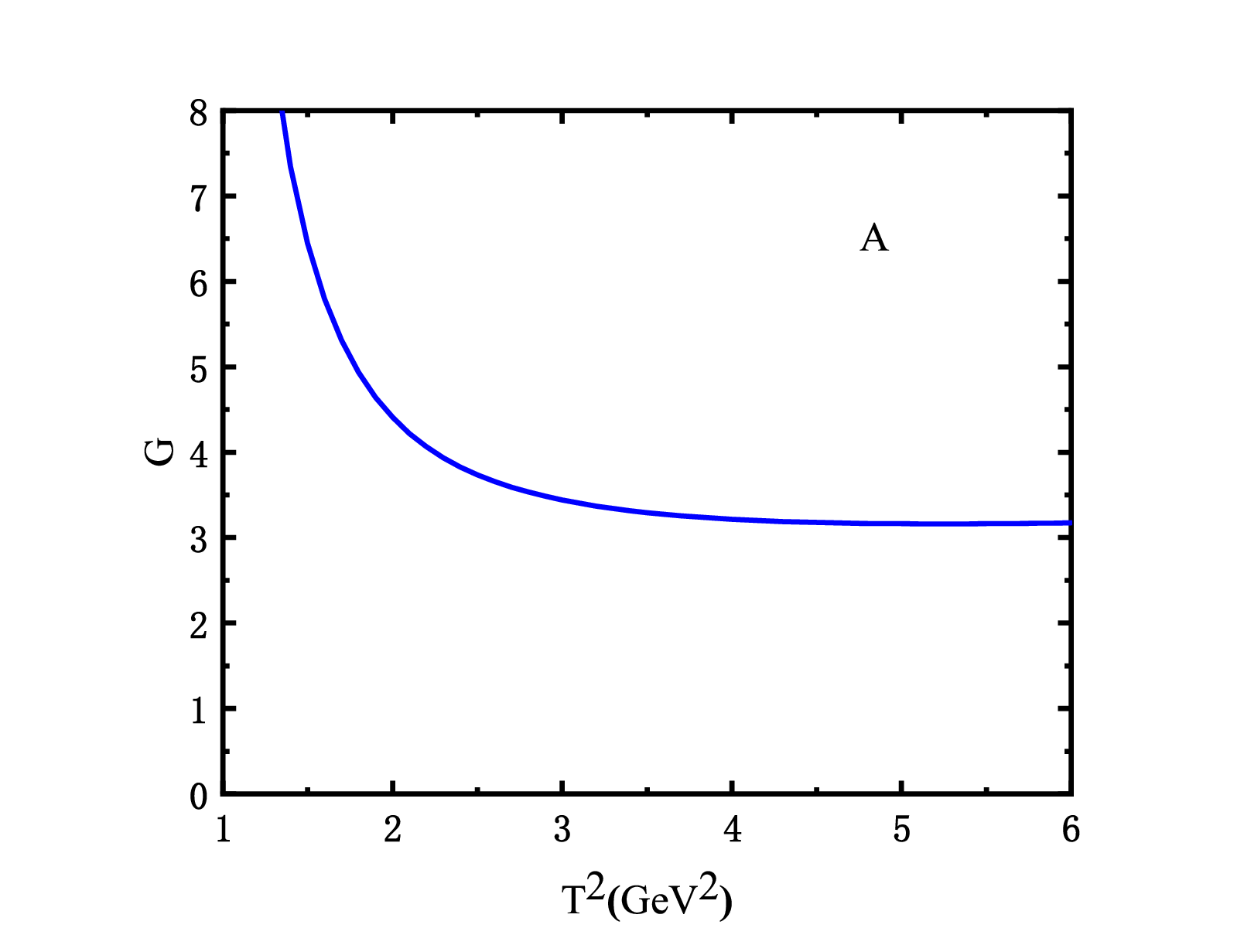}
\includegraphics[totalheight=5cm,width=7cm]{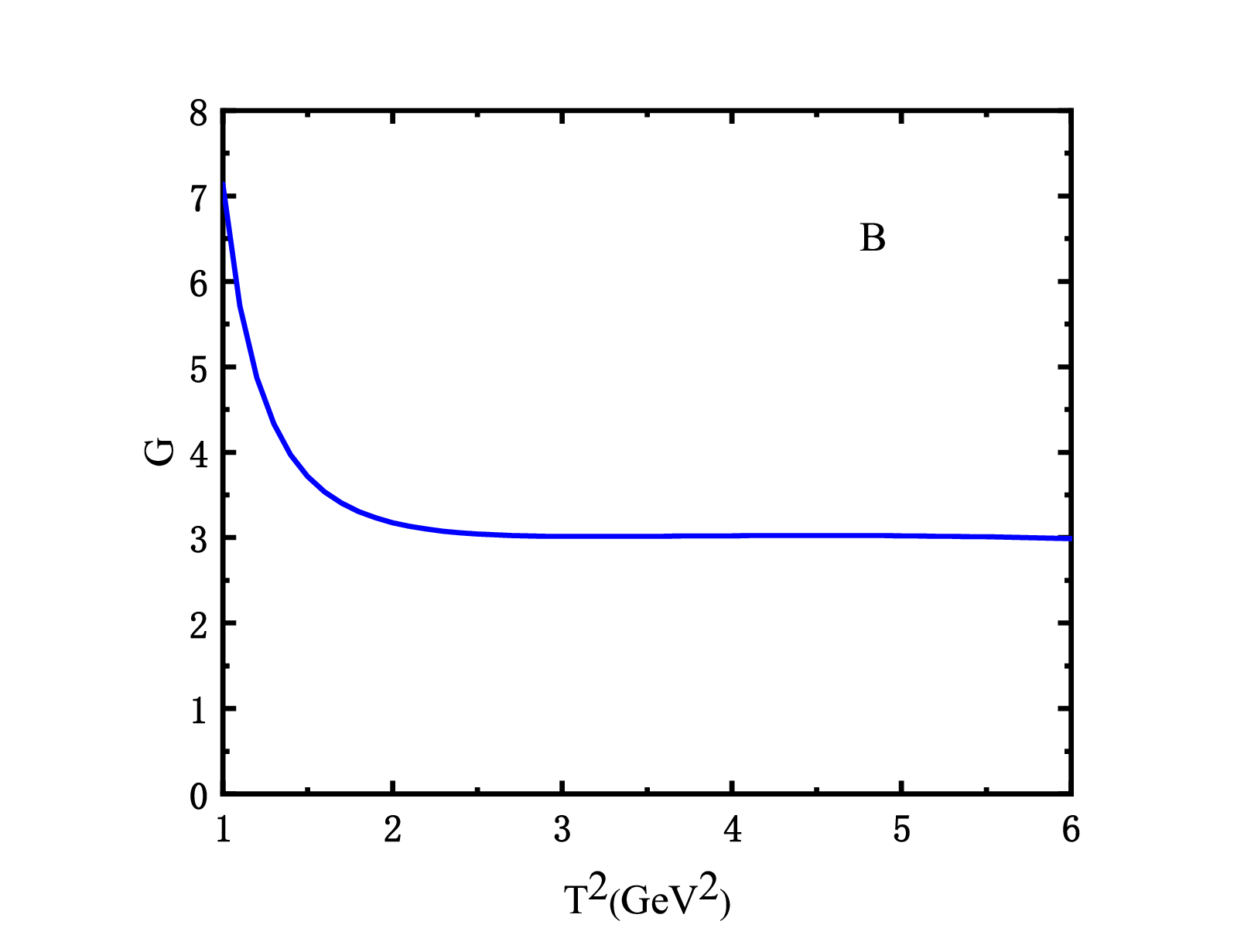}
\includegraphics[totalheight=5cm,width=7cm]{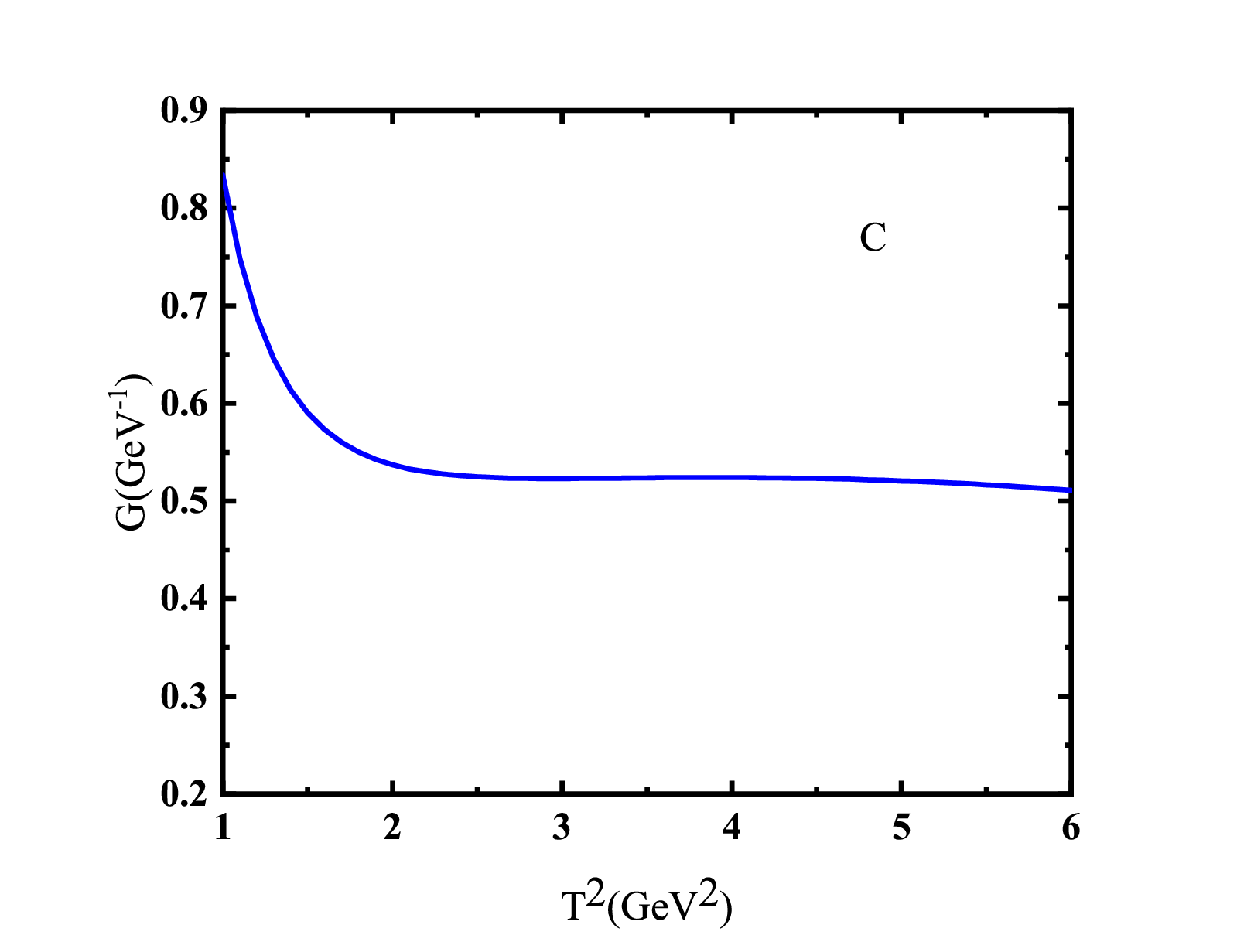}
\includegraphics[totalheight=5cm,width=7cm]{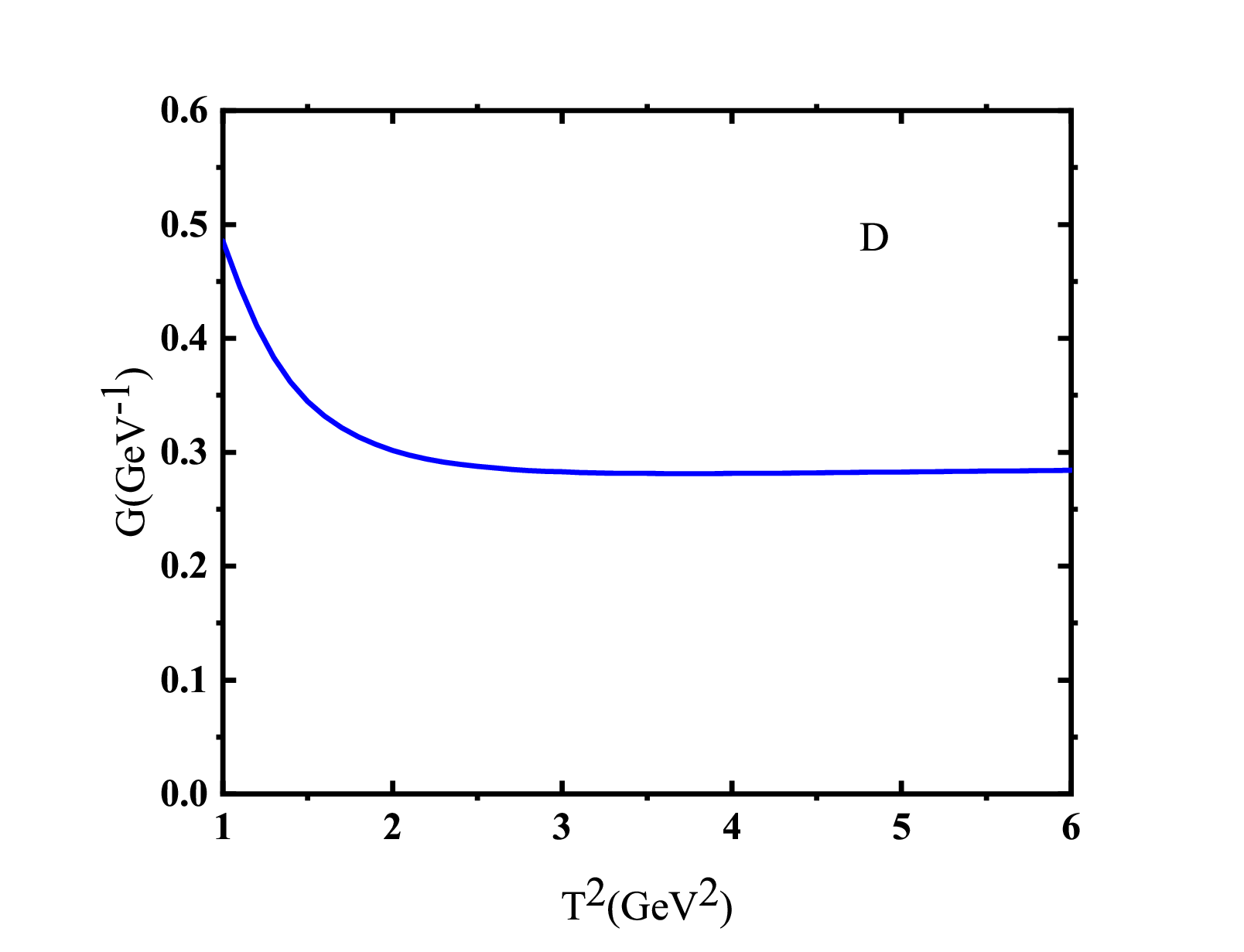}
\includegraphics[totalheight=5cm,width=7cm]{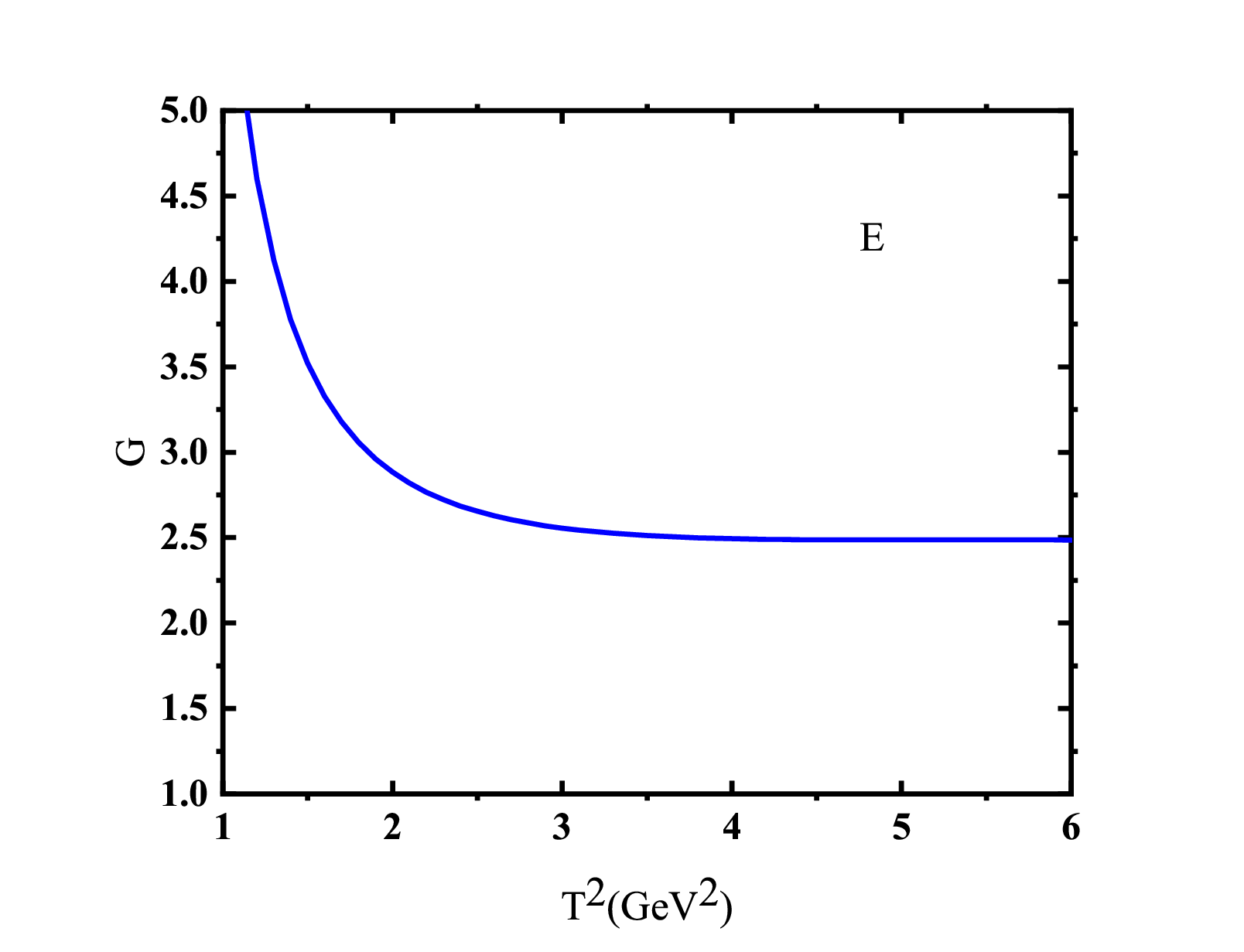}
\includegraphics[totalheight=5cm,width=7cm]{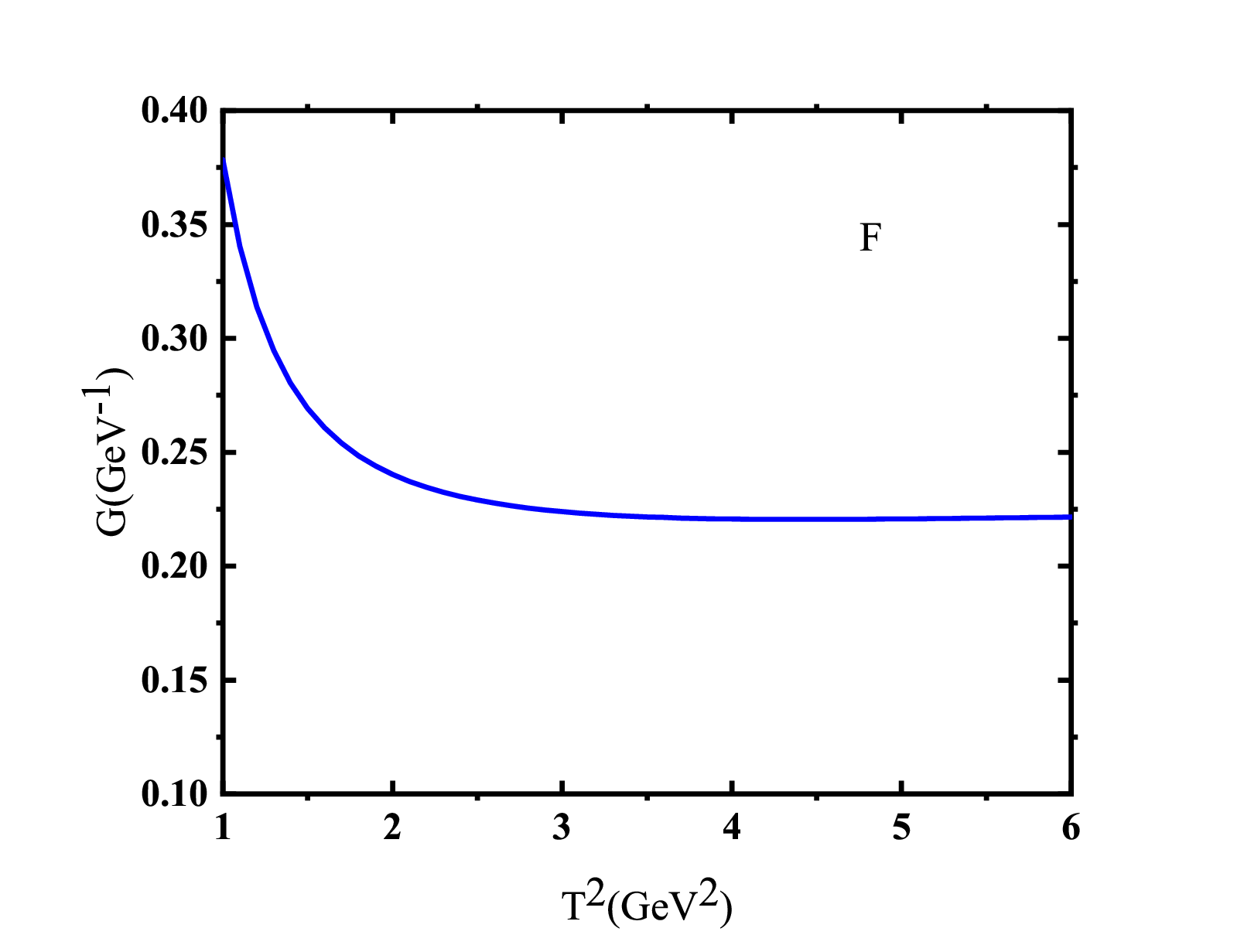}
\caption{The hadronic coupling constants with variations of the Borel parameters, where the $A$, $B$, $C$, $D$, $E$ and $F$ denote the hadronic coupling constants  $G_{\bar{D}_{s0}D_s^* S\widetilde{V}}$, $G_{\bar{D}_{s1}D_s S\widetilde{V}}$, $G_{\bar{D}_s D_s S\widetilde{V}}$, $G_{\bar{D}_s^* D_s^* S\widetilde{V}}$, $G_{\chi_{c0}\phi S\widetilde{V}}$ and $G_{J/\psi \phi S\widetilde{V}}$, respectively.}\label{hadron-coupling-fig}
\end{figure}

In Fig.\ref{hadron-coupling-fig}, the curves of the hadronic coupling constants $G_{\bar{D}_{s0}D_s^* S\widetilde{V}}$, $G_{\bar{D}_{s1}D_s S\widetilde{V}}$, $G_{\bar{D}_s D_s S\widetilde{V}}$, $G_{\bar{D}_s^* D_s^* S\widetilde{V}}$, $G_{\chi_{c0}\phi S\widetilde{V}}$ and $G_{J/\psi \phi S\widetilde{V}}$ are plotted with variations of the Borel parameters $T^2$ at large intervals  as an example. In the Borel windows, there appear flat platforms clearly, and thus we could  extract the hadronic  coupling constants reasonably.

The uncertainties of the hadronic coupling constants are analyzed routinely. They originate not only from the coupling constants but also from other input parameters.
We take the QCD sum rule for the channel $Y_{PA}\to \bar{D}D$ as an example, the uncertainties on  the hadron side can be written as  $\lambda_{PA}f_{\bar{D}}f_{D}G_{\bar{D}DPA} = \bar{\lambda}_{PA}\bar{f}_{\bar{D}}\bar{f}_{D}\bar{G}_{\bar{D}DPA}
+\delta\,\lambda_{PA}f_{\bar{D}}f_{D}G_{\bar{D}DPA}$, $C_{\bar{D}DPA} = \bar{C}_{\bar{D}DPA}+\delta C_{\bar{D}DPA}$, $\cdots$, where
\begin{eqnarray}
\delta\,\lambda_{PA}f_{\bar{D}}f_{D}G_{\bar{D}DPA} &=&\bar{\lambda}_{PA}\bar{f}_{\bar{D}}\bar{f}_{D}\bar{G}_{\bar{D}DPA}
\left( \frac{\delta f_{\bar{D}}}{\bar{f}_{\bar{D}}} +\frac{\delta f_{D}}{\bar{f}_{D}}+\frac{\delta \lambda_{PA}}{\bar{\lambda}_{PA}}
+\frac{\delta G_{\bar{D}DPA}}{\bar{G}_{\bar{D}DPA}}\right)\, ,
\end{eqnarray}
where the short overline denotes the central value. It can be approximately set as $\delta C_{\bar{D}DPA}=0$,  $\frac{\delta f_{\bar{D}}}{\bar{f}_{\bar{D}}} =\frac{\delta f_{D}}{\bar{f}_{D}}=\frac{\delta \lambda_{PA}}{\bar{\lambda}_{PA}}
=\frac{\delta G_{\bar{D}DPS}}{\bar{G}_{\bar{D}DPS}}$, $\cdots$ to avoid overestimating the uncertainties of the hadronic coupling constants.

After taking into account the relevant  uncertainties, the numerical values of the hadronic coupling constants can be obtained directly, which are shown explicitly in Table \ref{BorelP}.
Thereafter, the partial decay widths are obtained directly by the formula,
\begin{eqnarray}
\Gamma(Y\to F F^\prime) &=& \frac{|T|^2 p(m_Y,m_F,m_{F^\prime})}{24\pi m_Y^2} \, , \nonumber\\
|T|^2 &=& \Sigma |\langle F(p) F^\prime(q)|Y(p^\prime)\rangle|^2 \, ,
\end{eqnarray}
where $p(A,B,C)=\frac{\sqrt{[A^2-(B+C)^2][A^2-(B-C)^2]}}{2A}$, the $F$ and $F^\prime$ represent the final states. The explicit partial decay widths of different channels  are shown explicitly in Table \ref{Width-Part}.

\begin{table}
\begin{center}
\begin{tabular}{|c|c|c|c|c|c|c|c|c|}\hline\hline
Channels              &$T^2(\rm{GeV}^2)$ &$G $    \\ \hline

$\bar{D}D AA$         &$4.7-5.7$  &$(2.04\pm0.09)\times\rm{10^{-2}}\,\rm{GeV}^{-1}$   \\

$\bar{D}^*D AA$       &$4.2-5.2$  &$(1.92\pm0.07)\times\rm{10^{-2}}\,\rm{GeV}^{-2}$ \\

$\bar{D}^*D^* AA$     &$---$  &$0.0$      \\

$\bar{D}_0D^* AA$     &$4.0-5.0$  &$1.99\pm0.16$        \\

$\bar{D}_1D AA$       &$2.7-3.7$  &$0.83\pm0.08$   \\

$\eta_c\omega AA$     &$3.3-4.3$  &$(1.03\pm0.06)\times\rm{10^{-2}}\,\rm{GeV}^{-2}$     \\

$J/\psi\omega AA$     &$---$      &$0.0$     \\

$\chi_{c0}\omega AA$  &$4.9-5.9$  &$1.79\pm0.22$     \\

$\chi_{c1}\omega AA$  &$3.8-4.8$  &$(3.27\pm0.1)\times\rm{10^{-1}}\,\rm{GeV}^{-1}$     \\

$J/\psi f_0(500) AA$       &$3.0-4.0$  &$0.36\pm0.07$     \\ \hline

$\bar{D}_s D_s \widetilde{A}V$         &$2.2-3.2$  &$3.35\pm0.27$   \\

$\bar{D}^*_s D_s \widetilde{A}V$       &$4.8-5.8$  &$(2.52\pm0.11)\times\rm{10^{-2}}\,\rm{GeV}^{-1} $ \\

$\bar{D}^*_s D^*_s \widetilde{A}V$     &$4.1-5.1$  &$1.83\pm0.13$      \\

$\bar{D}_{s0}D^*_s \widetilde{A}V$     &$3.5-4.5$  &$0.29\pm0.02\,\rm{GeV}^{-1}$        \\

$\bar{D}_{s1}D_s \widetilde{A}V$       &$1.8-2.8$  &$6.34\pm0.65\,\rm{GeV}$   \\

$\eta_c\phi \widetilde{A}V$     &$2.3-3.3$  &$0.61\pm0.05\,\rm{GeV}^{-1}$     \\

$J/\psi\phi \widetilde{A}V$     &$---$      &$0.0$     \\

$\chi_{c0}\phi \widetilde{A}V$  &$3.0-4.0$  &$4.02\pm0.47\,\rm{GeV}$     \\

$\chi_{c1}\phi \widetilde{A}V$  &$4.7-5.7$  &$0.24\pm0.01\,\rm{GeV}^{-2}$     \\

$J/\psi f_0(980)\widetilde{A}V$       &$1.8-2.8$  &$6.19\pm0.78\,\rm{GeV}$     \\ \hline

$\bar{D}D PA$        &$2.3-3.3$  &$1.13\pm0.08$ \\

$\bar{D}^*DPA$       &$4.0-5.0$  &$(2.45\pm0.10)\times\rm{10^{-2}}\,\rm{GeV}^{-1}$ \\

$\bar{D}^*D^* PA$    &$4.1-5.1$  &$0.34\pm0.08$ \\

$\bar{D}_0D^*PA$     &$4.9-5.9$  &$1.25\pm0.06\,\rm{GeV}^{-1}$ \\

$\bar{D}_1DPA$       &$3.0-4.0$  &$3.23\pm0.31\,\rm{GeV}$ \\

$\eta_c\omega PA$    &$2.2-3.2$  &$0.69\pm0.05\,\rm{GeV}$     \\

$J/\psi\omega PA$    &$---$      &$0.0$     \\

$\chi_{c0}\omega PA$ &$2.6-3.6$  &$1.41\pm0.19\,\rm{GeV}$     \\

$\chi_{c1}\omega PA$ &$5.0-6.0$  &$0.38\pm0.02\,\rm{GeV}^{-2}$     \\

$J/\psi f_0(500) PA$      &$1.8-2.8$  &$1.05\pm0.18\,\rm{GeV}$     \\ \hline

$\bar{D}_s D_s S\widetilde{V}$        &$2.6-3.6$  &$0.52\pm0.03\,\rm{GeV}^{-1} $     \\

$\bar{D}^*_s D_s S\widetilde{V}$      &$4.6-5.6$  &$(1.35\pm0.05)\times 10^{-2}\,\rm{GeV}^{-2} $   \\

$\bar{D}^*_s D^*_s S\widetilde{V}$    &$3.3-4.3$  &$0.28\pm0.02\,\rm{GeV}^{-1}$     \\

$\bar{D}_{s0}D^*_s S\widetilde{V}$    &$4.7-5.7$  &$3.16\pm0.19 $  \\

$\bar{D}_{s1}D_s S\widetilde{V}$      &$3.9-4.9$  &$3.02\pm0.21 $   \\

$\eta_c\phi S\widetilde{V}$    &$4.6-5.6$  &$0.73\pm0.13\,\rm{GeV}^{-2} $   \\

$J/\psi\phi S\widetilde{V}$    &$3.8-4.8$  &$0.22\pm0.02\,\rm{GeV}^{-1} $   \\

$\chi_{c0}\phi S\widetilde{V}$ &$4.7-5.7$  &$2.49\pm0.23 $   \\

$\chi_{c1}\phi S\widetilde{V}$ &$2.7-3.7$  &$0.22\pm0.02\,\rm{GeV}^{-1} $   \\

$J/\psi f_0(980) S\widetilde{V}$      &$2.3-3.3$  &$1.36\pm0.17 $   \\

\hline\hline
\end{tabular}
\end{center}
\caption{ The Borel parameters (windows) $T^2$ and hadronic coupling constants $G$, where the "Channels" denotes the subscripts of the hadronic coupling constants defined in Eqs.\eqref{Coupling-1}-\eqref{Coupling-2} and in Appendix A. }\label{BorelP}
\end{table}

\begin{table}
\begin{center}
\begin{tabular}{|c|c|c|c|c|c|c|c|c|}\hline\hline
Channels                                               &$\Gamma(\rm{MeV})$ \\ \hline

$Y_{AA}\to \bar{D}^0D^0$, $\bar{D}^-D^+$   &$0.003\pm0.0$      \\

$Y_{AA}\to \frac{\bar{D}^{0*}D^0+\bar{D}^{0}D^{*0}}{\sqrt{2}}$, $\frac{\bar{D}^{-*}D^+ +\bar{D}^{-}D^{*+}}{\sqrt{2}}$   &$0.023\pm0.002 $ \\

$Y_{AA}\to \bar{D}^{*0}D^{*0}$, $\bar{D}^{*-}D^{*+}$  &$0.0$      \\

$Y_{AA}\to\frac{\bar{D}^0_0D^{*0}+\bar{D}^{*0}D^{0}_0}{\sqrt{2}}$, $\frac{\bar{D}^-_0D^{*+}+\bar{D}^{*-}D^{+}_0}{\sqrt{2}}$     &$6.03\pm0.97$        \\

$Y_{AA}\to\frac{\bar{D}^0_1D^{0}-\bar{D}^{0}D^{0}_1}{\sqrt{2}}$, $\frac{\bar{D}^-_1D^{+}-\bar{D}^{-}D^{+}_1}{\sqrt{2}}$  &$1.23\pm0.25$     \\

$Y_{AA}\to\eta_c\omega$     &$0.005\pm0.0$     \\

$Y_{AA}\to J/\psi\omega$    &$0.0$     \\

$Y_{AA}\to\chi_{c0}\omega$  &$7.07\pm1.74$     \\

$Y_{AA}\to\chi_{c1}\omega$  &$5.81\pm0.36$     \\

$Y_{AA}\to J/\psi f_0(500)$ &$0.31\pm0.12$     \\ \hline

$Y_{\widetilde{A}V}\to \bar{D}_s D_s$   &$52.04\pm8.39$      \\

$Y_{\widetilde{A}V}\to \frac{\bar{D}^{*}_s D_s +\bar{D}_s D^{*}_s}{\sqrt{2}}$   &$0.023\pm0.002$ \\

$Y_{\widetilde{A}V}\to \bar{D}^{*}_s D^{*}_s$  &$30.35\pm4.31$      \\

$Y_{\widetilde{A}V}\to\frac{\bar{D}_{s0} D^{*}_s+\bar{D}^{*}_s D_{s0}}{\sqrt{2}}$     &$3.96\pm0.57$        \\

$Y_{\widetilde{A}V}\to\frac{\bar{D}_{s1}D_s-\bar{D}_s D_{s1}}{\sqrt{2}}$ &$53.52\pm10.97$     \\

$Y_{\widetilde{A}V}\to\eta_c\phi$     &$12.24\pm1.92$     \\

$Y_{\widetilde{A}V}\to J/\psi\phi$    &$0.0$     \\

$Y_{\widetilde{A}V}\to\chi_{c0}\phi$  &$19.91\pm4.65$     \\

$Y_{\widetilde{A}V}\to\chi_{c1}\phi$  &$22.56\pm2.23$     \\

$Y_{\widetilde{A}V}\to J/\psi f_0(980)$ &$72.08\pm18.16$     \\ \hline

$Y_{PA}\to \bar{D}^0D^0$, $\bar{D}^-D^+$   &$8.50\pm1.20$ \\

$Y_{PA}\to \frac{\bar{D}^{0*}D^0+\bar{D}^{0}D^{*0}}{\sqrt{2}}$, $\frac{\bar{D}^{-*}D^+ +\bar{D}^{-}D^{*+}}{\sqrt{2}}$   &$0.035\pm0.003$ \\

$Y_{PA}\to \bar{D}^{*0}D^{*0}$, $\bar{D}^{*-}D^{*+}$  &$2.26\pm1.00$ \\

$Y_{PA}\to\frac{\bar{D}^0_0D^{*0}+\bar{D}^{*0}D^{0}_0}{\sqrt{2}}$, $\frac{\bar{D}^-_0D^{*+}+\bar{D}^{*-}D^{+}_0}{\sqrt{2}}$  &$80.49\pm7.73$ \\

$Y_{PA}\to\frac{\bar{D}^0_1D^{0}-\bar{D}^{0}D^{0}_1}{\sqrt{2}}$, $\frac{\bar{D}^-_1D^{+}-\bar{D}^{-}D^{+}_1}{\sqrt{2}}$   &$18.20\pm3.49$ \\

$Y_{PA}\to\eta_c\omega $                     &$22.52\pm3.51$     \\

$Y_{PA}\to J/\psi\omega $                    &$0.0$     \\

$Y_{PA}\to\chi_{c0}\omega$                   &$4.20\pm1.13$     \\

$Y_{PA}\to\chi_{c1}\omega $                  &$142.85\pm17.20$     \\

$Y_{PA}\to J/\psi f_0(500)$                  &$2.58\pm0.88$     \\ \hline

$Y_{S\widetilde{V}}\to \bar{D}_s D_s$     &$1.34\pm0.14 $     \\

$Y_{S\widetilde{V}}\to \frac{\bar{D}^{*}_s D_s +\bar{D}_s D^{*}_s}{\sqrt{2}}$  &$0.007\pm0.001 $   \\

$Y_{S\widetilde{V}}\to \bar{D}^{*}_s D^{*}_s$  &$0.80\pm0.09 $     \\

$Y_{S\widetilde{V}}\to\frac{\bar{D}_{s0} D^{*}_s+\bar{D}^{*}_s D_{s0}}{\sqrt{2}}$   &$14.19\pm1.71 $   \\

$Y_{S\widetilde{V}}\to\frac{\bar{D}_{s1}D_s-\bar{D}_s D_{s1}}{\sqrt{2}}$  &$12.85\pm1.79$   \\

$Y_{S\widetilde{V}}\to\eta_c\phi $      &$18.71\pm6.63 $   \\

$Y_{S\widetilde{V}}\to J/\psi\phi $     &$0.60\pm0.08 $   \\

$Y_{S\widetilde{V}}\to\chi_{c0}\phi $   &$8.20\pm1.52 $   \\

$Y_{S\widetilde{V}}\to \chi_{c1}\phi$   &$1.24\pm0.24 $   \\

$Y_{S\widetilde{V}}\to J/\psi f_0(980)$   &$3.54\pm0.89 $   \\

\hline\hline
\end{tabular}
\end{center}
\caption{ The partial decay widths of the vector tetraquark states $Y_{AA}$, $Y_{\widetilde{A}V}$, $Y_{PA}$ and $Y_{S\widetilde{V}}$. }\label{Width-Part}
\end{table}

Finally, we take the summations of all the partial decay widths to get the total  widths of those four tetraquark states,
\begin{eqnarray}\label{Widths}
\Gamma\left(Y_{PA}\right)&=&391.1\pm 21.4\, \rm{MeV}\, , \nonumber \\
\Gamma\left(Y_{AA}\right)&=&27.8\pm 2.3\, \rm{MeV}\, , \nonumber \\
\Gamma\left(Y_{\widetilde{A}V}\right)&=&266.7\pm 23.9\, \rm{MeV}\, , \nonumber \\
\Gamma\left(Y_{S\widetilde{V}}\right)&=&61.5\pm 7.3\, \rm{MeV}\, .
\end{eqnarray}

It can be clearly seen that the values of total  widths of the four states are quite different from each other. Therefore we can distinguish them easily in the  high energy experiments.
The prediction width $61.5\pm 7.3\, \rm{MeV}$ of the $Y_{S\widetilde{V}}$ is well compatible with the width $48\pm15\pm3\, \rm{MeV}$ from the Belle collaboration \cite{Belle-Y4660-2007}.
Moreover, our prediction of the width of the $Y_{S\widetilde{V}}$ is in excellent agreement with the average width of different experiments $55\pm9\, \rm{MeV}$ of the $Y(4660)$ from the Particle Data Group \cite{PDG}, which supports the assignment of the $Y(4660)$ as the $[sc]_S[\bar{s}\bar{c}]_{\widetilde{V}}-[sc]_{\widetilde{V}}[\bar{s}\bar{c}]_S$ tetraquark state.
While the widths $155.4\pm24.8\pm0.8$ \cite{BESIII-Y4660-2021} and $218.3\pm71.9$ \cite{BESIII-Y4660-2023} from the BESIII collaboration are larger than those from Belle and BaBar collaborations in magnitude, it indicates that the $Y(4660)$ maybe have several Fock components. The predictions $\Gamma\left(Y_{\widetilde{A}V}\right)=266.7\pm 23.9\, \rm{MeV}$ and $\Gamma\left(Y_{PA}\right)=391.1\pm 21.4\, \rm{MeV}$ are too large and the prediction $\Gamma\left(Y_{AA}\right)=27.8\pm 2.3\, \rm{MeV}$ is too small to match with the experimental width.
Thus the present study disfavors the assignment of the $Y(4660)$ to be $[sc]_{\widetilde{A}}[\bar{s}\bar{c}]_V+[sc]_V[\bar{s}\bar{c}]_{\widetilde{A}}$, $[uc]_P[\bar{u}\bar{c}]_A+[dc]_P[\bar{d}\bar{c}]_A-[uc]_A[\bar{u}\bar{c}]_P-[dc]_A[\bar{d}\bar{c}]_P$ or $[uc]_A[\bar{u}\bar{c}]_A+[dc]_A[\bar{d}\bar{c}]_A$  tetraquark states. They are meaningful to diagnose the vector exotic states and waiting to be examined in the future.

\section{Conclusion}
In this work, we take  four four-quark currents to explore the two-body strong decays of the $[sc]_{\widetilde{A}}[\bar{s}\bar{c}]_V+[sc]_V[\bar{s}\bar{c}]_{\widetilde{A}}$, $[sc]_S[\bar{s}\bar{c}]_{\widetilde{V}}-[sc]_{\widetilde{V}}[\bar{s}\bar{c}]_S$, $[uc]_P[\bar{u}\bar{c}]_A+[dc]_P[\bar{d}\bar{c}]_A-[uc]_A[\bar{u}\bar{c}]_P-[dc]_A[\bar{d}\bar{c}]_P$ and $[uc]_A[\bar{u}\bar{c}]_A+[dc]_A[\bar{d}\bar{c}]_A$ tetraquark states around $4.66\,\rm{GeV}$ with the quantum numbers $J^{PC}=1^{--}$  respectively   within the framework of the QCD sum rules.
We perform the operator product expansion up to the vacuum condensates of dimension 5 and match the QCD side with hadron side based on rigorous quark-hadron duality. The resulting  total widths of these states are quite different from each other.
The predicted width $61.5\pm 7.3\, \rm{MeV}$ of the $Y_{S\widetilde{V}}$ is in excellent  agreement with the experimental values of the $Y(4660)$, which favors the $[sc]_S[\bar{s}\bar{c}]_{\widetilde{V}}-[sc]_{\widetilde{V}}[\bar{s}\bar{c}]_S$ type tetraquark interpretation for the $Y(4660)$. While the predictions for the other tetraquark states serve as a guide for future experiments.

\section*{Appendix}
\appendix
\section{The hadronic coupling constants}\label{app-couplingconstants}
In this section, we write down the definitions of other  hadronic coupling constants.
\begin{eqnarray}
\langle \bar{D}^*(p)D^*(q)|Y_{PA}(p^\prime)\rangle&=&i\xi^*\cdot \xi^*(p-q)\cdot \varepsilon \,G_{\bar{D}^*D^*PA}\, , \nonumber\\
\langle \bar{D}_s^*(p)D_s^*(q)|Y_{\widetilde{A}V}(p^\prime)\rangle&=&-\xi^*\cdot \xi^*(p-q)\cdot \varepsilon \,G_{\bar{D}_s^*D_s^*\widetilde{A}V}\, , \nonumber\\
\langle \bar{D}^*(p)D^*(q)|Y_{AA}(p^\prime)\rangle&=&i\xi^*\cdot \xi^*(p-q)\cdot \varepsilon \,G_{\bar{D}^*D^*AA}\, , \nonumber\\
\langle \bar{D}_s^*(p)D_s^*(q)|Y_{S\widetilde{V}}(p^\prime)\rangle&=&-i\xi^*\cdot \xi^*(p-q)\cdot \varepsilon \,G_{\bar{D}_s^*D_s^*S\widetilde{V}}\, ,
\end{eqnarray}

\begin{eqnarray}
\langle \bar{D}_0(p)D^*(q)|Y_{PA}(p^\prime)\rangle&=&-i\xi^*\cdot \varepsilon p\cdot q \,G_{\bar{D}_0D^*PA}\, , \nonumber\\
\langle \bar{D}_{s0}(p)D_s^*(q)|Y_{\widetilde{A}V}(p^\prime)\rangle&=&\xi^*\cdot \varepsilon  \,G_{\bar{D}_{s0}D_s^*\widetilde{A}V}\, , \nonumber\\
\langle \bar{D}_0(p)D^*(q)|Y_{AA}(p^\prime)\rangle&=&i\xi^*\cdot \varepsilon p\cdot q \,G_{\bar{D}_0D^*AA}\, , \nonumber\\
\langle \bar{D}_0(p)D^*(q)|X_{AA}(p^\prime)\rangle&=&i\varepsilon^{\lambda\tau\rho\sigma} q_\lambda \xi^*_\tau p^\prime_\rho \varepsilon_\sigma  \,\bar{G}_{\bar{D}_0D^*AA}\, , \nonumber\\
\langle \bar{D}_{s0}(p)D_s^*(q)|Y_{S\widetilde{V}}(p^\prime)\rangle&=&-i\xi^*\cdot \varepsilon  \,G_{\bar{D}_{s0}D_s^*S\widetilde{V}}\, ,
\end{eqnarray}

\begin{eqnarray}
\langle \bar{D}_1(p)D(q)|Y_{PA}(p^\prime)\rangle&=&-\xi^*\cdot \varepsilon \,G_{\bar{D}_1DPA}\, , \nonumber \\
\langle \bar{D}_{s1}(p)D_s(q)|Y_{\widetilde{A}V}(p^\prime)\rangle&=&i\xi^*\cdot \varepsilon \,G_{\bar{D}_{s1}D_s\widetilde{A}V}\, , \nonumber \\
\langle \bar{D}_1(p)D(q)|Y_{AA}(p^\prime)\rangle&=&-i\xi^*\cdot \varepsilon \,G_{\bar{D}_1DAA}\, , \nonumber \\
\langle \bar{D}_1(p)D(q)|X_{AA}(p^\prime)\rangle&=&i\varepsilon^{\lambda\tau\rho\sigma}
p_\lambda \xi^*_\tau p^\prime_\rho \varepsilon_\sigma \,\bar{G}_{\bar{D}_1DAA}\, , \nonumber \\
\langle \bar{D}_{s1}(p)D_s(q)|Y_{S\widetilde{V}}(p^\prime)\rangle&=&-\xi^*\cdot \varepsilon \,G_{\bar{D}_{s1}D_s S\widetilde{V}}\, ,
\end{eqnarray}

\begin{eqnarray}
\langle \eta_c(p)\omega(q)|Y_{PA}(p^\prime)\rangle&=&i\varepsilon^{\lambda\tau\rho\sigma}
q_\lambda \xi^*_\tau p^\prime_\rho \varepsilon_\sigma \,G_{\eta_c \omega PA}\, , \nonumber \\
\langle \eta_c(p)\phi(q)|Y_{\widetilde{A}V}(p^\prime)\rangle&=&i\varepsilon^{\lambda\tau\rho\sigma}
q_\lambda \xi^*_\tau p^\prime_\rho \varepsilon_\sigma \,G_{\eta_c \phi \widetilde{A}V}\, ,\nonumber \\
\langle \eta_c(p)\omega(q)|Y_{AA}(p^\prime)\rangle&=&-i\varepsilon^{\lambda\tau\rho\sigma}
q_\lambda \xi^*_\tau p^\prime_\rho \varepsilon_\sigma \,G_{\eta_c \omega AA}\, , \nonumber \\
\langle \eta_c(p)\phi(q)|Y_{S\widetilde{V}}(p^\prime)\rangle&=&i\varepsilon^{\lambda\tau\rho\sigma}
q_\lambda \xi^*_\tau p^\prime_\rho \varepsilon_\sigma \,G_{\eta_c \phi S\widetilde{V}}\, , \nonumber \\
\langle \eta_c(p)\phi(q)|X_{S\widetilde{A}}(p^\prime)\rangle&=&i \xi^* \cdot \varepsilon  \,\bar{G}_{\eta_c \phi S\widetilde{A}}\, ,
\end{eqnarray}

\begin{eqnarray}
\langle J/\psi(p)\omega(q)|Y_{PA}(p^\prime)\rangle&=&\xi^*\cdot \xi^*(p-q)\cdot \varepsilon \,G_{J/\psi\omega PA}\, , \nonumber \\
\langle J/\psi(p)\phi(q)|Y_{\widetilde{A}V}(p^\prime)\rangle&=&\xi^*\cdot \xi^*(p-q)\cdot \varepsilon \,G_{J/\psi\phi\widetilde{A}V}\, , \nonumber \\
\langle J/\psi(p)\omega(q)|Y_{AA}(p^\prime)\rangle&=&i\xi^*\cdot \xi^*(p-q)\cdot \varepsilon \,G_{J/\psi\omega AA}\, , \nonumber \\
\langle J/\psi(p)\phi(q)|Y_{S\widetilde{V}}(p^\prime)\rangle&=&-i\xi^*\cdot \xi^*(p-q)\cdot \varepsilon \,G_{J/\psi\phi S\widetilde{V}}\, ,
\end{eqnarray}

\begin{eqnarray}
\langle \chi_{c0}(p)\omega(q)|Y_{PA}(p^\prime)\rangle&=&i\xi^*\cdot \varepsilon \,G_{\chi_{c0}\omega PA}\, , \nonumber \\
\langle \chi_{c0}(p)\phi(q)|Y_{\widetilde{A}V}(p^\prime)\rangle&=&-\xi^*\cdot \varepsilon \,G_{\chi_{c0}\phi\widetilde{A}V}\, , \nonumber \\
\langle \chi_{c0}(p)\omega(q)|Y_{AA}(p^\prime)\rangle&=&i\xi^*\cdot \varepsilon \,G_{\chi_{c0}\omega AA}\, , \nonumber \\
\langle \chi_{c0}(p)\omega(q)|X_{AA}(p^\prime)\rangle&=&i\varepsilon^{\lambda\tau\rho\sigma}
q_\lambda \xi^*_\tau p^\prime_\rho \varepsilon_\sigma \,\bar{G}_{\chi_{c0}\omega AA}\, , \nonumber \\
\langle \chi_{c0}(p)\phi(q)|Y_{S\widetilde{V}}(p^\prime)\rangle&=&-i\xi^*\cdot \varepsilon \,G_{\chi_{c0}\phi S\widetilde{V}}\, ,
\end{eqnarray}

\begin{eqnarray}
\langle \chi_{c1}(p)\omega(q)|Y_{PA}(p^\prime)\rangle&=&
\varepsilon^{\rho\sigma\lambda\tau}p_\rho \xi_\sigma^* \xi^*_\lambda \varepsilon_\tau p \cdot q\,G_{\chi_{c1}\omega PA}\, , \nonumber \\
\langle \chi_{c1}(p)\phi(q)|Y_{\widetilde{A}V}(p^\prime)\rangle&=&-i
\varepsilon^{\rho\sigma\lambda\tau}p_\rho \xi_\sigma^* \xi^*_\lambda \varepsilon_\tau p \cdot q\,G_{\chi_{c1}\phi\widetilde{A}V}\, , \nonumber \\
\langle \chi_{c1}(p)\omega(q)|Y_{AA}(p^\prime)\rangle&=&
-\varepsilon^{\rho\sigma\lambda\tau}p_\rho \xi_\sigma^* \xi^*_\lambda \varepsilon_\tau p \cdot q\,G_{\chi_{c1}\omega AA}\, , \nonumber \\
\langle \chi_{c1}(p)\phi(q)|Y_{S\widetilde{V}}(p^\prime)\rangle&=&
\varepsilon^{\rho\sigma\lambda\tau}p_\rho \xi_\sigma^* \xi^*_\lambda \varepsilon_\tau \,G_{\chi_{c1}\phi S\widetilde{V}}\, ,
\end{eqnarray}

\begin{eqnarray}
\langle J/\psi(p)f_0(500)(q)|Y_{PA}(p^\prime)\rangle&=&-i\xi^* \cdot \varepsilon  \,G_{J/\psi f_0(500)PA}\, , \nonumber \\
\langle J/\psi(p)f_0(980)(q)|Y_{\widetilde{A}V}(p^\prime)\rangle&=&\xi^* \cdot \varepsilon  \,G_{J/\psi f_0(980)\widetilde{A}V}\, , \nonumber \\
\langle J/\psi(p)f_0(500)(q)|Y_{AA}(p^\prime)\rangle&=&i\xi^* \cdot \varepsilon  \,G_{J/\psi f_0(500)AA}\, , \nonumber \\
\langle J/\psi(p)f_0(500)(q)|X_{AA}(p^\prime)\rangle&=&i\varepsilon^{\rho\sigma\lambda\tau}p_\rho \xi_\sigma^* p^\prime_\lambda \varepsilon_\tau  \,\bar{G}_{J/\psi f_0(500)AA}\, , \nonumber \\
\langle J/\psi(p)f_0(980)(q)|Y_{S\widetilde{V}}(p^\prime)\rangle&=&-i\xi^* \cdot \varepsilon  \,G_{J/\psi f_0(980)S\widetilde{V}}\, .
\end{eqnarray}

\section{The correlation functions}\label{app-correlationfunctions}
In this section, we present the correlation functions on the phenomenological side for the currents $J_{\mu\nu}^{AA}$, $J_{\mu}^{\widetilde{A}V}$ and $J_{\mu\nu}^{S\widetilde{V}}$.
\begin{eqnarray}
\Pi^{\bar{D}D AA}_{\mu\nu}(p,q)&=&\Pi_{\bar{D}D AA}(p^{\prime2},p^2,q^2)
\,\left[2(p_\mu q_\nu-p_\nu q_\mu)\right]+\cdots\, ,
\end{eqnarray}

\begin{eqnarray}
\Pi^{\bar{D}^*D AA}_{\alpha\mu\nu}(p,q)&=&
\Pi_{\bar{D}^*D AA}(p^{\prime2},p^2,q^2)
\,\left(\varepsilon_{\alpha\mu\lambda\tau}p^\lambda q^\tau q_\nu-\varepsilon_{\alpha\nu\lambda\tau}p^\lambda q^\tau q_\mu\right)+\cdots\, ,
\end{eqnarray}

\begin{eqnarray}
\Pi^{\bar{D}^*D^*AA}_{\alpha\beta\mu\nu}(p,q)&=&
\Pi_{\bar{D}^*D^*AA}(p^{\prime2},p^2,q^2)
\,\left[2g_{\alpha\beta}(p_\nu q_\mu-p_\mu q_\nu)\right]+\cdots\, ,
\end{eqnarray}

\begin{eqnarray}
\Pi^{\bar{D}_0D^*AA}_{\alpha\mu\nu}(p,q)&=&
\Pi_{\bar{D}_0D^*AA}(p^{\prime2},p^2,q^2)
\,\left(g_{\alpha\mu}q_\nu\right)+\cdots\, ,
\end{eqnarray}

\begin{eqnarray}
\Pi^{\bar{D}_1DAA}_{\alpha\mu\nu}(p,q)&=&
\Pi_{\bar{D}_1DAA}(p^{\prime2},p^2,q^2)
\,\left(-g_{\alpha\mu}p_\nu\right)+\cdots\, ,
\end{eqnarray}

\begin{eqnarray}
\Pi^{\eta_c\omega AA}_{\alpha\mu\nu}(p,q)&=&
\Pi_{\eta_c\omega AA}(p^{\prime2},p^2,q^2)
\,\left(\varepsilon_{\alpha\mu\lambda\tau}p^\lambda p_\nu q^\tau-\varepsilon_{\alpha\nu\lambda\tau}p^\lambda p_\mu q^\tau\right)+\cdots\, ,
\end{eqnarray}

\begin{eqnarray}
\Pi^{J/\psi\omega AA}_{\alpha\beta\mu\nu}(p,q)&=&
\Pi_{J/\psi\omega AA}(p^{\prime2},p^2,q^2)
\,\left[2g_{\alpha\beta}(p_\mu q_\nu-p_\nu q_\mu)\right]+\cdots\, ,
\end{eqnarray}

\begin{eqnarray}
\Pi^{\chi_{c0}\omega AA}_{\alpha\mu\nu}(p,q)&=&
\Pi_{\chi_{c0}\omega AA}(p^{\prime2},p^2,q^2)
\,\left(g_{\alpha\mu}q_\nu\right)+\cdots\, ,
\end{eqnarray}

\begin{eqnarray}
\Pi^{\chi_{c1}\omega AA}_{\alpha\beta\mu\nu}(p,q)&=&
\Pi_{\chi_{c1}\omega AA}(p^{\prime2},p^2,q^2)
\,i\left(\varepsilon_{\alpha\beta\mu\tau}p^\tau q_\nu-\varepsilon_{\alpha\beta\nu\tau}p^\tau q_\mu \right)+\cdots\, ,
\end{eqnarray}

\begin{eqnarray}
\Pi^{J/\psi f_0(500)AA}_{\alpha\mu\nu}(p,q)&=&
\Pi_{J/\psi f_0(500)AA}(p^{\prime2},p^2,q^2)
\,\left(g_{\alpha\mu}p_\nu\right)+\cdots\, ,
\end{eqnarray}

\begin{eqnarray}
\Pi^{\bar{D}_sD_s\widetilde{A}V}_{\mu}(p,q)&=&\Pi_{\bar{D}_sD_s\widetilde{A}V}(p^{\prime2},p^2,q^2)
\,i\left(p-q\right)_\mu+\cdots\, ,
\end{eqnarray}

\begin{eqnarray}
\Pi^{\bar{D}_s^*D_s\widetilde{A}V}_{\alpha\mu}(p,q)&=&
\Pi_{\bar{D}_s^*D_s\widetilde{A}V}(p^{\prime2},p^2,q^2)
\,\left(-i\varepsilon_{\alpha\mu\lambda\tau}p^\lambda q^\tau\right)+\cdots\, ,
\end{eqnarray}

\begin{eqnarray}
\Pi^{\bar{D}_s^*D_s^*\widetilde{A}V}_{\alpha\beta\mu}(p,q)&=&
\Pi_{\bar{D}_s^*D_s^*\widetilde{A}V}(p^{\prime2},p^2,q^2)
\,\left(-ig_{\alpha\beta}p_\mu\right)+\cdots\, ,
\end{eqnarray}

\begin{eqnarray}
\Pi^{\bar{D}_{s0}D_s^*\widetilde{A}V}_{\alpha\mu}(p,q)&=&
\Pi_{\bar{D}_{s0}D_s^*\widetilde{A}V}(p^{\prime2},p^2,q^2)
\,\left(-ig_{\alpha\mu}p \cdot q\right)+\cdots\, ,
\end{eqnarray}

\begin{eqnarray}
\Pi^{\bar{D}_{s1}D_s\widetilde{A}V}_{\alpha\mu}(p,q)&=&
\Pi_{\bar{D}_{s1}D_s\widetilde{A}V}(p^{\prime2},p^2,q^2)
\,\left(g_{\alpha\mu}\right)+\cdots\, ,
\end{eqnarray}

\begin{eqnarray}
\Pi^{\eta_c\phi\widetilde{A}V}_{\alpha\mu}(p,q)&=&
\Pi_{\eta_c\phi\widetilde{A}V}(p^{\prime2},p^2,q^2)
\,\left(-i\varepsilon_{\alpha\mu\lambda\tau}p^\lambda q^\tau\right)+\cdots\, ,
\end{eqnarray}

\begin{eqnarray}
\Pi^{J/\psi\phi\widetilde{A}V}_{\alpha\beta\mu}(p,q)&=&
\Pi_{J/\psi\phi\widetilde{A}V}(p^{\prime2},p^2,q^2)
\,\left(ig_{\alpha\beta}p_\mu\right)+\cdots\, ,
\end{eqnarray}

\begin{eqnarray}
\Pi^{\chi_{c0}\phi\widetilde{A}V}_{\alpha\mu}(p,q)&=&
\Pi_{\chi_{c0}\phi\widetilde{A}V}(p^{\prime2},p^2,q^2)
\,\left(ig_{\alpha\mu}\right)+\cdots\, ,
\end{eqnarray}

\begin{eqnarray}
\Pi^{\chi_{c1}\phi\widetilde{A}V}_{\alpha\beta\mu}(p,q)&=&
\Pi_{\chi_{c1}\phi\widetilde{A}V}(p^{\prime2},p^2,q^2)
\,\left(-\varepsilon_{\alpha\beta\mu\tau}p^\tau \,p \cdot q\right)+\cdots\, ,
\end{eqnarray}

\begin{eqnarray}
\Pi^{J/\psi f_0(980)\widetilde{A}V}_{\alpha\mu}(p,q)&=&
\Pi_{J/\psi f_0(980)\widetilde{A}V}(p^{\prime2},p^2,q^2)
\,\left(-ig_{\alpha\mu}\right)+\cdots\, ,
\end{eqnarray}

\begin{eqnarray}
\Pi^{\bar{D}_s D_s S\widetilde{V}}_{\mu\nu}(p,q)&=&\Pi_{\bar{D}_s D_s S\widetilde{V}}(p^{\prime2},p^2,q^2)
\,\left(-2\varepsilon_{\mu\nu\lambda\tau}p^\lambda q^\tau \right)+\cdots\, ,
\end{eqnarray}

\begin{eqnarray}
\Pi^{\bar{D}_s^* D_s S\widetilde{V}}_{\alpha\mu\nu}(p,q)&=&
\Pi_{\bar{D}_s^* D_s S\widetilde{V}}(p^{\prime2},p^2,q^2)
\,\,p \cdot q \,\left[ g_{\mu\alpha}(p-q)_\nu-g_{\nu\alpha}(p-q)_\mu\right]+\cdots\, ,
\end{eqnarray}

\begin{eqnarray}
\Pi^{\bar{D}_s^*D_s^*S\widetilde{V}}_{\alpha\beta\mu\nu}(p,q)&=&
\Pi_{\bar{D}_s^*D_s^*S\widetilde{V}}(p^{\prime2},p^2,q^2)
\,\left(-2g_{\alpha\beta}\varepsilon_{\mu\nu\lambda\tau} p^\lambda q^\tau \right)+\cdots\, ,
\end{eqnarray}

\begin{eqnarray}
\Pi^{\bar{D}_{s0}D_s^*S\widetilde{V}}_{\alpha\mu\nu}(p,q)&=&
\Pi_{\bar{D}_{s0}D_s^*S\widetilde{V}}(p^{\prime2},p^2,q^2)
\,\left(-\varepsilon_{\alpha\mu\nu\tau}p^\tau \right)+\cdots\, ,
\end{eqnarray}

\begin{eqnarray}
\Pi^{\bar{D}_{s1}D_s S\widetilde{V}}_{\alpha\mu\nu}(p,q)&=&
\Pi_{\bar{D}_{s1}D_s S\widetilde{V}}(p^{\prime2},p^2,q^2)
\,\left(i\varepsilon_{\alpha\mu\nu\tau}q^\tau \right)+\cdots\, ,
\end{eqnarray}

\begin{eqnarray}
\Pi^{\eta_c\phi S\widetilde{V}}_{\alpha\mu\nu}(p,q)&=&
\Pi_{\eta_c\phi S\widetilde{V}}(p^{\prime2},p^2,q^2)
\,\left(g_{\mu\alpha}p_\nu-g_{\nu\alpha}p_\mu \right)+\cdots\, ,
\end{eqnarray}

\begin{eqnarray}
\Pi^{J/\psi\phi S\widetilde{V}}_{\alpha\beta\mu\nu}(p,q)&=&
\Pi_{J/\psi\phi S\widetilde{V}}(p^{\prime2},p^2,q^2)
\,\left(2g_{\alpha\beta}\varepsilon_{\mu\nu\lambda\tau }p^\lambda q^\tau\right)+\cdots\, ,
\end{eqnarray}

\begin{eqnarray}
\Pi^{\chi_{c0}\phi S\widetilde{V}}_{\alpha\mu\nu}(p,q)&=&
\Pi_{\chi_{c0}\phi S\widetilde{V}}(p^{\prime2},p^2,q^2)
\,\left(-\varepsilon_{\alpha\mu\nu\tau}q^\tau\right)+\cdots\, ,
\end{eqnarray}

\begin{eqnarray}
\Pi^{\chi_{c1}\omega S\widetilde{V}}_{\alpha\beta\mu\nu}(p,q)&=&
\Pi_{\chi_{c1}\omega S\widetilde{V}}(p^{\prime2},p^2,q^2)
\, i  \left(g_{\alpha\mu}g_{\beta\nu}- g_{\alpha\nu}g_{\beta\mu} \right)+\cdots\, ,
\end{eqnarray}

\begin{eqnarray}
\Pi^{J/\psi f_0(980) S\widetilde{V}}_{\alpha\mu\nu}(p,q)&=&
\Pi_{J/\psi f_0(980) S\widetilde{V}}(p^{\prime2},p^2,q^2)
\,\left(-\varepsilon_{\alpha\mu\nu\tau}p^\tau\right)+\cdots\, .
\end{eqnarray}

\section{The spectral densities}\label{app-spectraldensities}
In this section, we present the explicit expressions of the QCD side of the QCD sum rules.

\subsection{The QCD side for the current $J_{\mu}^{PA}$}
\begin{eqnarray}
\Pi^{QCD}_{\bar{D}^*D PA}(T^2)&=&-\frac{m_c\langle\bar{q}g_s\sigma G q\rangle}{192\pi^2}\int_{m_c^2}^{s^0_D}du \left(3-\frac{m_c^2}{u}\right)\frac{1}{u} \exp\left(-\frac{u+m_c^2}{T^2} \right)\nonumber\\
&&+\frac{m_c\langle\bar{q}g_s\sigma G q\rangle}{64\pi^2}\int_{m_c^2}^{s^0_{D^*}}ds \left(1-\frac{m_c^2}{s}\right)\frac{1}{s} \exp\left(-\frac{s+m_c^2}{T^2} \right)\, ,
\end{eqnarray}

\begin{eqnarray}	 \Pi^{QCD}_{\bar{D}^*D^*PA}(T^2)&=&\frac{m_c}{128\pi^4}
\int_{m_c^2}^{s^0_{D^*}}ds	\int_{m_c^2}^{s^0_{D^*}}du \left(1-\frac{m_c^2}{s}\right)^2\left(1-\frac{m_c^2}{u}\right)^2 \left(2u+m_c^2\right) \, \exp\left(-\frac{s+u}{T^2} \right) \nonumber\\
&&+\frac{\langle\bar{q}q\rangle}{24\pi^2}\int_{m_c^2}^{s^0_{D^*}}ds \left(1-\frac{m_c^2}{s}\right)^2 \left(s-m_c^2\right)\,\exp\left(-\frac{s+m_c^2}{T^2} \right)\nonumber\\
&&+\frac{\langle\bar{q}g_s\sigma G q\rangle}{576\pi^2T^2}\left( 4+\frac{3m_c^2}{T^2}\right)\int_{m_c^2}^{s^0_{D^*}}du\left( 1-\frac{m_c^2}{u}\right)^2 \left(2u+m_c^2\right)\, \exp\left(-\frac{u+m_c^2}{T^2} \right)\nonumber\\
&&+\frac{m_c^4\langle\bar{q}g_s\sigma G q\rangle}{64\pi^2T^4}\int_{m_c^2}^{s^0_{D^*}}ds \left(1-\frac{m_c^4}{s^2}\right) \exp\left(-\frac{s+m_c^2}{T^2} \right)\nonumber\\
&&-\frac{m_c^2\langle\bar{q}g_s\sigma G q\rangle}{192\pi^2}\int_{m_c^2}^{s^0_{D^*}}ds \frac{1}{s}\left(2-\frac{4m_c^2}{s}\right)\, \exp\left(-\frac{s+m_c^2}{T^2} \right)\, ,
\end{eqnarray}

\begin{eqnarray}	 \Pi^{QCD}_{\bar{D}_0D^*PA}(T^2)&=&\frac{3m_c^2}{128\pi^4}
\int_{m_c^2}^{s^0_{D_0}}ds 	\int_{m_c^2}^{s^0_{D^*}}du \left(1-\frac{m_c^2}{s}\right)^2\left(1-\frac{m_c^4}{u^2}\right)  \, \exp\left(-\frac{s+u}{T^2} \right) \nonumber\\
&&-\frac{m_c\langle\bar{q}q\rangle}{16\pi^2}\int_{m_c^2}^{s^0_{D_0}}ds \left(1-\frac{m_c^2}{s}\right)^2 \,\exp\left(-\frac{s+m_c^2}{T^2} \right)\nonumber\\
&&+\frac{m_c\langle\bar{q}q\rangle}{16\pi^2}\int_{m_c^2}^{s^0_{D^*}}du \left(1-\frac{m_c^2}{u}\right)^2 \,\exp\left(-\frac{u+m_c^2}{T^2} \right)\nonumber\\
&&-\frac{m_c^3\langle\bar{q}g_s\sigma G q\rangle}{64\pi^2T^4} \int_{m_c^2}^{s^0_{D^*}}du \left( 1-\frac{m_c^2}{u}\right)^2 \, \exp\left(-\frac{u+m_c^2}{T^2} \right)\nonumber\\
&&+\frac{m_c\langle\bar{q}g_s\sigma G q\rangle}{192\pi^2T^2}\left(4+\frac{3m_c^2}{T^2}\right) \int_{m_c^2}^{s^0_{D_0}}ds \left(1-\frac{m_c^2}{s}\right)^2 \exp\left(-\frac{s+m_c^2}{T^2} \right)\nonumber\\
&&+\frac{m_c\langle\bar{q}g_s\sigma G q\rangle}{192\pi^2}\int_{m_c^2}^{s^0_{D^*}}du \left(1+\frac{m_c^4}{u^2}\right)\frac{1}{u-m_c^2}\, \exp\left(-\frac{u+m_c^2}{T^2} \right)	 \nonumber\\
&&+\frac{m_c^3\langle\bar{q}g_s\sigma G q\rangle}{64\pi^2}\int_{m_c^2}^{s^0_{D^*}}du \frac{1}{u^2}\, \exp\left(-\frac{u+m_c^2}{T^2} \right)	 \nonumber\\
&&-\frac{m_c\langle\bar{q}g_s\sigma G q\rangle}{192\pi^2}\int_{m_c^2}^{s^0_{D_0}}ds \left(3-\frac{2m_c^2}{s}\right)\frac{1}{s}\, \exp\left(-\frac{s+m_c^2}{T^2} \right)\, ,
\end{eqnarray}

\begin{eqnarray}	 \Pi^{QCD}_{\bar{D}_1D PA}(T^2)&=&\frac{1}{128\pi^4}\int_{m_c^2}^{s^0_{D_1}}ds
\int_{m_c^2}^{s^0_D}du \left(1-\frac{m_c^2}{s}\right)^2\left(1-\frac{m_c^2}{u}\right)^2 \, u\, \left(2s+m_c^2\right) \, \exp\left(-\frac{s+u}{T^2} \right) \nonumber\\
&&-\frac{m_c\langle\bar{q}q\rangle}{48\pi^2}\int_{m_c^2}^{s^0_{D_1}}ds \left(1-\frac{m_c^2}{s}\right)^2\left(2s+m_c^2\right) \,\exp\left(-\frac{s+m_c^2}{T^2} \right)\nonumber\\
&&+\frac{m_c\langle\bar{q}q\rangle}{16\pi^2}\int_{m_c^2}^{s^0_D}du \,u\,\left(1-\frac{m_c^2}{u}\right)^2 \,\exp\left(-\frac{u+m_c^2}{T^2} \right)\nonumber\\
&&-\frac{m_c^3\langle\bar{q}g_s\sigma G q\rangle}{64\pi^2T^4}\int_{m_c^2}^{s^0_D}du\, u\,\left( 1-\frac{m_c^2}{u}\right)^2 \, \exp\left(-\frac{u+m_c^2}{T^2} \right)\nonumber\\
&&-\frac{m_c\langle\bar{q}g_s\sigma G q\rangle}{192\pi^2T^2}\left(2-\frac{m_c^2}{T^2}\right)\int_{m_c^2}^{s^0_{D_1}}ds \left(1-\frac{m_c^2}{s}\right)^2\left(2s+m_c^2\right)\exp\left(-\frac{s+m_c^2}{T^2} \right)\nonumber\\
&&+\frac{m_c\langle\bar{q}g_s\sigma G q\rangle}{192\pi^2}\int_{m_c^2}^{s^0_D}du \left(3-\frac{2m_c^2}{u}\right)\, \exp\left(-\frac{u+m_c^2}{T^2} \right)	 \nonumber\\
&&-\frac{m_c^3\langle\bar{q}g_s\sigma G q\rangle}{192\pi^2}\int_{m_c^2}^{s^0_{D_1}}ds \frac{1}{s}\, \exp\left(-\frac{s+m_c^2}{T^2} \right)\, ,
\end{eqnarray}

\begin{eqnarray}
\Pi^{QCD}_{\eta_c\omega PA}(T^2)
&=&-\frac{m_c\langle\bar{q}q\rangle}{2\sqrt{2}\pi^2}\int_{4m_c^2}^{s^0_{\eta_c}}ds
\frac{\sqrt{\lambda(s,m_c^2,m_c^2)}}{s}  \, \exp\left(-\frac{s}{T^2} \right)\nonumber\\
&&+\frac{m_c\langle\bar{q}g_s\sigma G q\rangle}{6\sqrt{2}\pi^2T^2}\int_{4m_c^2}^{s^0_{\eta_c}}ds
\frac{\sqrt{\lambda(s,m_c^2,m_c^2)}}{s}  \, \exp\left(-\frac{s}{T^2} \right)\nonumber\\
&&-\frac{m_c\langle\bar{q}g_s\sigma G q\rangle}{24\sqrt{2}\pi^2}\int_{4m_c^2}^{s^0_{\eta_c}}ds
\frac{1}{\sqrt{s(s-4m_c^2)}}  \, \exp\left(-\frac{s}{T^2} \right)\, ,
\end{eqnarray}

\begin{eqnarray}
\Pi^{QCD}_{J/\psi\omega PA}(T^2)&=&0\, ,
\end{eqnarray}

\begin{eqnarray}
\Pi^{QCD}_{\chi_{c0}\omega PA}(T^2)&=&\frac{1}{32\sqrt{2}\pi^4}
\int_{4m_c^2}^{s^0_{\chi_{c0}}}ds
\int_{0}^{s^0_\omega}du \frac{\sqrt{\lambda(s,m_c^2,m_c^2)}}{s}\, u \left(s-4m_c^2\right)  \, \exp\left(-\frac{s+u}{T^2} \right)\, ,\nonumber\\
&&
\end{eqnarray}

\begin{eqnarray}
\Pi^{QCD}_{\chi_{c1}\omega PA}(T^2)&=&
-\frac{\langle\bar{q}q\rangle}{12\sqrt{2}\pi^2}\int_{4m_c^2}^{s^0_{\chi_{c1}}}ds
 \frac{\sqrt{\lambda(s,m_c^2,m_c^2)}}{s} \left(1+\frac{2m_c^2}{s}\right)  \exp\left(-\frac{s}{T^2} \right)\nonumber\\
&&+\frac{\langle\bar{q}g_s\sigma G q\rangle}{36\sqrt{2}\pi^2T^2}\int_{4m_c^2}^{s^0_{\chi_{c1}}}ds
\frac{\sqrt{\lambda(s,m_c^2,m_c^2)}}{s} \left(1+\frac{2m_c^2}{s}\right)  \exp\left(-\frac{s}{T^2} \right)\, ,	
\end{eqnarray}

\begin{eqnarray}
\Pi^{QCD}_{J/\psi f_0(500)PA}(T^2)&=&\frac{1}{32\sqrt{2}\pi^4}\int_{4m_c^2}^{s^0_{J/\psi}}ds
\int_{0}^{s^0_{f_0(500)}}du \frac{\sqrt{\lambda(s,m_c^2,m_c^2)}}{s}\,u\left(s+2m_c^2\right)  \exp\left(-\frac{s+u}{T^2} \right)\, ,\nonumber\\
&&
\end{eqnarray}

\subsection{The QCD side for the current $J_{\mu}^{\widetilde{A}V}$}
\begin{eqnarray}
\Pi^{QCD}_{\bar{D}_sD_s\widetilde{A}V}(T^2)&=&\frac{9}{64\sqrt{2}\pi^4}
\int_{m_c^2}^{s^0_{D_s}}ds	\int_{m_c^2}^{s^0_{D_s}}du \left(1-\frac{m_c^2}{s}\right)\left(1-\frac{m_c^2}{u}\right) \, \exp\left(-\frac{s+u}{T^2} \right)\nonumber\\
&&\frac{m_c\left(s-m_c^2\right)\left(u-m_c^2\right)+m_s\left(su+m_c^2(s+u)-3m_c^4\right)}{s}  \nonumber\\
&&-\frac{3m_c\langle\bar{s}s\rangle}{8\sqrt{2}\pi^2}\int_{m_c^2}^{s^0_{D_s}}ds \left(1-\frac{m_c^2}{s}\right) \frac{m_s\left(s+m_c^2\right)+m_c s-m_c^3}{s} \,\exp\left(-\frac{s+m_c^2}{T^2} \right)\nonumber\\
&&-\frac{3\langle\bar{s}s\rangle}{8\sqrt{2}\pi^2}\int_{m_c^2}^{s^0_{D_s}}du \left(1-\frac{m_c^2}{u}\right) \left(2m_sm_c+u-m_c^2\right) \,\exp\left(-\frac{u+m_c^2}{T^2} \right)\nonumber\\
&&+\frac{3m_s m_c\langle\bar{s}s\rangle}{16\sqrt{2}\pi^2}\int_{m_c^2}^{s^0_{D_s}}ds \left(1-\frac{m_c^2}{s}\right)^2 \left(1-\frac{s-m_c^2}{T^2}\right) \,\exp\left(-\frac{s+m_c^2}{T^2} \right)\nonumber\\
&&+\frac{3m_c^3\langle\bar{s}g_s\sigma G s\rangle}{32\sqrt{2}\pi^2T^4} \int_{m_c^2}^{s^0_{D_s}}ds \left(1-\frac{m_c^2}{s}\right) \frac{m_s\left(s+m_c^2\right)+m_cs-m_c^3}{s}\, \exp\left(-\frac{s+m_c^2}{T^2} \right)\nonumber\\
&&+\frac{3m_c^2\langle\bar{s}g_s\sigma G s\rangle}{32\sqrt{2}\pi^2T^4} \int_{m_c^2}^{s^0_{D_s}}du \left(1-\frac{m_c^2}{u}\right) \left(2m_sm_c+u-m_c^2\right)\, \exp\left(-\frac{u+m_c^2}{T^2} \right)\nonumber\\
&&+\frac{m_s m_c^3\langle\bar{s}g_s\sigma G s\rangle}{32\sqrt{2}\pi^2T^6} \int_{m_c^2}^{s^0_{D_s}}ds \left(1-\frac{m_c^2}{s}\right)^2 \left(s-m_c^2\right)\, \exp\left(-\frac{s+m_c^2}{T^2} \right)\nonumber\\
&&-\frac{3m_c\langle\bar{s}g_s\sigma G s\rangle}{16\sqrt{2}\pi^2T^2} \int_{m_c^2}^{s^0_{D_s}}ds \left(1-\frac{m_c^2}{s}\right) \frac{m_s\left(s+m_c^2\right)+m_cs-m_c^3}{s}\, \exp\left(-\frac{s+m_c^2}{T^2} \right)\nonumber\\
&&+\frac{m_c^2\langle\bar{s}g_s\sigma G s\rangle}{32\sqrt{2}\pi^2} \int_{m_c^2}^{s^0_{D_s}}ds \frac{m_sm_c+s-m_c^2}{s^2}\, \exp\left(-\frac{s+m_c^2}{T^2} \right)\nonumber\\
&&+\frac{\langle\bar{s}g_s\sigma G s\rangle}{192\sqrt{2}\pi^2} \int_{m_c^2}^{s^0_{D_s}}du \frac{6m_sm_cu+3u^2-4m_c^2u+m_c^4}{u^2} \, \exp\left(-\frac{u+m_c^2}{T^2} \right)\nonumber\\
&&+\frac{m_c\langle\bar{s}g_s\sigma G s\rangle}{32\sqrt{2}\pi^2}\int_{m_c^2}^{s^0_{D_s}}ds \frac{m_s\left(s^2+m_c^4\right)+m_c^3s-m_c^5}{s^2\left(s-m_c^2\right)} \exp\left(-\frac{s+m_c^2}{T^2} \right)\nonumber\\
&&+\frac{\langle\bar{s}g_s\sigma G s\rangle}{192\sqrt{2}\pi^2}\int_{m_c^2}^{s^0_{D_s}}du \exp\left(-\frac{u+m_c^2}{T^2} \right)\nonumber\\
&&\frac{6m_sm_c^3u+6m_sm_cu^2+m_c^6-m_c^4u-3m_c^2u^2+3u^3}{u^2\left(u-m_c^2\right)} \, ,
\end{eqnarray}

\begin{eqnarray}
\Pi^{QCD}_{\bar{D}_s^*D_s\widetilde{A}V}(T^2)&=&-\frac{m_c\langle\bar{s}g_s\sigma G s\rangle}{96\sqrt{2}\pi^2} \int_{m_c^2}^{s^0_{D_s}}du \frac{2m_sm_c+3m_c^2-u}{u^2}  \exp\left(-\frac{u+m_c^2}{T^2} \right)\nonumber\\
&&-\frac{m_c\langle\bar{s}g_s\sigma G s\rangle}{96\sqrt{2}\pi^2}\int_{m_c^2}^{s^0_{D_s^*}}ds \frac{2m_sm_c+3m_c^2+s}{s^2} \exp\left(-\frac{s+m_c^2}{T^2} \right)\nonumber\\
&&-\frac{\langle\bar{s}g_s\sigma G s\rangle}{96\sqrt{2}\pi^2}\int_{m_c^2}^{s^0_{D_s}}du \frac{5m_s\left(u^2+m_c^4\right)}{u^2\left(u-m_c^2\right)} \, \exp\left(-\frac{u+m_c^2}{T^2} \right)	 \nonumber\\
&&-\frac{\langle\bar{s}g_s\sigma G s\rangle}{96\sqrt{2}\pi^2}\int_{m_c^2}^{s^0_{D_s^*}}ds \frac{5m_s\left(s^2+m_c^4\right)}{s^2\left(s-m_c^2\right)}\, \exp\left(-\frac{s+m_c^2}{T^2} \right) \, ,
\end{eqnarray}

\begin{eqnarray}
\Pi^{QCD}_{\bar{D}_s^*D_s^*\widetilde{A}V}(T^2)&=&-\frac{1}{64\sqrt{2}\pi^4}
\int_{m_c^2}^{s^0_{D_s^*}}ds	\int_{m_c^2}^{s^0_{D_s^*}}du \left(1-\frac{m_c^2}{s}\right)\left(1-\frac{m_c^2}{u}\right) \, \exp\left(-\frac{s+u}{T^2} \right)\nonumber\\
&&\frac{m_s\left(m_c^6+m_c^4(s+7u)-m_c^2u(5s+2u)-2su^2\right) -m_c\left(s-m_c^2\right)\left(u-m_c^2\right)\left(2u+m_c^2\right)}{su}  \nonumber\\
&&-\frac{m_c\langle\bar{s}s\rangle}{8\sqrt{2}\pi^2}\int_{m_c^2}^{s^0_{D_s^*}}ds \left(1-\frac{m_c^2}{s}\right) \frac{m_s\left(s+m_c^2\right)+m_c\left(s-m_c^2\right)}{s} \,\exp\left(-\frac{s+m_c^2}{T^2} \right)\nonumber\\
&&-\frac{\langle\bar{s}s\rangle}{24\sqrt{2}\pi^2}\int_{m_c^2}^{s^0_{D_s^*}}du \left(1-\frac{m_c^2}{u}\right) \frac{6m_sm_cu+\left(u-m_c^2\right)\left(2u+m_c^2\right)}{u} \,\exp\left(-\frac{u+m_c^2}{T^2} \right)\nonumber\\
&&-\frac{3m_s m_c\langle\bar{s}s\rangle}{8\sqrt{2}\pi^2T^2} \int_{m_c^2}^{s^0_{D_s^*}}ds \left(1-\frac{m_c^2}{s}\right)^2 s \,\exp\left(-\frac{s+m_c^2}{T^2} \right)\nonumber\\
&&+\frac{m_c^3\langle\bar{s}g_s\sigma G s\rangle}{32\sqrt{2}\pi^2T^4} \int_{m_c^2}^{s^0_{D_s^*}}ds \left(1-\frac{m_c^2}{s}\right) \frac{m_s\left(s+m_c^2\right)+m_c\left(s-m_c^2\right)}{s}\, \exp\left(-\frac{s+m_c^2}{T^2} \right)\nonumber\\
&&+\frac{\langle\bar{s}g_s\sigma G s\rangle}{288\sqrt{2}\pi^2T^2}\left(4+\frac{3m_c^2}{T^2}\right) \int_{m_c^2}^{s^0_{D_s^*}}du \left(1-\frac{m_c^2}{u}\right) \, \exp\left(-\frac{u+m_c^2}{T^2} \right)\nonumber\\
&&\frac{6m_sm_cu+\left(u-m_c^2\right)\left(2u+m_c^2\right)}{u}\nonumber\\
&&+\frac{m_s m_c\langle\bar{s}g_s\sigma G s\rangle}{96\sqrt{2}\pi^2T^2} \left(1+\frac{m_c^2}{T^2}-\frac{m_c^4}{T^4}\right) \int_{m_c^2}^{s^0_{D_s^*}}ds \left(1-\frac{m_c^2}{s}\right)^2 \, \exp\left(-\frac{s+m_c^2}{T^2} \right)\nonumber\\
&&+\frac{m_s m_c^3\langle\bar{s}g_s\sigma G s\rangle}{288\sqrt{2}\pi^2T^6} \int_{m_c^2}^{s^0_{D_s^*}}du \left(1-\frac{m_c^2}{u}\right)^2 \left(2u+m_c^2\right) \, \exp\left(-\frac{u+m_c^2}{T^2} \right)\nonumber\\
&&+\frac{m_c^2\langle\bar{s}g_s\sigma G s\rangle}{96\sqrt{2}\pi^2} \int_{m_c^2}^{s^0_{D_s^*}}ds \frac{m_sm_c+s+4m_c^2}{s^2}\, \exp\left(-\frac{s+m_c^2}{T^2} \right)\nonumber\\
&&+\frac{\langle\bar{s}g_s\sigma G s\rangle}{96\sqrt{2}\pi^2} \int_{m_c^2}^{s^0_{D_s^*}}du \frac{m_sm_c+m_c^2}{u} \, \exp\left(-\frac{u+m_c^2}{T^2} \right)\nonumber\\
&&+\frac{5m_c\langle\bar{s}g_s\sigma G s\rangle}{96\sqrt{2}\pi^2}\int_{m_c^2}^{s^0_{D_s^*}}ds \frac{m_s\left(s^2+m_c^4\right)}{s^2\left(s-m_c^2\right)} \exp\left(-\frac{s+m_c^2}{T^2} \right)\nonumber\\
&&+\frac{\langle\bar{s}g_s\sigma G s\rangle}{96\sqrt{2}\pi^2}\int_{m_c^2}^{s^0_{D_s^*}}du \frac{5m_sm_c\left(u+m_c^2\right)}{u\left(u-m_c^2\right)} \exp\left(-\frac{u+m_c^2}{T^2} \right) \, ,
\end{eqnarray}

\begin{eqnarray}
\Pi^{QCD}_{J/\psi\omega\widetilde{A}V}(T^2)&=&0\, ,
\end{eqnarray}

\begin{eqnarray}
\Pi^{QCD}_{\chi_{c0}\omega\widetilde{A}V}(T^2)&=&\frac{3}{32\sqrt{2}\pi^4}
\int_{4m_c^2}^{s^0_{\chi_{c0}}}ds	\int_{0}^{s^0_\phi}du \frac{\sqrt{\lambda(s,m_c^2,m_c^2)}}{s}\,u\left(s-4m_c^2\right)  \exp\left(-\frac{s+u}{T^2} \right) \nonumber\\
&&-\frac{3m_s\langle\bar{s}s\rangle}{4\sqrt{2}\pi^2} \int_{4m_c^2}^{s^0_{\chi_{c0}}}ds \frac{\sqrt{\lambda(s,m_c^2,m_c^2)}}{s}   \left(s-4m_c^2\right) \exp\left(-\frac{s}{T^2} \right)\nonumber\\
&&+\frac{m_s\langle\bar{s}g_s\sigma G s\rangle}{48\sqrt{2}\pi^2}\int_{4m_c^2}^{s^0_{\chi_{c0}}}ds
\frac{3s-16m_c^2}{\sqrt{s(s-4m_c^2)}} \exp\left(-\frac{s}{T^2} \right)\, ,
\end{eqnarray}

\begin{eqnarray}
\Pi^{QCD}_{\bar{D}_{s0}D_s^*\widetilde{A}V}(T^2)&=&\frac{3m_c^2}{64\sqrt{2}\pi^4}
\int_{m_c^2}^{s^0_{D_{s0}}}ds	\int_{m_c^2}^{s^0_{D_s^*}}du \left(1-\frac{m_c^2}{s}\right) \left(1-\frac{m_c^2}{u}\right) \exp\left(-\frac{s+u}{T^2} \right) \nonumber\\
&&\frac{2m_sm_c(s-u)+\left(s-m_c^2\right)\left(u-m_c^2\right)}{su} \nonumber\\
&&+\frac{\langle\bar{s}s\rangle}{8\sqrt{2}\pi^2}\int_{m_c^2}^{s^0_{D_{s0}}}ds \left(1-\frac{m_c^2}{s}\right) \frac{m_s\left(s+m_c^2\right)-m_c\left(s-m_c^2\right)}{s} \,\exp\left(-\frac{s+m_c^2}{T^2} \right)\nonumber\\
&&+\frac{\langle\bar{s}s\rangle}{8\sqrt{2}\pi^2}\int_{m_c^2}^{s^0_{D_s^*}}du \left(1-\frac{m_c^2}{u}\right) \frac{m_s\left(u+m_c^2\right)+m_c\left(u-m_c^2\right)}{u} \,\exp\left(-\frac{u+m_c^2}{T^2} \right)\nonumber\\
&&+\frac{m_s m_c^2\langle\bar{s}s\rangle}{16\sqrt{2}\pi^2T^2} \int_{m_c^2}^{s^0_{D_{s0}}}ds \left(1-\frac{m_c^2}{s}\right)^2 \,\exp\left(-\frac{s+m_c^2}{T^2} \right)\nonumber\\
&&-\frac{m_s m_c^2\langle\bar{s}s\rangle}{16\sqrt{2}\pi^2T^2} \int_{m_c^2}^{s^0_{D_s^*}}du \left(1-\frac{m_c^2}{u}\right)^2 \,\exp\left(-\frac{u+m_c^2}{T^2} \right)\nonumber\\
&&-\frac{\langle\bar{s}g_s\sigma G s\rangle}{96\sqrt{2}\pi^2T^2} \left(4+\frac{3m_c^2}{T^2}\right) \int_{m_c^2}^{s^0_{D_{s0}}}ds \left(1-\frac{m_c^2}{s}\right)\, \exp\left(-\frac{s+m_c^2}{T^2} \right)\nonumber\\
&&\frac{m_s\left(s+m_c^2\right)-m_c\left(s-m_c^2\right)}{s}\nonumber\\
&&-\frac{m_c^2\langle\bar{s}g_s\sigma G s\rangle}{32\sqrt{2}\pi^2T^4} \int_{m_c^2}^{s^0_{D_s^*}}du \left(1-\frac{m_c^2}{u}\right) \frac{m_s\left(u+m_c^2\right)+m_c\left(u-m_c^2\right)}{u} \, \exp\left(-\frac{u+m_c^2}{T^2} \right)\nonumber\\
&&-\frac{m_s m_c^4\langle\bar{s}g_s\sigma G s\rangle}{96\sqrt{2}\pi^2T^6} \int_{m_c^2}^{s^0_{D_{s0}}}ds \left(1-\frac{m_c^2}{s}\right)^2 \, \exp\left(-\frac{s+m_c^2}{T^2} \right)\nonumber\\
&&+\frac{m_s m_c^4\langle\bar{s}g_s\sigma G s\rangle}{96\sqrt{2}\pi^2T^6} \int_{m_c^2}^{s^0_{D_s^*}}du \left(1-\frac{m_c^2}{u}\right)^2 \, \exp\left(-\frac{u+m_c^2}{T^2} \right)\nonumber\\
&&-\frac{m_c\langle\bar{s}g_s\sigma G s\rangle}{96\sqrt{2}\pi^2} \int_{m_c^2}^{s^0_{D_{s0}}}ds \frac{3m_sm_c-s-2m_c^2}{s^2} \, \exp\left(-\frac{s+m_c^2}{T^2} \right)\nonumber\\
&&-\frac{m_c\langle\bar{s}g_s\sigma G s\rangle}{96\sqrt{2}\pi^2} \int_{m_c^2}^{s^0_{D_s^*}}du \frac{3m_sm_c+u+2m_c^2}{u^2} \, \exp\left(-\frac{u+m_c^2}{T^2} \right)\nonumber\\
&&-\frac{\langle\bar{s}g_s\sigma G s\rangle}{96\sqrt{2}\pi^2} \int_{m_c^2}^{s^0_{D_{s0}}}ds \frac{5m_s\left(s^2+m_c^4\right)}{s^2\left(s-m_c^2\right)} \, \exp\left(-\frac{s+m_c^2}{T^2} \right)\nonumber\\
&&-\frac{\langle\bar{s}g_s\sigma G s\rangle}{96\sqrt{2}\pi^2} \int_{m_c^2}^{s^0_{D_s^*}}du \frac{5m_s\left(u^2+m_c^4\right)}{u^2\left(u-m_c^2\right)} \, \exp\left(-\frac{u+m_c^2}{T^2} \right) \, ,
\end{eqnarray}

\begin{eqnarray}
\Pi^{QCD}_{\chi_{c1}\omega\widetilde{A}V}(T^2)&=&\frac{m_s}{32\sqrt{2}\pi^4}
\int_{4m_c^2}^{s^0_{\chi_{c1}}}ds	\int_{0}^{s^0_\phi}du \frac{\sqrt{\lambda(s,m_c^2,m_c^2)}}{s} \left(1+\frac{2m_c^2}{s}\right)  \exp\left(-\frac{s+u}{T^2} \right) \nonumber\\
&&-\frac{\langle\bar{s}s\rangle}{12\sqrt{2}\pi^2}\int_{4m_c^2}^{s^0_{\chi_{c1}}}ds
\frac{\sqrt{\lambda(s,m_c^2,m_c^2)}}{s} \left(1+\frac{2m_c^2}{s}\right) \exp\left(-\frac{s}{T^2} \right)\nonumber\\
&&+\frac{\langle\bar{s}g_s\sigma G s\rangle}{36\sqrt{2}\pi^2T^2}\int_{4m_c^2}^{s^0_{\chi_{c1}}}ds
\frac{\sqrt{\lambda(s,m_c^2,m_c^2)}}{s} \left(1+\frac{2m_c^2}{s}\right) \exp\left(-\frac{s}{T^2} \right)\, ,
\end{eqnarray}

\begin{eqnarray}
\Pi^{QCD}_{\bar{D}_{s1}D_s\widetilde{A}V}(T^2)&=&\frac{3}{64\sqrt{2}\pi^4}
\int_{m_c^2}^{s^0_{D_{s1}}}ds	\int_{m_c^2}^{s^0_{D_s}}du \left(1-\frac{m_c^2}{s}\right) \left(1-\frac{m_c^2}{u}\right)  \, \exp\left(-\frac{s+u}{T^2} \right) \nonumber\\
&&\frac{\left(s-m_c^2\right)\left(2s+m_c^2\right)\left(u-m_c^2\right)-6m_s m_c\left(m_c^4-2m_c^2s+m_c s(3u-2s)\right)}{s}  \nonumber\\
&&+\frac{m_c\langle\bar{s}s\rangle}{8\sqrt{2}\pi^2}\int_{m_c^2}^{s^0_{D_{s1}}}ds \left(1-\frac{m_c^2}{s}\right) \frac{6m_sm_cs-\left(s-m_c^2\right)\left(2s+m_c^2\right)}{s} \,\exp\left(-\frac{s+m_c^2}{T^2} \right)\nonumber\\
&&+\frac{3m_c\langle\bar{s}s\rangle}{8\sqrt{2}\pi^2}\int_{m_c^2}^{s^0_{D_s}}du \left(1-\frac{m_c^2}{u}\right) \left(2m_sm_c+u-m_c^2\right) \,\exp\left(-\frac{u+m_c^2}{T^2} \right)\nonumber\\
&&+\frac{m_s\langle\bar{s}s\rangle}{16\sqrt{2}\pi^2} \left(1+\frac{m_c^2}{T^2}\right) \int_{m_c^2}^{s^0_{D_{s1}}}ds \left(1-\frac{m_c^2}{s}\right)^2 \left(2s+m_c^2\right) \,\exp\left(-\frac{s+m_c^2}{T^2} \right)\nonumber\\
&&-\frac{3m_s m_c^2\langle\bar{s}s\rangle}{16\sqrt{2}\pi^2 T^2} \int_{m_c^2}^{s^0_{D_s}}du \left(1-\frac{m_c^2}{u}\right)^2 u \,\exp\left(-\frac{u+m_c^2}{T^2} \right)\nonumber\\
&&+\frac{m_c\langle\bar{s}g_s\sigma G s\rangle}{32\sqrt{2}\pi^2T^2} \left(2-\frac{m_c^2}{T^2}\right) \int_{m_c^2}^{s^0_{D_{s1}}}ds \left(1-\frac{m_c^2}{s}\right) \, \exp\left(-\frac{s+m_c^2}{T^2} \right)\nonumber\\
&&\frac{6m_sm_cs-\left(s-m_c^2\right)\left(2s+m_c^2\right)}{s}\nonumber\\
&&-\frac{3m_c^3\langle\bar{s}g_s\sigma G s\rangle}{32\sqrt{2}\pi^2T^4} \int_{m_c^2}^{s^0_{D_s}}du \left(1-\frac{m_c^2}{u}\right)^2 \left(2m_sm_c+u-m_c^2\right) \, \exp\left(-\frac{u+m_c^2}{T^2} \right)\nonumber\\
&&-\frac{m_s m_c^4\langle\bar{s}g_s\sigma G s\rangle}{96\sqrt{2}\pi^2 T^6} \int_{m_c^2}^{s^0_{D_{s1}}}ds \left(1-\frac{m_c^2}{s}\right)^2 \left(2s+m_c^2\right) \, \exp\left(-\frac{s+m_c^2}{T^2} \right)\nonumber\\
&&+\frac{m_s\langle\bar{s}g_s\sigma G s\rangle}{32\sqrt{2}\pi^2 T^2} \left(-1-\frac{m_c^2}{T^2}+\frac{m_c^4}{T^4}\right) \int_{m_c^2}^{s^0_{D_s}}du \left(1-\frac{m_c^2}{u}\right)^2 u \, \exp\left(-\frac{u+m_c^2}{T^2} \right)\nonumber\\
&&+\frac{m_c^2\langle\bar{s}g_s\sigma G s\rangle}{96\sqrt{2}\pi^2} \int_{m_c^2}^{s^0_{D_{s1}}}ds \frac{3m_s s-m_c\left(s-2m_c^2\right)}{s^2} \, \exp\left(-\frac{s+m_c^2}{T^2} \right)\nonumber\\
&&-\frac{m_c\langle\bar{s}g_s\sigma G s\rangle}{192\sqrt{2}\pi^2} \int_{m_c^2}^{s^0_{D_s}}du \frac{6m_sm_cu+2u\left(3u-2m_c^2\right)}{u^2} \, \exp\left(-\frac{u+m_c^2}{T^2} \right)\nonumber\\
&&-\frac{m_c^2\langle\bar{s}g_s\sigma G s\rangle}{96\sqrt{2}\pi^2} \int_{m_c^2}^{s^0_{D_{s1}}}ds \frac{3m_s\left(s+m_c^2\right)}{s\left(s-m_c^2\right)} \, \exp\left(-\frac{s+m_c^2}{T^2} \right)\nonumber\\
&&-\frac{m_c\langle\bar{s}g_s\sigma G s\rangle}{192\sqrt{2}\pi^2} \int_{m_c^2}^{s^0_{D_s}}du \frac{6m_sm_c\left(u+m_c^2\right)}{u\left(u-m_c^2\right)} \, \exp\left(-\frac{u+m_c^2}{T^2} \right)\, ,
\end{eqnarray}

\begin{eqnarray}
\Pi^{QCD}_{\eta_c\phi\widetilde{A}V}(T^2)&=& \frac{3m_sm_c}{16\sqrt{2}\pi^4}\int_{4m_c^2}^{s^0_{\eta_c}}ds
\int_{0}^{s^0_\phi}du \,\frac{\sqrt{\lambda(s,m_c^2,m_c^2)}}{s} \exp\left(-\frac{s+u}{T^2} \right)\nonumber\\
&&-\frac{m_c\langle\bar{s}s\rangle}{\sqrt{2}\pi^2}
\int_{4m_c^2}^{s^0_{\eta_c}}ds \frac{\sqrt{\lambda(s,m_c^2,m_c^2)}}{s}   \exp\left(-\frac{s}{T^2} \right)\nonumber\\
&&+\frac{m_c\langle\bar{s}g_s\sigma G s\rangle}{3\sqrt{2}\pi^2T^2} \int_{4m_c^2}^{s^0_{\eta_c}}ds \frac{\sqrt{\lambda(s,m_c^2,m_c^2)}}{s} \exp\left(-\frac{s}{T^2} \right)\nonumber\\
&&+\frac{m_c\langle\bar{s}g_s\sigma G s\rangle}{4\sqrt{2}\pi^2}\int_{4m_c^2}^{s^0_{\eta_c}}ds
\frac{1}{\sqrt{s(s-4m_c^2)}}   \exp\left(-\frac{s}{T^2} \right)\, ,
\end{eqnarray}

\begin{eqnarray}
\Pi^{QCD}_{J/\psi f_0(980)\widetilde{A}V}(T^2)&=&\frac{3}{32\sqrt{2}\pi^4}\int_{4m_c^2}^{s^0_{J/\psi}}
ds	\int_{0}^{s^0_{f_0(980)}}du \frac{\sqrt{\lambda(s,m_c^2,m_c^2)}}{s}\,u\left(s+2m_c^2\right)  \exp\left(-\frac{s+u}{T^2} \right)\nonumber\\
&&+\frac{m_s\langle\bar{s}s\rangle}{2\sqrt{2}\pi^2} \int_{4m_c^2}^{s^0_{J/\psi}}ds \frac{\sqrt{\lambda(s,m_c^2,m_c^2)}}{s}   \left(s+2m_c^2\right) \exp\left(-\frac{s}{T^2} \right)\nonumber\\
&&-\frac{m_s m_c^2\langle\bar{s}g_s\sigma G s\rangle}{6\sqrt{2}\pi^2} \int_{4m_c^2}^{s^0_{J/\psi}}ds
\frac{1}{\sqrt{s(s-4m_c^2)}}  \exp\left(-\frac{s}{T^2} \right)\nonumber\\
&&+\frac{m_s\langle\bar{s}g_s\sigma G s\rangle}{4\sqrt{2}\pi^2 T^2} \int_{4m_c^2}^{s^0_{J/\psi}}ds
\frac{\sqrt{\lambda(s,m_c^2,m_c^2)}}{s} \left(s+2m_c^2\right) \exp\left(-\frac{s}{T^2} \right)\, ,
\end{eqnarray}

\subsection{The QCD side for the current $J_{\mu\nu}^{AA}$}
\begin{eqnarray}	
\Pi^{QCD}_{\bar{D}D AA}(T^2)&=&-\frac{m_c^3\langle\bar{q}g_s\sigma G q\rangle}{48\pi^2}\int_{m_c^2}^{s^0_{D}}ds \frac{1}{s^2}\, \exp\left(-\frac{s+m_c^2}{T^2} \right)\, ,
\end{eqnarray}

\begin{eqnarray}
\Pi^{QCD}_{\bar{D}^*D AA}(T^2)&=&-\frac{m_c^2\langle\bar{q}g_s\sigma G q\rangle}{96\pi^2}\int_{m_c^2}^{s^0_D}du \frac{1}{u^2} \exp\left(-\frac{u+m_c^2}{T^2} \right)\, ,
\end{eqnarray}

\begin{eqnarray}	
\Pi^{QCD}_{\bar{D}^*D^*AA}(T^2)&=&0\, ,
\end{eqnarray}

\begin{eqnarray}	
\Pi^{QCD}_{\bar{D}_0D^*AA}(T^2)&=&\frac{3m_c}{128\pi^4}
\int_{m_c^2}^{s^0_{D_0}}ds 	\int_{m_c^2}^{s^0_{D^*}}du \left(1-\frac{m_c^2}{s}\right)^2\left(1-\frac{m_c^2}{u}\right)^2 s \, \exp\left(-\frac{s+u}{T^2} \right) \nonumber\\
&&-\frac{\langle\bar{q}q\rangle}{16\pi^2}\int_{m_c^2}^{s^0_{D_0}}ds \left(1-\frac{m_c^2}{s}\right)^2 s \,\exp\left(-\frac{s+m_c^2}{T^2} \right)\nonumber\\
&&+\frac{m_c^2\langle\bar{q}q\rangle}{16\pi^2}\int_{m_c^2}^{s^0_{D^*}}du \left(1-\frac{m_c^2}{u}\right)^2 \,\exp\left(-\frac{u+m_c^2}{T^2} \right)\nonumber\\
&&+\frac{\langle\bar{q}g_s\sigma G q\rangle}{192\pi^2T^2}\left(4+\frac{3m_c^2}{T^2}\right) \int_{m_c^2}^{s^0_{D_0}}ds \left(1-\frac{m_c^2}{s}\right)^2 s \exp\left(-\frac{s+m_c^2}{T^2} \right)\nonumber\\
&&+\frac{m_c^2\langle\bar{q}g_s\sigma G q\rangle}{64\pi^2T^2} \left(2-\frac{m_c^2}{T^2}\right) \int_{m_c^2}^{s^0_{D^*}}du \left( 1-\frac{m_c^2}{u}\right)^2 \, \exp\left(-\frac{u+m_c^2}{T^2} \right)\nonumber\\
&&+\frac{m_c^2\langle\bar{q}g_s\sigma G q\rangle}{192\pi^2} \int_{m_c^2}^{s^0_{D_0}}ds \frac{1}{s} \exp\left(-\frac{s+m_c^2}{T^2} \right)\nonumber\\
&&+\frac{m_c^4\langle\bar{q}g_s\sigma G q\rangle}{192\pi^2}\int_{m_c^2}^{s^0_{D^*}}du \frac{1}{u^2}\, \exp\left(-\frac{u+m_c^2}{T^2} \right)\, ,
\end{eqnarray}

\begin{eqnarray}	
\Pi^{QCD}_{\bar{D}_1D AA}(T^2)&=&\frac{3m_c}{128\pi^4}\int_{m_c^2}^{s^0_{D_1}}ds
\int_{m_c^2}^{s^0_D}du \left(1-\frac{m_c^2}{s}\right)^2\left(1-\frac{m_c^2}{u}\right)^2 u \, \exp\left(-\frac{s+u}{T^2} \right) \nonumber\\
&&-\frac{m_c^2\langle\bar{q}q\rangle}{16\pi^2}\int_{m_c^2}^{s^0_{D_1}}ds \left(1-\frac{m_c^2}{s}\right)^2 \,\exp\left(-\frac{s+m_c^2}{T^2} \right)\nonumber\\
&&+\frac{\langle\bar{q}q\rangle}{16\pi^2}\int_{m_c^2}^{s^0_D}du \left(1-\frac{m_c^2}{u}\right)^2 u \,\exp\left(-\frac{u+m_c^2}{T^2} \right)\nonumber\\
&&-\frac{m_c^2\langle\bar{q}g_s\sigma G q\rangle}{64\pi^2T^2} \left(2-\frac{m_c^2}{T^2}\right) \int_{m_c^2}^{s^0_{D_1}}ds \left(1-\frac{m_c^2}{s}\right)^2 \exp\left(-\frac{s+m_c^2}{T^2} \right)\nonumber\\
&&-\frac{\langle\bar{q}g_s\sigma G q\rangle}{192\pi^2T^2} \left(4+\frac{3m_c^2}{T^2}\right) \int_{m_c^2}^{s^0_D}du \left( 1-\frac{m_c^2}{u}\right)^2 u \, \exp\left(-\frac{u+m_c^2}{T^2} \right)\nonumber\\
&&-\frac{m_c^2\langle\bar{q}g_s\sigma G q\rangle}{192\pi^2}\int_{m_c^2}^{s^0_D}du  \frac{1}{u} \, \exp\left(-\frac{u+m_c^2}{T^2} \right)	\nonumber\\
&&-\frac{m_c^4\langle\bar{q}g_s\sigma G q\rangle}{192\pi^2}\int_{m_c^2}^{s^0_{D_1}}ds \frac{1}{s^2}\, \exp\left(-\frac{s+m_c^2}{T^2} \right)\, ,
\end{eqnarray}

\begin{eqnarray}
\Pi^{QCD}_{\eta_c\omega AA}(T^2)
&=&-\frac{m_c^2\langle\bar{q}g_s\sigma G q\rangle}{12\sqrt{2}\pi^2}\int_{4m_c^2}^{s^0_{\eta_c}}ds
\frac{1}{\sqrt{s(s-4m_c^2)}}  \, \exp\left(-\frac{s}{T^2} \right)\, ,
\end{eqnarray}

\begin{eqnarray}
\Pi^{QCD}_{J/\psi\omega AA}(T^2)&=&0\, ,
\end{eqnarray}

\begin{eqnarray}
\Pi^{QCD}_{\chi_{c0}\omega AA}(T^2)&=&-\frac{\langle\bar{q}q\rangle}{4\sqrt{2}\pi^2} \int_{4m_c^2}^{s^0_{\chi_{c0}}}ds
\frac{\sqrt{\lambda(s,m_c^2,m_c^2)}}{s} \left(s-4m_c^2\right)  \exp\left(-\frac{s}{T^2} \right)\nonumber\\
&&+\frac{\langle\bar{q}g_s\sigma G q\rangle}{12\sqrt{2}\pi^2T^2} \int_{4m_c^2}^{s^0_{\chi_{c0}}}ds
\frac{\sqrt{\lambda(s,m_c^2,m_c^2)}}{s} \left(s-4m_c^2\right)  \exp\left(-\frac{s}{T^2} \right)\nonumber\\
&&+\frac{m_c^2\langle\bar{q}g_s\sigma G q\rangle}{24\sqrt{2}\pi^2} \int_{4m_c^2}^{s^0_{\chi_{c0}}}ds
\frac{1}{\sqrt{s(s-4m_c^2)}} \exp\left(-\frac{s}{T^2} \right)\, ,
\end{eqnarray}

\begin{eqnarray}
\Pi^{QCD}_{\chi_{c1}\omega AA}(T^2)&=& \frac{1}{24\sqrt{2}\pi^4}\int_{4m_c^2}^{s^0_{\chi_{c1}}}ds \int_{0}^{s^0_{\omega}}du \frac{\sqrt{\lambda(s,m_c^2,m_c^2)}}{s} u\left(7s-40m_c^2\right)  \exp\left(-\frac{s+u}{T^2} \right)\, ,	\nonumber\\
&&
\end{eqnarray}

\begin{eqnarray}
\Pi^{QCD}_{J/\psi f_0(500)AA}(T^2)&=&\frac{3m_c}{32\sqrt{2}\pi^4}\int_{4m_c^2}^{s^0_{J/\psi}}ds
\int_{0}^{s^0_{f_0(500)}}du \frac{\sqrt{\lambda(s,m_c^2,m_c^2)}}{s} u \exp\left(-\frac{s+u}{T^2} \right)\, ,
\end{eqnarray}

\subsection{The QCD side for the current $J_{\mu\nu}^{S\widetilde{V}}$}
\begin{eqnarray}
\Pi^{QCD}_{\bar{D}_s D_s S\widetilde{V}}(T^2)&=&\frac{3m_c}{128\sqrt{2}\pi^4} \int_{m_c^2}^{s^0_{D_s}}ds \int_{m_c^2}^{s^0_{D_s}}du \left(1-\frac{m_c^2}{s}\right) \left(1-\frac{m_c^2}{u}\right) \exp\left(-\frac{s+u}{T^2}\right)\nonumber\\
&&\frac{2m_s\left(su-m_c^4\right)+m_c\left(s-m_c^2\right)\left(u-m_c^2\right)}{su}\nonumber\\
&&-\frac{\langle\bar{s}s\rangle}{8\sqrt{2}\pi^2}\int_{m_c^2}^{s^0_{D_s}}ds \left(1-\frac{m_c^2}{s}\right) \frac{m_s\left(s+m_c^2\right)+m_c\left(s-m_c^2\right)}{s} \exp\left(-\frac{s+m_c^2}{T^2} \right)\nonumber\\
&&+\frac{m_c^2\langle\bar{s}g_s\sigma G  s\rangle}{32\sqrt{2}\pi^2 T^4} \int_{m_c^2}^{s^0_{D_s}}ds \left(1-\frac{m_c^2}{s}\right) \frac{m_s\left(s+m_c^2\right)+m_c\left(s-m_c^2\right)}{s} \exp\left(-\frac{s+m_c^2}{T^2} \right)\nonumber\\
&&-\frac{m_c\langle\bar{s}g_s\sigma G s\rangle}{96\sqrt{2}\pi^2}\int_{m_c^2}^{s^0_{D_s}}ds \frac{m_s m_c+2s-m_c^2}{s^2} \exp\left(-\frac{s+m_c^2}{T^2} \right)\, ,
\end{eqnarray}

\begin{eqnarray}
\Pi^{QCD}_{\bar{D}_s^*D_s S\widetilde{V}}(T^2)&=&
-\frac{m_c^4\langle\bar{s}g_s\sigma G s\rangle}{48\sqrt{2}\pi^2}\int_{m_c^2}^{s^0_{D_s^*}}ds \frac{1}{s^3} \exp\left(-\frac{s+m_c^2}{T^2} \right)\, ,
\end{eqnarray}

\begin{eqnarray}
\Pi^{QCD}_{\bar{D}_s^*D_s^* S\widetilde{V}}(T^2)&=&\frac{3m_c}{256\sqrt{2}\pi^4} \int_{m_c^2}^{s^0_{D_s^*}}ds \int_{m_c^2}^{s^0_{D_s^*}}du \left(1-\frac{m_c^2}{s}\right)^2 \frac{m_s\left(u^2-m_c^4\right)+m_c\left(u-m_c^2\right)^2}{u^2} \exp\left(-\frac{s+u}{T^2}\right)\nonumber\\
&&-\frac{\langle\bar{s}s\rangle}{32\sqrt{2}\pi^2}\int_{m_c^2}^{s^0_{D_s^*}}ds \frac{2m_s\left(s^2-m_c^4\right)+m_c\left(s-m_c^2\right)^2}{s^2} \exp\left(-\frac{s+m_c^2}{T^2} \right)\nonumber\\
&&+\frac{\langle\bar{s}g_s\sigma G s\rangle}{128\sqrt{2}\pi^2 T^2} \left(1+\frac{m_c^2}{T^2}\right) \int_{m_c^2}^{s^0_{D_s^*}}ds \frac{2m_s\left(s^2-m_c^4\right)+m_c\left(s-m_c^2\right)^2}{s^2} \exp\left(-\frac{s+m_c^2}{T^2} \right)\nonumber\\
&&+\frac{\langle\bar{s}g_s\sigma G s\rangle}{384\sqrt{2}\pi^2 T^2} \int_{m_c^2}^{s^0_{D_s^*}}ds \frac{m_s\left(s^2-m_c^4\right)+2m_c\left(s-m_c^2\right)^2}{s^2} \exp\left(-\frac{s+m_c^2}{T^2} \right)\nonumber\\
&&-\frac{m_c\langle\bar{s}g_s\sigma G s\rangle}{384\sqrt{2}\pi^2} \int_{m_c^2}^{s^0_{D_s^*}}ds \frac{1}{s} \left(1-\frac{m_c^2}{s}\right) \exp\left(-\frac{s+m_c^2}{T^2} \right)\nonumber\\
&&+\frac{m_s m_c^2\langle\bar{s}g_s\sigma G s\rangle}{384\sqrt{2}\pi^2} \int_{m_c^2}^{s^0_{D_s^*}}du \frac{1}{u^2} \exp\left(-\frac{u+m_c^2}{T^2} \right)\nonumber\\
&&-\frac{\langle\bar{s}g_s\sigma G s\rangle}{768\sqrt{2}\pi^2}\int_{m_c^2}^{s^0_{D_s^*}}ds \frac{2m_s\left(s^2+m_c^4\right)+3m_c^3\left(s-m_c^2\right)}{s^2\left(s-m_c^2\right)} \exp\left(-\frac{s+m_c^2}{T^2} \right)\nonumber\\
&&-\frac{m_c\langle\bar{s}g_s\sigma G s\rangle}{1536\sqrt{2}\pi^2} \int_{m_c^2}^{s^0_{D_s^*}}du \frac{2m_sm_c\left(3u-m_c^2\right)+2u^2-m_c^2 u+m_c^4}{u^2\left(u-m_c^2\right)} \exp\left(-\frac{u+m_c^2}{T^2} \right) \, ,
\end{eqnarray}

\begin{eqnarray}
\Pi^{QCD}_{\bar{D}_{s0}D_s^*S\widetilde{V}}(T^2)&=&\frac{1}{64\sqrt{2}\pi^4}
\int_{m_c^2}^{s^0_{D_{s0}}}ds	\int_{m_c^2}^{s^0_{D_s^*}}du \left(1-\frac{m_c^2}{s}\right)\left(1-\frac{m_c^2}{u}\right) \exp\left(-\frac{s+u}{T^2} \right) \nonumber\\
&&\frac{m_s\left[m_c^6+m_c^4(s-5u)+m_c^2u(7s-2u)-2su^2\right]+m_c\left(2u+m_c^2\right)\left(s-m_c^2\right)\left(u-m_c^2\right)} {su} \nonumber\\
&&+\frac{m_c\langle\bar{s}s\rangle}{8\sqrt{2}\pi^2}\int_{m_c^2}^{s^0_{D_{s0}}}ds \left(1-\frac{m_c^2}{s}\right) \frac{m_s\left(s+m_c^2\right)-m_c\left(s-m_c^2\right)}{s} \exp\left(-\frac{s+m_c^2}{T^2} \right)\nonumber\\
&&-\frac{\langle\bar{s}s\rangle}{24\sqrt{2}\pi^2}\int_{m_c^2}^{s^0_{D_s^*}}du \left(1-\frac{m_c^2}{u}\right) \frac{6m_sm_cu+\left(2u+m_c^2\right)\left(u-m_c^2\right)}{u} \,\exp\left(-\frac{u+m_c^2}{T^2} \right)\nonumber\\
&&+\frac{m_sm_c^3\langle\bar{s}s\rangle}{16\sqrt{2}\pi^2T^2}\int_{m_c^2}^{s^0_{D_{s0}}}ds \left(1-\frac{m_c^2}{s}\right)^2 \exp\left(-\frac{s+m_c^2}{T^2} \right)\nonumber\\
&&-\frac{m_sm_c\langle\bar{s}s\rangle}{48\sqrt{2}\pi^2T^2}\int_{m_c^2}^{s^0_{D_s^*}}du \left(1-\frac{m_c^2}{u}\right)^2 \left(2u+m_c^2\right) \,\exp\left(-\frac{u+m_c^2}{T^2} \right)\nonumber\\
&&-\frac{m_c^3\langle\bar{s}g_s\sigma G s\rangle}{32\sqrt{2}\pi^2T^4}\int_{m_c^2}^{s^0_{D_{s0}}}ds \left(1-\frac{m_c^2}{s}\right) \frac{m_s\left(s+m_c^2\right)-m_c\left(s-m_c^2\right)}{s} \exp\left(-\frac{s+m_c^2}{T^2} \right)\nonumber\\
&&-\frac{m_c^2\langle\bar{s}g_s\sigma G s\rangle}{96\sqrt{2}\pi^2T^4} \int_{m_c^2}^{s^0_{D_s^*}}du \left(1-\frac{m_c^2}{u}\right) \frac{6m_sm_cu+\left(2u+m_c^2\right)\left(u-m_c^2\right)}{u} \exp\left(-\frac{u+m_c^2}{T^2} \right)\nonumber\\
&&+\frac{m_sm_c\langle\bar{s}g_s\sigma G s\rangle}{96\sqrt{2}\pi^2T^2} \left(1+\frac{m_c^2}{T^2}-\frac{m_c^4}{T^4}\right) \int_{m_c^2}^{s^0_{D_{s0}}}ds \left(1-\frac{m_c^2}{s}\right)^2 \exp\left(-\frac{s+m_c^2}{T^2} \right)\nonumber\\
&&+\frac{m_sm_c^3\langle\bar{s}g_s\sigma G s\rangle}{288\sqrt{2}\pi^2T^6} \int_{m_c^2}^{s^0_{D_s^*}}du \left(1-\frac{m_c^2}{u}\right)^2 \left(2u+m_c^2\right) \exp\left(-\frac{u+m_c^2}{T^2} \right)\nonumber\\
&&+\frac{m_c^2\langle\bar{s}g_s\sigma G s\rangle}{96\sqrt{2}\pi^2} \int_{m_c^2}^{s^0_{D_{s0}}}ds \frac{m_sm_c-2s+m_c^2}{s^2} \exp\left(-\frac{s+m_c^2}{T^2} \right)\nonumber\\
&&+\frac{m_c\langle\bar{s}g_s\sigma G s\rangle}{96\sqrt{2}\pi^2} \int_{m_c^2}^{s^0_{D_s^*}}du \frac{2m_s u+m_c^3}{u^2} \exp\left(-\frac{u+m_c^2}{T^2} \right)\, ,
\end{eqnarray}

\begin{eqnarray}
\Pi^{QCD}_{J/\psi\phi S\widetilde{V}}(T^2)&=&\frac{9m_sm_c}{256\sqrt{2}\pi^4}\int_{4m_c^2}^{s^0_{J/\psi}}ds
\int_{0}^{s^0_\phi}du \,\frac{\sqrt{\lambda(s,m_c^2,m_c^2)}}{s} \exp\left(-\frac{s+u}{T^2} \right)\nonumber\\
&&-\frac{3m_c\langle\bar{s}s\rangle}{32\sqrt{2}\pi^2} \int_{4m_c^2}^{s^0_{J/\psi}}ds \frac{\sqrt{\lambda(s,m_c^2,m_c^2)}}{s} \exp\left(-\frac{s}{T^2} \right)\nonumber\\
&&+\frac{4m_c\langle\bar{s}g_s\sigma G s\rangle}{128\sqrt{2}\pi^2T^2}\int_{4m_c^2}^{s^0_{J/\psi}}ds
\frac{\sqrt{\lambda(s,m_c^2,m_c^2)}}{s} \exp\left(-\frac{s}{T^2} \right)\nonumber\\
&&-\frac{5m_c\langle\bar{s}g_s\sigma G s\rangle}{384\sqrt{2}\pi^2}\int_{4m_c^2}^{s^0_{J/\psi}}ds
\frac{1}{\sqrt{s(s-4m_c^2)}}  \exp\left(-\frac{s}{T^2} \right)\, ,
\end{eqnarray}

\begin{eqnarray}
\Pi^{QCD}_{\bar{D}_{s1}D_s S\widetilde{V}}(T^2)&=&\frac{1}{64\sqrt{2}\pi^4} \int_{m_c^2}^{s^0_{D_{s1}}}ds	 \int_{m_c^2}^{s^0_{D_s}}du \left(1-\frac{m_c^2}{s}\right)\left(1-\frac{m_c^2}{u}\right) \exp\left(-\frac{s+u}{T^2} \right) \nonumber\\
&&\frac{m_c\left(2s+m_c^2\right)\left(s-m_c^2\right)\left(u-m_c^2\right)-m_s\left[m_c^6+m_c^4(u-5s)+m_c^2s(7u-2s)-2s^2u\right]} {su} \nonumber\\
&&+\frac{\langle\bar{s}s\rangle}{24\sqrt{2}\pi^2}\int_{m_c^2}^{s^0_{D_{s1}}}ds \left(1-\frac{m_c^2}{s}\right) \frac{6m_sm_cs-\left(2s+m_c^2\right)\left(s-m_c^2\right)}{s} \,\exp\left(-\frac{s+m_c^2}{T^2} \right)\nonumber\\
&&+\frac{m_c\langle\bar{s}s\rangle}{8\sqrt{2}\pi^2}\int_{m_c^2}^{s^0_{D_s}}du \left(1-\frac{m_c^2}{u}\right) \frac{m_s\left(u+m_c^2\right)+m_c\left(u-m_c^2\right)}{u} \exp\left(-\frac{u+m_c^2}{T^2} \right)\nonumber\\
&&+\frac{m_sm_c\langle\bar{s}s\rangle}{48\sqrt{2}\pi^2T^2}\int_{m_c^2}^{s^0_{D_{s1}}}ds \left(1-\frac{m_c^2}{s}\right)^2 \left(2s+m_c^2\right) \,\exp\left(-\frac{s+m_c^2}{T^2} \right)\nonumber\\
&&-\frac{m_sm_c^3\langle\bar{s}s\rangle}{16\sqrt{2}\pi^2T^2}\int_{m_c^2}^{s^0_{D_s}}du \left(1-\frac{m_c^2}{u}\right)^2 \exp\left(-\frac{u+m_c^2}{T^2} \right)\nonumber\\
&&-\frac{m_c^2\langle\bar{s}g_s\sigma G s\rangle}{96\sqrt{2}\pi^2T^4} \int_{m_c^2}^{s^0_{D_{s1}}}ds \left(1-\frac{m_c^2}{s}\right) \frac{6m_sm_c s-\left(2s+m_c^2\right)\left(s-m_c^2\right)}{s} \exp\left(-\frac{s+m_c^2}{T^2} \right)\nonumber\\
&&-\frac{m_c^3\langle\bar{s}g_s\sigma G s\rangle}{32\sqrt{2}\pi^2T^4}\int_{m_c^2}^{s^0_{D_s}}du \left(1-\frac{m_c^2}{u}\right) \frac{m_s\left(u+m_c^2\right)+m_c\left(u-m_c^2\right)}{u} \exp\left(-\frac{u+m_c^2}{T^2} \right)\nonumber\\
&&-\frac{m_sm_c^3\langle\bar{s}g_s\sigma G s\rangle}{288\sqrt{2}\pi^2T^6} \int_{m_c^2}^{s^0_{D_{s1}}}ds \left(1-\frac{m_c^2}{s}\right)^2 \left(2s+m_c^2\right) \exp\left(-\frac{s+m_c^2}{T^2} \right)\nonumber\\
&&-\frac{m_sm_c\langle\bar{s}g_s\sigma G s\rangle}{96\sqrt{2}\pi^2T^2} \left(1+\frac{m_c^2}{T^2}-\frac{m_c^4}{T^4}\right) \int_{m_c^2}^{s^0_{D_s}}du \left(1-\frac{m_c^2}{u}\right)^2 \exp\left(-\frac{u+m_c^2}{T^2} \right)\nonumber\\
&&+\frac{m_c\langle\bar{s}g_s\sigma G s\rangle}{96\sqrt{2}\pi^2} \int_{m_c^2}^{s^0_{D_{s1}}}ds \frac{2m_s s-m_c^3}{s^2} \exp\left(-\frac{s+m_c^2}{T^2} \right)\nonumber\\
&&+\frac{m_c^2\langle\bar{s}g_s\sigma G s\rangle}{96\sqrt{2}\pi^2} \int_{m_c^2}^{s^0_{D_s}}du \frac{m_sm_c+2u-m_c^2}{u^2} \exp\left(-\frac{u+m_c^2}{T^2} \right)\, ,
\end{eqnarray}

\begin{eqnarray}
\Pi^{QCD}_{\eta_c\phi S\widetilde{V}}(T^2)&=&\frac{m_c}{16\sqrt{2}\pi^4}\int_{4m_c^2}^{s^0_{\eta_c}}ds
\int_{0}^{s^0_\phi}du \,u\,\frac{\sqrt{\lambda(s,m_c^2,m_c^2)}}{s} \exp\left(-\frac{s+u}{T^2} \right)\nonumber\\
&&-\frac{m_sm_c\langle\bar{s}s\rangle}{2\sqrt{2}\pi^2}\int_{m_c^2}^{s^0_{\eta_c}}ds \,\frac{\sqrt{\lambda(s,m_c^2,m_c^2)}}{s} \,\exp\left(-\frac{s}{T^2} \right)\nonumber\\
&&-\frac{m_sm_c\langle\bar{s}g_s\sigma G s\rangle}{24\sqrt{2}\pi^2}\int_{m_c^2}^{s^0_{\eta_c}}ds \,\frac{1}{\sqrt{s\left(s-4m_c^2\right)}} \,\exp\left(-\frac{s}{T^2} \right)\, ,
\end{eqnarray}

\begin{eqnarray}
\Pi^{QCD}_{\chi_{c1}\phi S\widetilde{V}}(T^2)&=&\frac{1}{48\sqrt{2}\pi^4}\int_{4m_c^2}^{s^0_{\chi_{c1}}}ds
\int_{0}^{s^0_\phi}du \frac{\sqrt{\lambda(s,m_c^2,m_c^2)}}{s}\,u\left(s-4m_c^2\right)   \exp\left(-\frac{s+u}{T^2} \right)\nonumber\\
&&-\frac{m_s\langle\bar{s}s\rangle}{6\sqrt{2}\pi^2} \int_{4m_c^2}^{s^0_{\chi_{c1}}}ds \frac{\sqrt{\lambda(s,m_c^2,m_c^2)}}{s} \left(s-4m_c^2\right)  \exp\left(-\frac{s}{T^2} \right)\nonumber\\
&&+\frac{m_sm_c^2\langle\bar{s}g_s\sigma G s\rangle}{24\sqrt{2}\pi^2}\int_{4m_c^2}^{s^0_{\chi_{c1}}}ds
\frac{1}{\sqrt{s(s-4m_c^2)}}   \exp\left(-\frac{s}{T^2} \right)\, ,
\end{eqnarray}

\begin{eqnarray}
\Pi^{QCD}_{\chi_{c0}\phi S\widetilde{V}}(T^2)&=&\frac{3m_s}{32\sqrt{2}\pi^4}\int_{4m_c^2}^{s^0_{\chi_{c0}}}ds
\int_{0}^{s^0_\phi}du \,\frac{\sqrt{\lambda(s,m_c^2,m_c^2)}}{s} \left(s-4m_c^2\right) \exp\left(-\frac{s+u}{T^2} \right)\nonumber\\
&&-\frac{\langle\bar{s}s\rangle}{4\sqrt{2}\pi^2} \int_{4m_c^2}^{s^0_{\chi_{c0}}}ds \frac{\sqrt{\lambda(s,m_c^2,m_c^2)}}{s} \left(s-4m_c^2\right)  \exp\left(-\frac{s}{T^2} \right)\nonumber\\
&&+\frac{\langle\bar{s}g_s\sigma G s\rangle}{12\sqrt{2}\pi^2T^2}\int_{4m_c^2}^{s^0_{\chi_{c0}}}ds
\frac{\sqrt{\lambda(s,m_c^2,m_c^2)}}{s}\left(s-4m_c^2 \right)   \exp\left(-\frac{s}{T^2} \right)\nonumber\\
&&-\frac{\langle\bar{s}g_s\sigma G s\rangle}{24\sqrt{2}\pi^2}\int_{4m_c^2}^{s^0_{\chi_{c0}}}ds
\frac{s-5m_c^2}{\sqrt{s(s-4m_c^2)}}  \exp\left(-\frac{s}{T^2} \right)\, ,
\end{eqnarray}

\begin{eqnarray}
\Pi^{QCD}_{J/\psi f_0(980) S\widetilde{V}}(T^2)&=&\frac{3m_c}{32\sqrt{2}\pi^4}\int_{4m_c^2}^{s^0_{J/\psi}}ds
\int_{0}^{s^0_{f_0(980)}}du \,u\frac{\sqrt{\lambda(s,m_c^2,m_c^2)}}{s} \exp\left(-\frac{s+u}{T^2} \right)\nonumber\\
&&+\frac{m_sm_c\langle\bar{s}s\rangle}{2\sqrt{2}\pi^2} \int_{4m_c^2}^{s^0_{J/\psi}}ds \frac{\sqrt{\lambda(s,m_c^2,m_c^2)}}{s} \exp\left(-\frac{s}{T^2} \right)\nonumber\\
&&+\frac{m_sm_c\langle\bar{s}g_s\sigma G s\rangle}{12\sqrt{2}\pi^2}\int_{4m_c^2}^{s^0_{J/\psi}}ds
\frac{s-m_c^2}{s\sqrt{s(s-4m_c^2)}} \exp\left(-\frac{s}{T^2} \right)\nonumber\\
&&+\frac{m_sm_c\langle\bar{s}g_s\sigma G s\rangle}{4\sqrt{2}\pi^2T^2} \int_{4m_c^2}^{s^0_{J/\psi}}ds \frac{\sqrt{\lambda(s,m_c^2,m_c^2)}}{s} \exp\left(-\frac{s}{T^2} \right)\, ,
\end{eqnarray}
$\lambda(a,b,c)=a^2+b^2+c^2-2ab-2bc-2ca$.

\section*{Acknowledgements}
This  work is supported by National Natural Science Foundation, Grant Number  12575083.


\begin{thebibliography}{99}

\bibitem{PDG}  S. Navas et al, Phys. Rev. {\bf D110} (2024) 030001.

\bibitem{Belle-Y4660-2007} X. L. Wang et al., Phys. Rev. Lett. {\bf 99} (2007) 142002.

\bibitem{BaBar-Y4660-2012} J. P. Lees et al, Phys. Rev. {\bf D89}   (2014) 111103.

\bibitem{Belle-Y4660-2014} X. L. Wang et al, Phys. Rev. {\bf D91}   (2015) 112007.

\bibitem{BESIII-Y4660-2021} M. Ablikim et al, Phys. Rev. {\bf D104}   (2021) 052012.

\bibitem{BESIII-Y4660-2023} M. Ablikim et al, Phys. Rev. Lett. {\bf 130} (2023) 121901.

\bibitem{Belle-Y4630-2008} G. Pakhlova  et al,  Phys. Rev. Lett. {\bf 101} (2008) 172001.

\bibitem{FKGuo-4630} F. K. Guo, J. Haidenbauer, C. Hanhart and U. G. Meissner, Phys. Rev. {\bf D82} (2010) 094008.

\bibitem{Polosa-4660} G. Cotugno, R. Faccini , A. D. Polosa and C. Sabelli, Phys. Rev. Lett. {\bf 104} (2010) 132005.

\bibitem{DVBugg-4630} D. V. Bugg, J. Phys. {\bf G36} (2009) 075002.


\bibitem{BESIII-Y4630-2023} M. Ablikim et al, Phys. Rev. Lett. {\bf 131} (2023) 191901.

\bibitem{Belle-Y4660-2019} S. Jia et al, Phys. Rev. {\bf D100}   (2019) 111103.

\bibitem{Belle-Y4660-2020} S. Jia et al, Phys. Rev. {\bf D101}   (2020) 091101.

\bibitem{FKGuo-4660} F. K. Guo, C. Hanhart and U. G. Meissner, Phys. Lett. {\bf B665} (2008) 26.

\bibitem{Nielsen-4660-mole} R. M. Albuquerque, M. Nielsen and R. Rodrigues da Silva, Phys. Rev. {\bf D84} (2011) 116004.

\bibitem{Wang-CTP-4660} Z. G. Wang and X. H. Zhang, Commun. Theor. Phys. {\bf 54} (2010) 323.

\bibitem{Nielsen-4660} R. M. Albuquerque and M. Nielsen, Nucl. Phys. {\bf A815} (2009) 53; Erratum: [Nucl. Phys. {\bf A857} (2011) 48].

\bibitem{Ebert-4660} D. Ebert, R. N. Faustov and V. O. Galkin, Eur. Phys. J. {\bf C58} (2008) 399.

\bibitem{ChenZhu} W. Chen and S. L. Zhu, Phys. Rev. {\bf D83} (2011) 034010.

\bibitem{ZhangHuang-PRD} J. R. Zhang and M. Q. Huang, Phys. Rev. {\bf D83} (2011) 036005.

\bibitem{ESantopinto-PRD} M. N. Anwar, J. Ferretti and E. Santopinto, Phys. Rev. {\bf D98} (2018) 094015.

\bibitem{Azizi-4660} H. Sundu, S. S. Agaev and K. Azizi, Phys. Rev. {\bf D98} (2018) 054021.

\bibitem{Maiani-4660} L. Maiani, F. Piccinini, A. D. Polosa and V. Riquer, Phys. Rev. {\bf D89} (2014) 114010.

\bibitem{WangY4360Y4660-1803} Z. G. Wang, Eur. Phys. J. {\bf C78} (2018)  518.

\bibitem{Wang-tetra-formula}  Z. G. Wang, Eur. Phys. J. {\bf C74} (2014)  2874.

\bibitem{WangEPJC-1601} Z. G. Wang, Eur. Phys. J. {\bf C76} (2016)  387.

\bibitem{WZG-NPB-cucd-Vector} Z. G. Wang, Nucl. Phys. {\bf B973} (2021) 115592.

\bibitem{WZG-NPB-cscs-Vector} Z. G. Wang, Nucl. Phys. {\bf B1002} (2024) 116514.

\bibitem{NLi-CPC} N. Li, H. Z. He, W. Liang, Q. F. L{\"u}, D. Y. Chen and Y. B. Dong, Chin. Phys. {\bf C47} (2023) 063102.



\bibitem{Hadro-Charm} S. Dubynskiy and M. B. Voloshin, Phys. Lett. {\bf B666} (2008) 344.

\bibitem{MLYan-4660} G. J. Ding, J. J. Zhu and M. L. Yan, Phys. Rev. {\bf D77} (2008) 014033.

\bibitem{BSZou-PRD} M. N. Anwar, Y. Lu and B. S. Zou, Phys. Rev. {\bf D95} (2017) 114031.

\bibitem{XHZhong-PRD} Q. Deng, R. H. Ni, Q. Li and X. H. Zhong, Phys. Rev. {\bf D110} (2024) 056034.

\bibitem{LuZhongZhu-4660} L. Y. Xiao, X. Z. Weng, Q. F. L{\"u}, X. H. Zhong and S. L. Zhu, Eur. Phys. J. {\bf C78} (2018) 605.

\bibitem{KTChao-4660} B. Q. Li and K. T. Chao, Phys. Rev. {\bf D79} (2009) 094004.

\bibitem{XLiu-4660} J. Z. Wang, R. Q. Qian, X. Liu and T. Matsuki, Phys. Rev. {\bf D101} (2020) 034001.

\bibitem{CFQiao-4660} B. D. Wan, L. Tang and C. F. Qiao, Eur. Phys. J. {\bf C80} (2020)  121.

\bibitem{CFQiao-4260} C. F. Qiao, Phys. Lett. {\bf B639} (2006) 263.

\bibitem{CFQiao-baryonium} C. F. Qiao, J. Phys.  {\bf G35} (2008) 075008.

\bibitem{JHe-baryonium} L. Q. Song, D. Song, J. T. Zhu and J. He, Phys. Lett. {\bf B835} (2022) 137586.

\bibitem{XLiu-baryonium} Z. Liu, H. Xu, Z. W. Liu and X. Liu, arXiv:2603.09404 [hep-ph].


\bibitem{WZG-baryonium} X. W. Wang, Z. G. Wang and G. L. Yu, Eur. Phys. J. {\bf A57} (2021) 275.


\bibitem{WZG-review} Z. G. Wang, Front. Phys. {\bf 21} (2026) 016300.

\bibitem{WangZG-landau-PRD} Z.~G.~Wang, Phys. Rev. {\bf D101} (2020)  074011.

\bibitem{Nielsen-JPG-Review}
R. M. Albuquerque, J. M. Dias, K. P. Khemchandani, A. M. Torres, F. S. Navarra, M. Nielsen and C. M. Zanetti, J. Phys. {\bf G46} (2019)  093002.


\bibitem{BaBar-Y4260-2005} B. Aubert et al,  Phys. Rev. Lett. {\bf 95} (2005) 142001.

\bibitem{BESIII-Y4260pis23-2017}  M. Ablikim et al, Phys. Rev. {\bf D96}  (2017) 032004.

\bibitem{BESIII-Y4260pis23-2021}  M. Ablikim et al,	Phys. Rev. {\bf D 104} (2021) 052012.



\bibitem{WZG-CPC-cucd-Pwave} Z. G. Wang, Chin. Phys. {\bf C48} (2024) 103103.

\bibitem{WZG-4260-Pwave} Z. G. Wang, Eur. Phys. J. {\bf C78} (2018) 933.

\bibitem{WZG-vector-Pwave} Z. G. Wang, Eur. Phys. J. {\bf C79} (2019) 29.

\bibitem{WZG-Y4660-Decay} Z. G. Wang, Eur. Phys. J. {\bf C79} (2019)  184.



\bibitem{SVZ79}  M. A. Shifman, A. I. Vainshtein and V. I. Zakharov, Nucl. Phys. {\bf B147} (1979) 385; Nucl. Phys. {\bf B147} (1979) 448.

\bibitem{Reinders85} L. J. Reinders, H. Rubinstein and S. Yazaki, Phys. Rept. {\bf 127} (1985) 1.


\bibitem{WZG-ZJX-Zc-Decay}   Z. G. Wang and J. X. Zhang,  Eur. Phys. J. {\bf C78} (2018)  14.


\bibitem{WZG-YXS-cccc-AAPPS} Z. G. Wang and X. S. Yang, AAPPS Bull. {\bf 34} (2024) 5.

\bibitem{WZG-YXS-cccc-CPC} X. S. Yang and Z. G. Wang, Chin. Phys. {\bf C49} (2025) 063108.

\bibitem{Narison-mix} S. Narison and R. Tarrach, Phys. Lett. {\bf 125 B} (1983) 217.

\bibitem{WangZG-HuangT-PRD-2014} Z. G. Wang and T. Huang, Phys. Rev. {\bf D89} (2014) 054019.

\bibitem{WangZG-HuangT-EPJC-2014}  Z. G. Wang and T. Huang,  Eur. Phys. J.
{\bf C74} (2014)  2891.

\bibitem{WangZG-EPJC-2014} Z. G. Wang, Eur. Phys. J. {\bf C74} (2014)  2963.

\bibitem{WangZG-HuangT-NPA-2014} Z. G. Wang and T. Huang, Nucl. Phys. {\bf A930} (2014) 63.

\bibitem{Narison-SB-2024}
R. M. Albuquerque, S. Narison, A. Rabemananjara and D. Rabetiarivony, Nucl. Part. Phys. Proc. {\bf 343} (2024) 61.

\bibitem{Narison-SB-2023}
R. M. Albuquerque, S. Narison and D. Rabetiarivony, Nucl. Part. Phys. Proc. {\bf 324-329} (2023) 54.

\bibitem{KTChao-JHEP-2024} R. H. Wua, C. Y. Wang, C. Meng, Y. Q. Ma and K. T. Chao,  JHEP {\bf 06} (2024) 216.


\bibitem{Colangelo-Review}  P. Colangelo and A. Khodjamirian, hep-ph/0010175.

\bibitem{Becirevic} D. Becirevic, G. Duplancic, B. Klajn, B. Melic and F. Sanfilippo,  Nucl. Phys. {\bf B883} (2014) 306.

\bibitem{Charmonium-PRT} V. A. Novikov, L. B. Okun, M. A. Shifman, A. I. Vainshtein, M. B. Voloshin and V. I. Zakharov, Phys. Rept. {\bf 41} (1978) 1.

\bibitem{WZG-heavy-decay} Z. G. Wang, Eur. Phys. J. {\bf C75} (2015) 427.

\bibitem{PBall-decay-Kv} P. Ball and G. W. Jones,  JHEP {\bf 0703} (2007) 069.

\bibitem{ChengHY-2022} H. Y. Cheng, C. W. Chiang and Z. Q. Zhang, Phys. Rev. {\bf D105} (2022) 033006.

\bibitem{WZG-EPJC-scalar} Z. G. Wang,   Eur. Phys. J. {\bf C76} (2016)  427.

\bibitem{Wang-f980-decay}  Z. G. Wang, W. M. Yang and S. L. Wan, Eur. Phys. J. {\bf C37} (2004)  223.

\end{thebibliography}
\end{document}